\documentclass[10pt,aps,prb,superscriptaddress,floatfix,showpacs,longbibliography,notitlepage]{revtex4-1}
\usepackage{graphicx}
\usepackage{amsfonts}
\usepackage{amssymb}
\usepackage{amsmath}
\usepackage{txfonts}
\usepackage{lipsum}
\usepackage{color}
\usepackage{wasysym}
\usepackage{hyperref}
\usepackage{bbold}
\usepackage{etoolbox} 
\graphicspath{{figures/}}
\usepackage{feynmp-auto}
\usepackage{ulem}

\usepackage{mathtools}
\usepackage{multirow}
\usepackage{hhline}

\usepackage{tikz}

\usepackage{mathrsfs} 

\DeclareMathOperator{\sgn}{sgn}

\DeclareMathOperator{\Imag}{Im}

\newcommand{\kv}{\mathbf{k}}

\usepackage{letltxmacro}
\LetLtxMacro{\oldsqrt}{\sqrt}
\renewcommand{\sqrt}[2][\mkern8mu]{\mkern-6mu\mathop{}\oldsqrt[#1]{#2}}

\begin{document}
\title{
Local Plaquette Physics as Key Ingredient of High-Temperature Superconductivity in Cuprates
}
\author{Michael Danilov}
\affiliation{Institute of Theoretical Physics, University of Hamburg, 20355 Hamburg, Germany}

 \author{Erik G. C. P. van Loon}
 \email{erik.van\_loon@teorfys.lu.se}
 \affiliation{Institut f{\"u}r Theoretische Physik, Universit{\"a}t Bremen, Otto-Hahn-Allee 1, 28359 Bremen, Germany}
 \affiliation{Bremen Center for Computational Materials Science, Universit{\"a}t Bremen, Am Fallturm 1a, 28359 Bremen, Germany}
 \affiliation{Department of Physics, Lund University, Professorsgatan 1, 223 63, Lund, Sweden}

\author{Sergey Brener}
\affiliation{Institute of Theoretical Physics, University of Hamburg, 20355 Hamburg, Germany}
\affiliation{The Hamburg Centre for Ultrafast Imaging, Luruper Chaussee 149, 22761 Hamburg, Germany}

\author{Sergei Iskakov}
\affiliation{Department of Physics, University of Michigan, Ann Arbor, Michigan 48109, USA}
\affiliation{Theoretical Physics and Applied Mathematics Department, Ural Federal University, Mira Str.19, 620002, Yekaterinburg, Russia}

\author{Mikhail I. Katsnelson}
\affiliation{Radboud University, Institute for Molecules and Materials, 6525AJ Nijmegen, The Netherlands}
\affiliation{Theoretical Physics and Applied Mathematics Department, Ural Federal University, Mira Str.19, 620002, Yekaterinburg, Russia}

\author{Alexander I. Lichtenstein}
\email{alichten@physnet.uni-hamburg.de}
\affiliation{Institute of Theoretical Physics, University of Hamburg, 20355 Hamburg, Germany}
\affiliation{The Hamburg Centre for Ultrafast Imaging, Luruper Chaussee 149, 22761 Hamburg, Germany}
\affiliation{Theoretical Physics and Applied Mathematics Department, Ural Federal University, Mira Str.19, 620002, Yekaterinburg, Russia}

\begin{abstract}
A major pathway towards understanding complex systems is given by exactly solvable reference systems that contain the essential physics of the system.
For the $t-t'-U$ Hubbard model, the four-site plaquette is known to have a 
point in the $U-\mu$ space where states with electron occupations $N=2, 3, 4$ per plaquette are degenerate [Phys. Rev. B {\bf 94}, 125133 (2016)]. 
We show that such a critical point in the lattice causes an instability in the particle-particle singlet d-wave channel and manifests some of the essential elements of the cuprate superconductivity. For this purpose we design an efficient superperturbation theory -- based on the dual fermion approach -- with the critical plaquette as the reference system.
Thus, the perturbation theory already contains the relevant d-wave fluctuations from the beginning via the two-particle correlations of the plaquette. We find that d-wave superconductivity remains a leading instability channel under reasonably broad range of parameters. The next-nearest-neighbour hopping $t'$ is shown to play a crucial role in a formation of strongly bound electronic bipolarons whose coherence at lower temperature results in superconductivity. The physics of the pseudogap within the developed picture is also discussed. 
\end{abstract}

\maketitle

\section{Introduction}

After 35 years since the discovery of the high-temperature superconductivity~\cite{Muller_HTSC}, there is still no consensus on the nature of the mechanism of d-wave pairing in cuprates~\cite{HTSC_view,Scalapino_RMP,Scalapino_bound,Keimer_rev,Deveraux_Tp,Zhang_noSC}.  Nevertheless, new experimental findings clearly point to the existence of a quantum critical point around a hole doping of $\delta\approx0.24$~\cite{Taillefer_rev,Hussey20,Hussey21}. This concentration separates the exotic bad-metal state for smaller doping from Fermi-liquid behaviour for larger hole concentration with ``normal'' Fermi-surface described, at least qualitatively, by conventional density-functional theory~\cite{OKA}. Moreover, the carrier density obtained from Hall effect measurements in large-doping regime is equal to its nominal value $n_H\approx1-\delta$ while for smaller doping the bad-metal behaviour appears with Fermi-arcs, ``enigmatic pseudogap phase'' and ($n_H\approx\delta$) at high temperature~\cite{Taillefer_rev,Taillefer_FS}. Recent investigations of highly overdoped cuprates show that this ``strange metal phase'' is located around $\delta_c\approx0.24$ point~\cite{Hussey20}. For hole concentrations less than $\delta_c$ superconducting pairs come entirely from the region of incoherent electrons at the antinode region ($X$-point) of the Brillouin zone (Planckian dissipators)~\cite{Hussey20}. 

Electronic specific heat measurements for many different cuprate superconductors reveal in a normal phase a huge peak in the electron Density of States (DOS) at the Fermi energy at $\delta_c\approx0.24$ with a strong evidence of the presence of a Quantum Critical Point (QCP) at this hole doping~\cite{Taillefer_rev}. 
Taking into account this critical concentration as a fingerprint of high-Tc materials, we formulate a simple strong coupling theory of the electronic instability based on cluster dual-fermion superperturbation theory~\cite{DF_Rev,HH_CDF}.

First-principle electronic structure calculations~\cite{OKA} suggest that a single-band tight-binding model with next nearest neighbour (NNN) hopping and on-site Coulomb interaction, the so-called ``$t-t'-U$'' Hubbard model, has all ingredients to describe high-Tc phenomena.
Moreover, the case of $t'/t=-0.15$ corresponds to the LSCO-cuprate family while one expects $t'/t=-0.3$
 to describe cuprate families with higher $T_c$ such as e.g. YBCO and Tl2201 ~\cite{Pavarini}.
We developed an efficient second-order perturbation theory starting from a 2$\times$2 plaquette, where $\delta=0.25$ corresponds to a highly degenerate point for $U/t\approx 6$ ~\cite{Harland16}.
In a close analogy with the Kondo model, where the degeneracy of the two spin states of a magnetic impurity plays a crucial role in the anomalous low-energy properties, the special properties of the degenerate states of the plaquette can 
reveal the nature of the anomalous behavior of the interacting Hubbard model on a two-dimensional lattice.

The first attempt to discuss the plaquette physics as the main ingredient of the high-Tc theory was done with the cluster dynamical mean-field theory (DMFT) scheme~\cite{dwCDMFT}, and later Altman and Auerbach analytically explained the importance of plaquette two-hole states with $d_{x^2-y^2}$ symmetry~\cite{Auerbach}. Nevertheless they did not consider the possibility of a degenerate ground state of the plaquette~\cite{Harland16} with a correspondingly divergent perturbation series similar to the perturbative theory in Kondo problem~\cite{hewson}. In some sense, the degeneracy of the ground states with $N=2,3,4$ electrons per plaquette in the critical point plays the same role as the degeneracy between spin-up and spin-down states in the conventional Kondo effect and is crucially important for the pseudogap formation~\cite{Harland16}.
If we treat the Kondo problem in dual perturbation from the atomic limit~\cite{Krivenko} then the local four-point vertex is divergent at low temperarure, while the Green's function is finite. In the case of degenerate plaquette both Green's function for reference system and vertex are divergent
for low temperature.

\begin{figure}[t!]
 \centering
\includegraphics[width=0.6\textwidth]{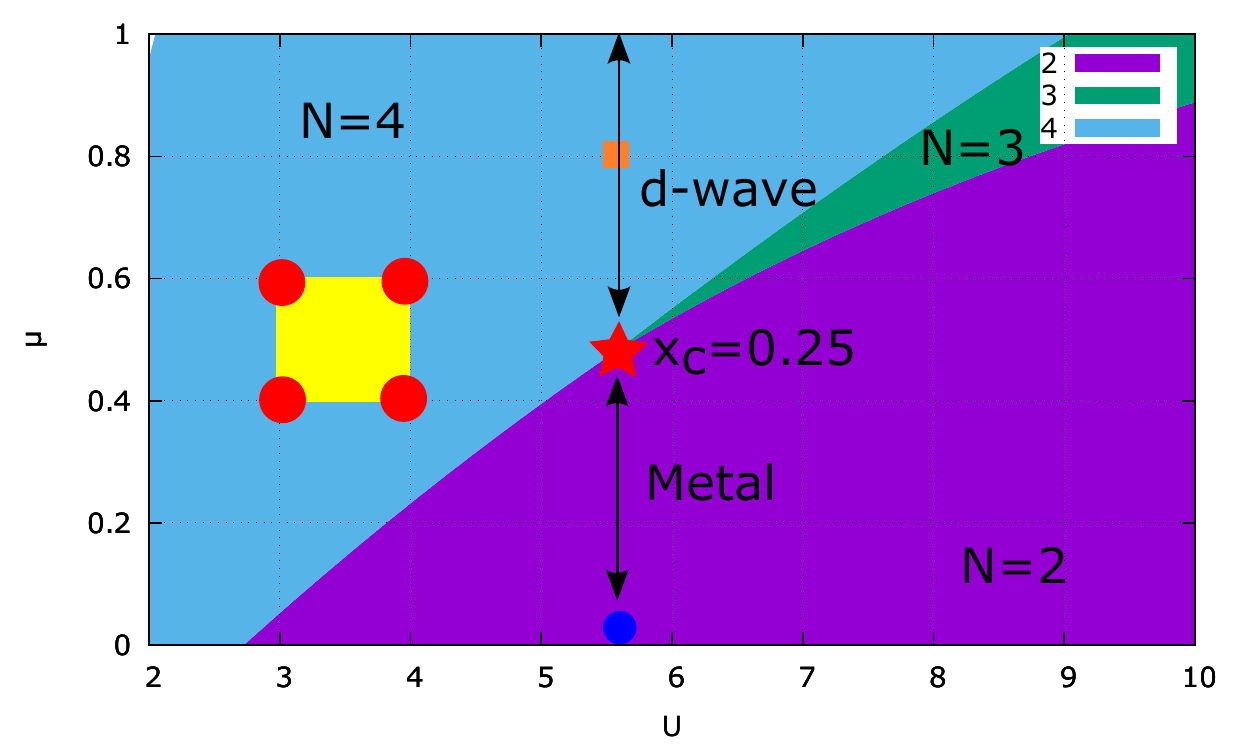}
 \caption{
 Phase diagram of 2$\times$2 plaquette for different particle sectors ($N=2,3,4$) and zero temperature with the degenerate point marked by star. 
 The circle and and square display the shifted chemical potentials for a test comparison.
 The region of $d_{x^2-y^2}$ superconducting phase and normal metal for square lattice are also marked.
 }
\label{fig::2x2phase}
\end{figure}

We will start with this six-fold degenerate ground state of a 2$\times$2 plaquette with $t'/t$ fixed to $-0.15$ depicted as a star-point in the Fig. \ref{fig::2x2phase}. Since we use here periodic boundary conditions the critical Coulomb interaction for plaquette degenerate point becomes $U/t=5.56$ in contrast with the case of isolated plaquette~\cite{Harland16}. This is in a very good agreement with the value of the Coulomb interaction $U/t=5.6$ that was found in the diagrammatic Monte Carlo calculations~\cite{Wei_point} in a search of pseudogap formation, and the value of $U/t\approx6$ pointed out in the recent review~\cite{Taillefer_rev} as the most reasonable value of the effective Hubbard interaction for cuprates.  Note also that periodic boundary conditions effectively double $t'$ compared to $t$, which explains the chosen value of the NNN hopping twice smaller than in Ref.~\cite{Harland16}. 
At a special value of the chemical potential~\cite{Harland16} $\mu\approx0.48$ the ground state for the half-filled $N=4$ antiferromagnetic singlet is degenerate with the singlet for $N=2$ electrons and with two doublets from $N=3$ sector. For these values of the parameters the plaquette state corresponds to the hole doping of $\delta_c=0.25$. If we start from such a degenerate point as a reference system, any perturbation theory for the lattice will be highly divergent.

We will also consider reference systems differing from the degenerate point in the value of the chemical potential. For smaller $\mu\approx 0$ (marked with the circle in Fig. \ref{fig::2x2phase}) the lattice would tend to a metallic behavior, for larger $\mu\approx 0.8$ (marked with the square) the perturbation for the lattice results in a superconducting $d_{x^2-y^2}$ instability.

Facing such a complex system that is hard to solve exactly, it is frequently useful to consider solvable reference systems instead. This strategy is the basis for variational and mean-field approaches, among others. 
The simplest exactly solvable model is based on infinite-dimensional case with a plaquette as elementary unit (cluster DMFT)~\cite{Harland20}. However, nonlocal correlations effects should be relevant for the low-dimensioanl systems which means that we have to go beyond this limiting case.
In electronic systems, the dual fermion~\cite{RKL08} approach provides a recipe for using arbitrary local reference systems~\cite{Brener_refsys}, with a way to incorporate nonlocal corrections in a systematic fashion. There is a large amount of freedom in choosing this reference system which can be used to capture essential physics of the full system under investigation. 

In the case of the doped $t-t'-U$ Hubbard model, d-wave superconducting fluctuations are known to be important, and a four-site plaquette is the minimal reference system that contains their spatial structure and additionally has an important degenerate point~\cite{Harland16,Bagrov_4x4Network}. Below we show that this degenerate point also induces clear signatures in the two-particle correlation function, which is the basic building block of the dual fermion perturbation theory. 

The central question for a reliable theory of the high-Tc cuprates can be formulated in the following manner: what is the mechanism of superconducting coupling and which minimal model explains the key experimental observations such as nodal-antinodal dichotomy and pseudogap formation in the underdoped regime, strange metal behaviour, etc.?
An important part of that question is: what is the minimal length scale needed to understand these phenomena? For the Mott insulating phase, a single atom with Coulomb interaction, coupled to a (dynamical) bath, is qualitatively sufficient. Extending this to a single bond explains how antiferromagnetic exchange interactions between local moments emerge. It has been argued, starting from Ref.~\cite{dwCDMFT}, that a plaquette consisting of $2\times 2$ sites is the minimal unit when thinking about d-wave superconductivity: it is sufficiently large to express the phase difference in the horizontal and vertical direction that characterizes d-wave superconductivity. The $t-t'-U$ plaquette is known to have a  critical line of degenerate states in parameter space of ($U, t', \mu$)~\cite{Bagrov_4x4Network}.  We will argue that, similar to how the generation of antiferromagnetic exchange on a single bond forms the starting point for antiferromagnetism, this plaquette degeneracy plays a central role in the origin of d-wave superconductivity.

\section{Dual Fermion approach with a general reference system}
\label{sec:DF} 

We start with a general lattice fermion model with the local Hubbard-like interaction vertex $U$. Generalisation to the multi-orbital case with general interactions is straightforward~\cite{leshouches}.
The general strategy of the dual fermion approach is related to formally exact separation of the local and non-local correlations effects. We introduce auxiliary dual fermionic fields which will couple local correlated impurities or clusters back to the original lattice~\cite{RKL08}.

Using the path-integral formalism 
the partition function of a general fermionic lattice system can be written in the form
of the functional integral over Grassmann variables  $[c^{ \ast } ,c ]$ :

\begin{equation}
Z =\int \mathcal{D} [c^{ \ast } ,c]\exp ( -S_{L} [c^{ \ast } ,c]) 
\label{eqn::action}
\end{equation}

The original action of interacting lattice fermions $S_{L}$ can be expressed as the sum of the one-electron contribution and the interaction term. The former is most conveniently represented in Matsubara and momentum space, using the Fourier transformed hopping matrix $\hat{t}_{\mathbf{k}}$ (in the single-orbital case, this provides the energy spectrum), whereas the Hubbard interaction $U$ is local and instantaneous and is therefore treated in imaginary time and real space. Any type of local multi-orbital interaction is allowed.
\begin{align}
S_{L} [c^{ \ast } ,c] = -\sum _{\mathbf{k} \nu  \sigma }c_{\mathbf{k} \nu  \sigma }^{ \ast }  \left [i \nu  +\mu  -\hat{t}_{\mathbf{k}}\right ] c_{\mathbf{k} \nu  \sigma }^{\,} +\sum _{i} \int_0^\beta d\tau \;Un_{i\tau\uparrow }^{ \ast } n_{i\tau\downarrow }. 
\label{eqn::df::slat}
\end{align}
Here and in the following, $\nu  =(2 n +1) \pi /\beta$ ($\omega  =2 n \pi /\beta $), with $n \in \mathbf{Z}$, are the fermionic (bosonic) Matsubara frequencies,  $\beta $  is the inverse temperature, $\tau $ is the imaginary time in the interval $\left [0 ,\beta \right )$, $\mu $ is the chemical potential, the index $i$ labels the lattice sites, $m$ can refers to different orbitals inside plaquette ($\hat{t}$ can be a matrix in orbital space), $\sigma $ is the spin projection and the $\mathbf{k}$-vectors are quasimomenta. In order to keep the notation simple, it is useful to introduce the combined index $\left \vert 1\right \rangle  \equiv \left \vert i ,m ,\sigma  ,\tau \right \rangle $ while assuming summation over repeated indices. 
Summation over Matsubara frequencies $\nu$ assume normalization factor $1/\beta$ and the $\mathbf{k}$ integration normalized by volume of Brillouin zone.
Translational invariance is assumed for simplicity in the following, although a real space formulation is possible~\cite{Takemori18}.

\begin{figure}[t!]
 \centering
\includegraphics[width=0.6\textwidth]{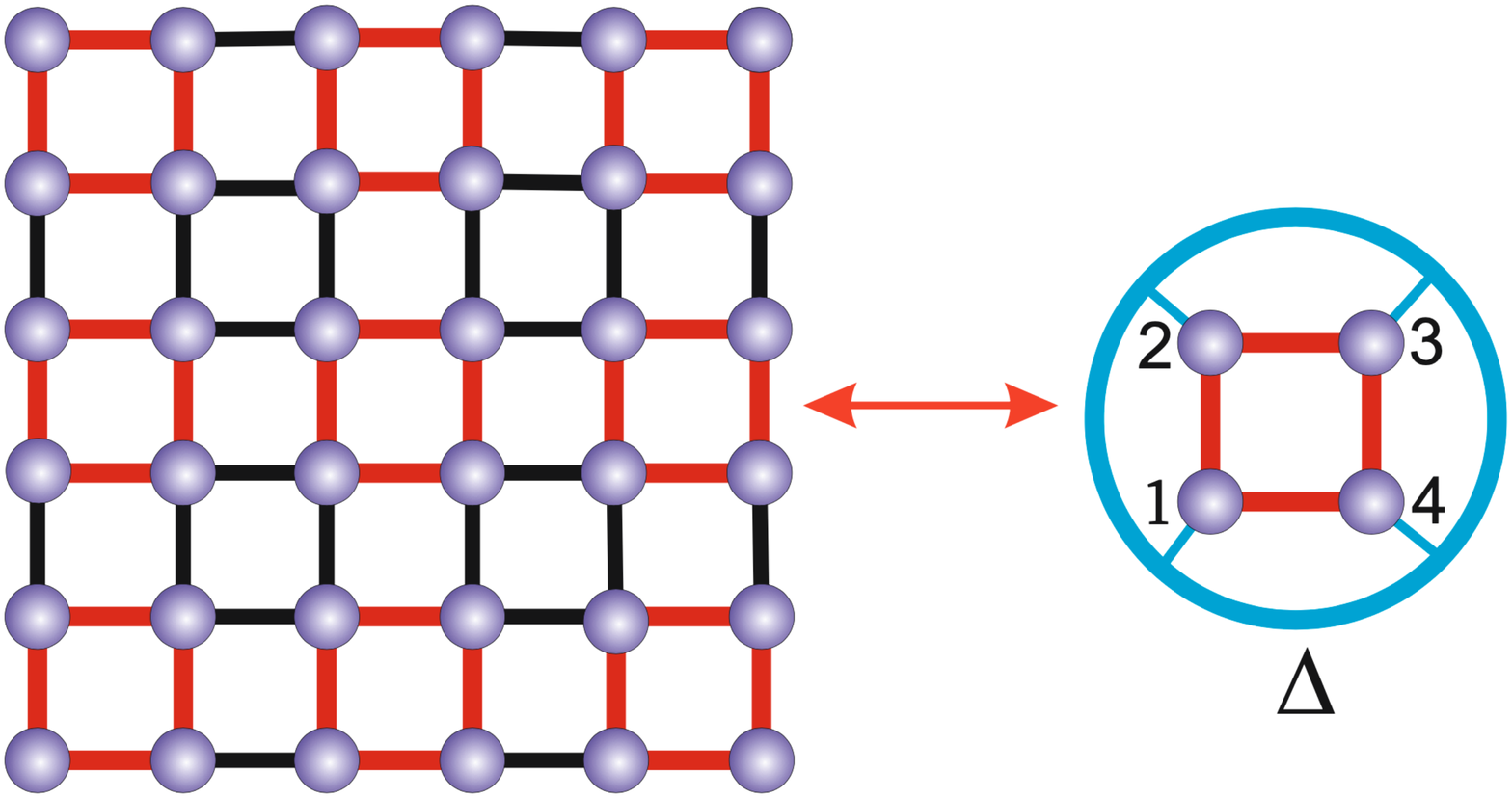}
 \caption{
 Schematic representation of a plaquette reference system for the square lattice.
 }
\label{fig::refCDMFT}
\end{figure}

In order to formulate an expansion around a suitable
 reference action, as illustrated in Fig.~\ref{fig::refCDMFT}, a quantum cluster problem  is introduced by a general frequency dependent hybridization function $\hat{\Delta}_{\nu }$ and the same local interaction,
\begin{align} 
S_{\Delta}[c_{i}^{ \ast } ,c_{i}] = -\sum _{\nu \, ,\sigma }c_{i\nu  \sigma }^{ \ast } \left [i \nu  +\mu  - \hat{\Delta}_{\nu }\right ] c_{i\nu  \sigma } +\sum_{\nu }U n_{i \nu \uparrow }^{ \ast } n_{i \nu \downarrow }.
\label{eqn::df::simp}
\end{align}
$\hat{ \Delta}_{\nu }$ in this notation is the plaquette-local effective "hybridization" matrix which describes hoppings inside the cluster as well connections to an auxiliary fermionic bath. Note that $\hat{\Delta}_\nu$ is allowed to contain instantaneous parts, i.e., finite asymptotic for $\nu\longrightarrow\infty$.
The main motivation for rewriting the lattice action in terms of a quantum cluster model is that such a reference system can be solved numerically exactly for a given hybridization function using Exact Diagonalization (ED) or continuous time Quantum Monte Carlo (CT-QMC)~\cite{CTQMC}. In this work, we use an isolated cluster as a reference model. In that case, $\Delta$ is completely instantaneous and the model is solvable by ED.

Using the locality of the hybridization function $ \hat{\Delta}_{\nu }$, the lattice action Eq. (\ref{eqn::df::slat}) can be rewritten exactly in terms of the individual impurity models and the effective one-electron coupling $(\hat{\Delta} _{\nu }- \hat{t}_{\mathbf{k}} )$ between different impurities  (or plaquettes):

\begin{align}
S_{L} [c^{ \ast } ,c] =\sum _{i}S_{\Delta } [c_{i}^{ \ast } ,c_{i}] +\sum _{\mathbf{k} \nu  \sigma }c_{\mathbf{k} \nu  \sigma }^{ \ast } \left (\hat{t}_{\mathbf{k}} - \hat{\Delta} _{\nu } \right ) c_{\mathbf{k} \nu  \sigma } 
\label{eqn::df::action_rew}
\end{align}
Although we can solve an individual impurity model exactly, in the present formulation the effect of spatial correlations due to the second term in Eq.(\ref{eqn::df::action_rew})
is still problematic, since the impurity action is non-Gaussian and one cannot use the Wick's theorem. 
The main idea of the dual fermion transformation is the change of variables from strongly correlated fermions $(c^{ \ast } ,c)$ to weakly correlated ``dual''  Grassmann fields $(d^{ \ast } ,d)$ in the path integral representation for the partition function from Eq.~\eqref{eqn::action}, followed by a simple perturbation treatment. 
The new "dual" variables are introduced through the following Hubbard-Stratonovich  (HS)-transformation~\cite{Stratonovich_HS,Hubbard_HS} with the following single-particle matrix $\tilde{t}_{\mathbf{k}\nu } = \left( \hat{t}_{\mathbf{k}} - \hat{\Delta}_{\nu} \right)$ .

\begin{equation}
e^{-c_{1}^{ \ast }\;\tilde{t }_{12}\;c_{2} } =\det \left [-\tilde{t}\right ]\int \mathcal{D}\left [d^{ \ast } ,d\right ]e^{ d_{1}^{ \ast }\;\tilde{t}_{12}^{ -1}\;d_{2} -d_{1}^{ \ast }c_{1} -c_{1}^{ \ast }d_{1} }
\label{hs_transfo}
\end{equation}.

We can immediately see this HS-transformation ``localizes'' the  $[c_{i}^{ \ast } ,c_{j}]$ fermions: while on the left hand-side they are still ``hopping'' through the lattice, on the right-hand side they are localized on one site $[c_{i}^{ \ast } ,c_{i}]$.

Compared to the original dual fermion scheme~\cite{RKL08}, we perform the Hubbard-Stratonovich decoupling here without any scaling factors related
with local Green's function in order to reduce the number of matrix multiplications in the final algorithm~\cite{DF2}. In this way, the notation of the formalism becomes closer to the original strong-coupling expansion~\cite{Sarker_1988,Pairault_PRL,Pairault_EPJB,Dupuis_Pairault,Dupuis}. Nevertheless, we would like to stress that the dual fermion theory includes a freedom to choose an arbitrary hybridisation function $\hat{\Delta}_{\nu}$.

With this reference system, the lattice partition function becomes
\begin{align}
\frac{Z}{Z_{d} }=\int \mathcal{D} [c^{ \ast } ,c , d^{ \ast } ,d]\exp \left ( -S [c^{ \ast } ,c , d^{ \ast } ,d]\right ) 
\end{align}
with $Z_{d} =\det \left [-\tilde{t}\right ]$.
The lattice action transforms to
\begin{align}
S [c^{ \ast } ,c , d^{ \ast } ,d] =\sum _{i}S_{\Delta }^{i} -\sum _{ \mathbf{k} ,\nu  ,\sigma }d_{\mathbf{k} \nu  \sigma }^{ \ast }  \left ( \hat{t}_{k}-\hat{\Delta}_{\nu }\right )^{ -1} d_{\mathbf{k}\nu   \sigma }
\end{align}
Hence the coupling between sites is transferred to a local coupling to the auxiliary fermions:
\begin{align}
S_{\Delta }^{i} [c_{i}^{ \ast } ,c_{i} ,d_{i}^{ \ast } ,d_{i}] =S_{\Delta } [c_{i}^{ \ast } ,c_{i}] +\sum _{ \nu  ,\sigma }\left (d_{i \nu  \sigma }^{ \ast }\; c_{i \nu  \sigma } +c_{i \nu  \sigma }^{ \ast }\;  d_{i \nu  \sigma }\right ) 
\label{eqn::df::Ssite}
\end{align}

For the last term we use the invariance of the trace so that the sum over all states labeled by $\mathbf{k}$ could be replaced by the equivalent summation over all sites by a change of basis in the second term. The crucial point is that the
coupling to the auxiliary fermions is purely local and $S_{\Delta }^{i}$ decomposes into a sum of local terms. The lattice fermions can therefore be integrated out from $S_{\Delta }^{i}$ for each site $i$ separately. This completes the change of variables:
 
 \begin{align} 
\frac{1}{Z_{\Delta}} \int \mathcal{D} [c^{ \ast } ,c] \exp  \left ( -S_{\Delta }^{i} [ ,c_{i}^{ \ast } ,c_{i} ,d_{i}^{ \ast }d_{i}]\right ) = 
\exp  \left ( -\sum _{\nu \, \sigma }d_{i \nu  \sigma }^{ \ast }\; g_{\nu } d_{i \nu  \sigma } -V_{i} [d_{i}^{ \ast }d_{i}]\right ) 
 \label{eqn::df::def_V}
 \end{align}

where $Z_{\Delta}$ is partition function of impurity action Eq. (\ref{eqn::df::simp}) and $g_{\nu } $ is the exact impurity Green function

\begin{align}
g_{1 2} &  = - \langle c_{1} c_{2}^{ \ast } \rangle _{\Delta } =\frac{1}{Z_{\Delta}} \int \mathcal{D} [c^{ \ast } ,c]\; c_{1} c_{2}^{ \ast }\; e^{-S_{\Delta }[c^{ \ast } ,c] } 
\label{eqn::gimp}
\end{align}

The above equation may be viewed as the defining equation for the dual potential $V [d^{ \ast } ,d]$. The  choice of the dual transformation in the form of Eq.(\ref{hs_transfo}), without the traditional renormalization of the $d$-fields by a factor of $g_{\nu}^{-1}$, ensures a particularly simple form of this potential. The price that we pay for this simple form is the unconventional dimensionality of the dual Green's function and self-energy, but it proves to be very convenient for numerical multiorbital/cluster calculations.
An explicit expression is found by expanding both sides of Eq.\ (\ref{eqn::df::def_V})
and equating the resulting expressions order by order. Formally this can be done up to all orders and in this sense the transformation to the dual fermions is
exact. For most applications, the dual potential is approximated by the first non-trivial interaction vertex:

\begin{equation}
V [d^{ \ast } ,d] =\frac{1}{4} \sum _{1234  } \gamma^P_{1 2 3 4} d_{1}^{ \ast } d_{2}^{ \ast } d_{3} d_{4}
\label{eqn::gvetrex}
\end{equation}
where for the local vertex the combined index $1 \equiv \{m\nu  \sigma \}$ comprises orbital degrees of freedom (or cluster sites), frequency, and  spin. $\gamma $ is the exact, fully antisymmetric, reducible two-particle vertex of the local quantum impurity problem, in the particle-particle notation (denoted by the index $P$). The absense of normalization in the HS-transformation~(\ref{hs_transfo}) leads to the impurity ``legs'' remain ``unamputated''. Normally this procedure implies division by the single-particle Green's functions. In the multiorbital case, this division involves a potentially unstable matrix inversion, which is avoided by the present choice of normalization.
The vertex is then given by the connected part of the local two-particle correlations function
\begin{align}
\gamma_{1 2 3 4} &  =\kappa_{1 2 3 4} -\kappa_{1 2 3 4}^{0} 
\label{eqn::df::gamma4}
\end{align}

with the two-particle Green's function, $\kappa$, of the local reference system being defined in particle-particle notation as :

\begin{align}
\kappa^P_{1 2 3 4} &  = \langle c_{1} c_{2} c_{4}^{ \ast } c_{3}^{ \ast } \rangle _{\Delta } =\frac{1}{Z_{\Delta}} \int \mathcal{D} [c^{ \ast } ,c] \;c_{1} c_{2} c_{4}^{ \ast } c_{3}^{ \ast } \;
e^{-S_{\Delta }[c^{ \ast } ,c] }  
\label{eqn::chi}
\end{align}

The disconnected part, $\kappa^0$, of the plaquette two-particle Green function reads

\begin{align}
\kappa_{1 2 3 4}^{0} &  =g_{13}g_{24} -g_{14}g_{23} \label{eqn::df::chi0}
\end{align}

The single- and two-particle Green functions can be calculated using the CT-QMC Monte Carlo algorithms \cite{CTQMC}.
After integrating out the lattice fermions, the dual action depends on the new variables only
\begin{equation}
\tilde{S} [d^{ \ast } ,d] = -\sum _{ \mathbf{k}\, \nu \sigma }d_{ \mathbf{k} \nu \sigma }^{ \ast } \; \tilde{G}_{0\mathbf{k}\nu }^{ -1}\; d_{ \mathbf{k} \nu \sigma } +\sum _{i}V_{i} [d_{i}^{ \ast } ,d_{i}],
\label{eqn::df::dual_action}
\end{equation}
here the bare dual Green function is of the form
\begin{equation}
\tilde{G}_{\mathbf{k}\nu }^{0}  =\left [\left (\hat{t}_{\mathbf{k}} - \hat{\Delta} _{\nu }\right )^{ -1} -g_{\nu }\right ]^{ -1} .
\label{eqn::df::Gdbare}
\end{equation}

Action Eq.(\ref{eqn::df::dual_action}) allows us to calculate the dual self-energy, $\tilde{\Sigma }_{{\mathbf{k}}\nu}$ with a level of approximation of our choice. Once this is done, the results are transformed back using an exact relation between the dual and the lattice Green's functions (Appendix \ref{App_Exact}). 

The lattice self-energy is the sum of the reference contribution $\Sigma^{0}$ (i.e the self-energy of the impurity or the cluster) and the correction $\Sigma ^{\prime } $ which is related to the dual self-energy  $\tilde{\Sigma }$ in the following manner~\cite{DF_Rev}
\begin{align}
\Sigma_{{\mathbf{k}}\nu} &=\Sigma^0_\nu+\Sigma^{\prime }_{{\mathbf{k}}\nu} \nonumber\\
\Sigma^{\prime }_{{\mathbf{k}}\nu}&=g^{-1}_\nu-(g_\nu+{\tilde{\Sigma }}_{{\mathbf{k}}\nu})^{-1}
\label{DF_SE}
\end{align}

For numerical calculations it is more convenient not to calculate the lattice self-energy, but  to use directly a simple connection between the dual self-energy and the lattice Green's function\cite{RKL08}
\begin{equation}
G_{{\mathbf{k}}\nu}=\left[ \left( g_\nu+{\tilde{\Sigma}_{{\mathbf{k}}\nu} }\right) ^{-1}-\tilde{t}_{\mathbf{k}\nu } \right]^{-1}.
\label{DF_GF}
\end{equation}
where ${\tilde{\Sigma}_{{\mathbf{k}}\nu} }$ is calculated via diagrammatic perturbation scheme using the $\tilde{G}^{ -1}_{0\mathbf{k}\nu}$ matrix and plaquette vertex $\gamma _{ 1 2 3 4}$.
The properly rescaled dual self energy plays the role of a T-matrix for the the reference Green's function $g$.
With this relation, the calculation only involves single and two-particle correlation functions of the reference system and no ``amputated'' quantities. By avoiding many matrix inversions, this makes it suitable for multi-orbital systems. 
The case of the "bare dual fermions" ${\tilde{\Sigma}_{{\mathbf{k}}\nu} }=0$ is equivalent to the
cluster perturbation theory \cite{Valenti_CPT}.

\section{Perturbation in Dual Space}

The cluster dual fermion perturbation theory (Fig.~\ref{fig::refCDMFT}) starts with the interaction between dual fermions. We use here the particle-hole notation for the local vertex and write explicit spin indices and Matsubara frequency structure 
of the connected two particle Green's function\cite{RKL08,hafermann_thesis} as follows:
\begin{equation}
-\gamma_{1234}^{\sigma \sigma ^{\prime }}((\nu ,\nu ^{\prime },\omega)
=\left<c^{\phantom{\ast} }_{1\sigma }(\nu)c_{2\sigma }^{\ast }(\nu+\omega)c^{\phantom{\ast} }_{3\sigma ^{\prime }}(\nu'+\omega)c_{4\sigma
^{\prime }}^{\ast }(\nu')\right > _{\Delta } - \beta g^{\sigma}_{12}(\nu) g^{\sigma^{\prime}}_{34}(\nu')\delta_{\omega 0 } +  
\beta g^{\sigma}_{14}(\nu) g^{\sigma}_{32}(\nu+\omega) \delta_{\nu \nu^{\prime} } \delta_{\sigma \sigma^{\prime} }.
\end{equation}
In Matsubara space, the vertex depends on two fermionic $(\nu ,\nu ^{\prime })$ and one bosonic 
($\omega $)  frequencies.
For the sake of completeness and the reader's convenience we mention that the connection between the particle-particle and the particle-hole notation reads $\gamma_{1234}(\nu,\nu',\omega)=\gamma^P_{1342}(\nu,\nu',\nu+\nu'+\omega)$ with the particle-particle frequency notation being $\kappa^P_{1234}(\nu,\nu',\omega)=\langle c^{\phantom{\ast} }_{1}(\nu)c^{\phantom{\ast} }_2(\omega-\nu)c^{\ast}_4(\omega-\nu')c^{\ast}_3(\nu')\rangle_{\Delta}$. 
Thus, the bare vertex of the dual fermion perturbation theory is the full connected correlation function of the reference system. The present vertex differs from the usual dual fermion expression due to the different rescaling factor of the Hubbard-Stratonovich field. Here, we avoid amputation of the legs of the vertex, which requires division by Green's functions at all external points. 

It is useful to symmetrize the vertex into charge density ($d$) and magnetic ($m$) channels: 
\begin{equation*}
\gamma_{1234}^{d/m}(\nu ,\nu ^{\prime },\omega ) =\gamma_{1234}^{\uparrow \uparrow }(\nu, \nu ^{\prime },\omega )\pm \gamma_{1234}^{\uparrow\downarrow }(\nu ,\nu ^{\prime },\omega )
\end{equation*}

\begin{figure}[t!]
 \centering
 \includegraphics[width=0.13\textwidth,angle=0]{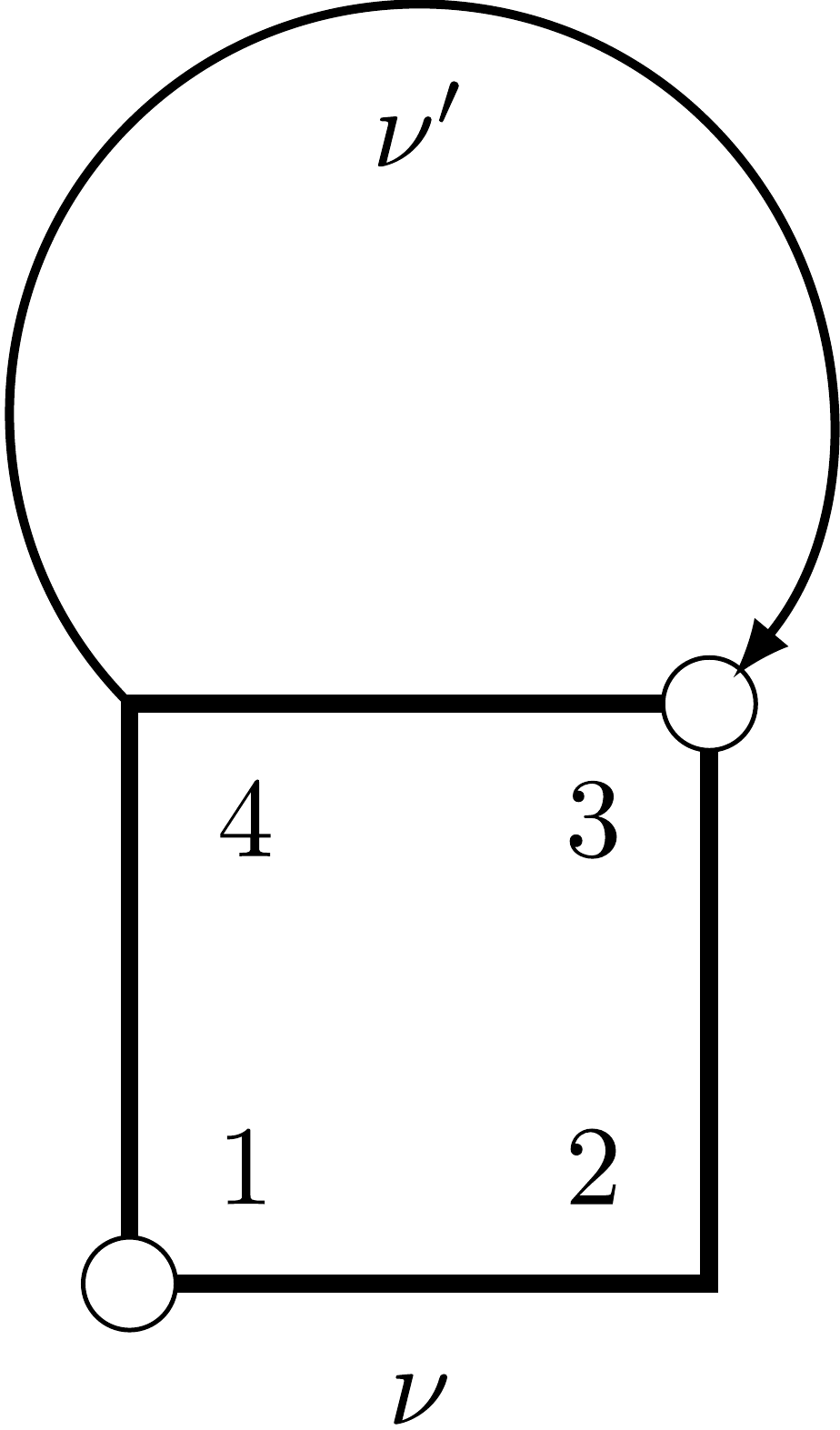}
 \hskip 1.5cm
  \includegraphics[width=0.50\textwidth]{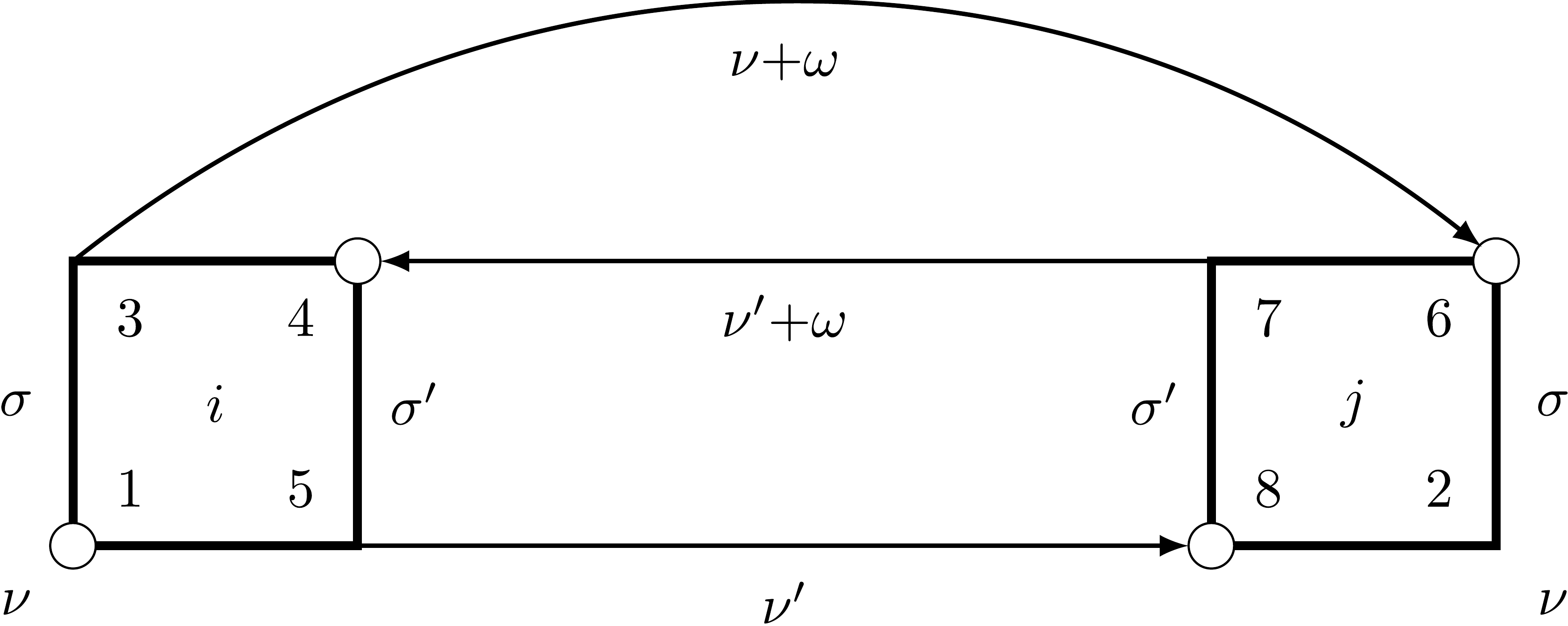}
 \caption{
 Feynman diagram for the first order (left) and the second order (right) dual fermion perturbation for the  self-energy $\widetilde{\Sigma }$: a line represents the non-local $\widetilde{G}_{43}$ and a box is 
 the local $\gamma_{1234}$.
 }
\label{fig::12order}
\end{figure}

Now we can write the first-order dual fermion self-energy which is local in plaquette space  (Fig.~\ref{fig::12order}):
\begin{eqnarray}
\tilde{\Sigma }_{12}^{(1)i}(\nu)=\sum_{\nu ^{\prime}, 3,4  }\gamma_{1234}^{d}(\nu,
\nu ^{\prime },0)\tilde{G}^{ii}_{43}(\nu ^{\prime})
\label{df:1order}
\end{eqnarray}

The second order Feynman diagram for DF-perturbation (Fig.~\ref{fig::12order}) in real space ($\bf{R}_{ij}$) has density- and magnetic-channel contributions 
with corresponding constants ($c_d=-\frac{1}{4} $ and $c_m=-\frac{3}{4} $) :

\begin{align}
\tilde{\Sigma }_{12}^{(2)ij}(\nu ) =&\sum_{\nu ^{\prime }\omega } \sum_{3-8 } \sum_{\alpha=d,m } c_\alpha
\gamma ^{\alpha,i }_{1345} (\nu ,\nu ^{\prime } ,\omega) 
\tilde{G}^{ij}_{36}(\nu +\omega )\tilde{G}^{ji}_{74}(\nu ^{\prime }+\omega )
\tilde{G}^{ij}_{58}(\nu ^{\prime }) 
\gamma^{\alpha,j }_{8762} (\nu ^{\prime },\nu  ,\omega)    
\label{eqn::sigma2}
\end{align}

In principle, one can go beyond the second order perturbation expansion and include dual ladder diagrams  \cite{LDFA,hafermann_thesis}, dual parquet diagrams\cite{DF_Parq} or a stochastic sum of all dual diagrams with the two-particle vertex $\gamma_{1234}$, using diagrammatic Monte Carlo in dual space \cite{DiagDFQMC1,DiagDFQMC2,DiagDBQMC}. In addition, the diagrammatic series can be made self-consistent, using dual skeleton diagrams and ``bold'' lines. Finally, one can also update the reference system (and obtain a frequency dependent $\Delta$) with quite involved numerical approach. But as the main goal of the present work is not to present quantitatively reliable results but rather to highlight the connection between the degenerate reference system and the superconducting fluctuations we will mostly stick to the second-order consideration.
The calculations shown here were performed using a Fortran implementation of dual fermions that uses the equivalence of the four sites in the plaquette to speed up the vertex calculation. The results were checked against an open source implementation of the second-order dual fermion perturbation~\cite{DF2,DF2github}, based on TRIQS~\cite{triqs} and with pomerol~\cite{pomerol} as an impurity solver as well as cross-checked with the momentum-space cluster dual-fermion scheme~\cite{Iskakov1}.

\section{Results for Plaquette dual scheme }

We study the optimally doped square lattice Hubbard model, with nearest neighbour hopping $t$ and NNN hopping $t'$. 
As illustrated in Fig.~\ref{fig::refCDMFT}, the original lattice can be reconsidered as a lattice of $2\times 2$ plaquettes.
Every unit cell of the plaquette lattice contains 4 atoms of the original lattice, as shown on the left-hand side of  Fig.~\ref{fig::refCDMFT}. The plaquette lattice has the following  $4\times4$ hopping matrix (see Fig.~\ref{fig::refCDMFT}),

\begin{align}
t_{\mathbf{k}}=\left( 
\begin{array}{cccc}
\varepsilon  & tK^{0+} & t'L^{-+} & tK^{-0} \\ 
tK^{0-} & \varepsilon  & tK^{-0} & t'L^{--} \\ 
t'L^{+-} & tK^{+0} & \varepsilon  & tK^{0-} \\ 
tK^{+0} & t'L^{++} & tK^{0+} & \varepsilon 
\end{array}
\right) 
\label{fig::tk}
\end{align}
where the functions $K_{\mathbf{k}}^{mn} $ and  $L_{\mathbf{k}}^{mn} $, with $m,n \in \{-1,0,+1\}$, are defined as
\begin{align*}
K_{\mathbf{k}}^{mn} &=1+e^{i(mk_{x}+nk_{y})} \\
L_{\mathbf{k}}^{mn} &=1+e^{i(mk_{x}+nk_{y})}+e^{imk_{x}}+e^{ink_{y}}
\end{align*}

We will use a single plaquette as the reference system.
Compared to the single-site dual fermion formalism, this plaquette reference system already encompasses the short-ranged correlations that are essential in this system.

In the dual fermion approach, there is a general freedom of choosing the most appropriate reference system. 
One way to construct a plaquette reference system would be to simply remove all black links in Fig.~\ref{fig::refCDMFT} (and attach the remaining sites to a bath). This is equivalent to the self-consistent cluster-DMFT scheme~\cite{dwCDMFT} and corresponds to averaging over the supercell Brillouin zone. 
This scheme, however, eliminates exactly half of the nearest-neighbor hoppings and three quarters of the next-nearest-neighbor hoppings. 

Here we choose another path and consider plaquettes with periodic boundary conditions as a static reference system. In terms of the supercell Brillouin zone, this corresponds to achieving self-consistency for $k=0$ only, instead of the momentum average. 
The intra-plaquette hopping reads
\begin{align}
\Delta_0 \equiv t_{\mathbf{k}=0}=\left( 
\begin{array}{cccc}
\varepsilon_0  & 2t & 4t'_0 & 2t \\ 
2t & \varepsilon_0  & 2t & 4t'_0 \\ 
4t'_0 & 2t & \varepsilon_0  & 2t \\ 
2t & 4t'_0 & 2t & \varepsilon_0 
\end{array}
\right) .
\label{fig::t0}
\end{align}
Note that we include the possibility of using a different chemical potential $\mu_0=-\varepsilon_0$ in the reference system, compared to that of the lattice model $\mu=-\varepsilon$ to adjust the hole dopping to about $\delta=0.15$. We fix the nearest neighbour hopping $t$ but retain the freedom of adjusting the next nearest neighbour hopping $t'$ in the dual fermion transformation. For example this may be used to reduce the factor 4 for the $t'$ hoppings for the periodic boundary conditions for 2$\times$ 2 plaquette if we chose $t'_0=t'/2$.

With the plaquette as the reference system, one can use the exact diagonalization approach to calculate the dual Green's function and the plaquette vertex function\cite{Superpert}.
We choose the optimal parameters for the High-T$_c$ cuprates  where the ground state of the plaquette is six-fold degenerate\cite{Harland16}  with
$U=5.56$,  $t=-1$, $t'_0=0.15$ $\mu_0=0.48$ with $t'=0.15$ or $0.3$ and $\mu=0.7$ or $1.5$ correspondingly to keep the optimal doping $\delta\approx0.15$ in the lattice. We investigate different temperatures as low as possible until the dual perturbation theory breaks down due to the divergence in the plaquette vertex function at the degenerate point in the limit $T\rightarrow 0$.

\begin{figure}[t!]
\includegraphics[width=0.7\linewidth]{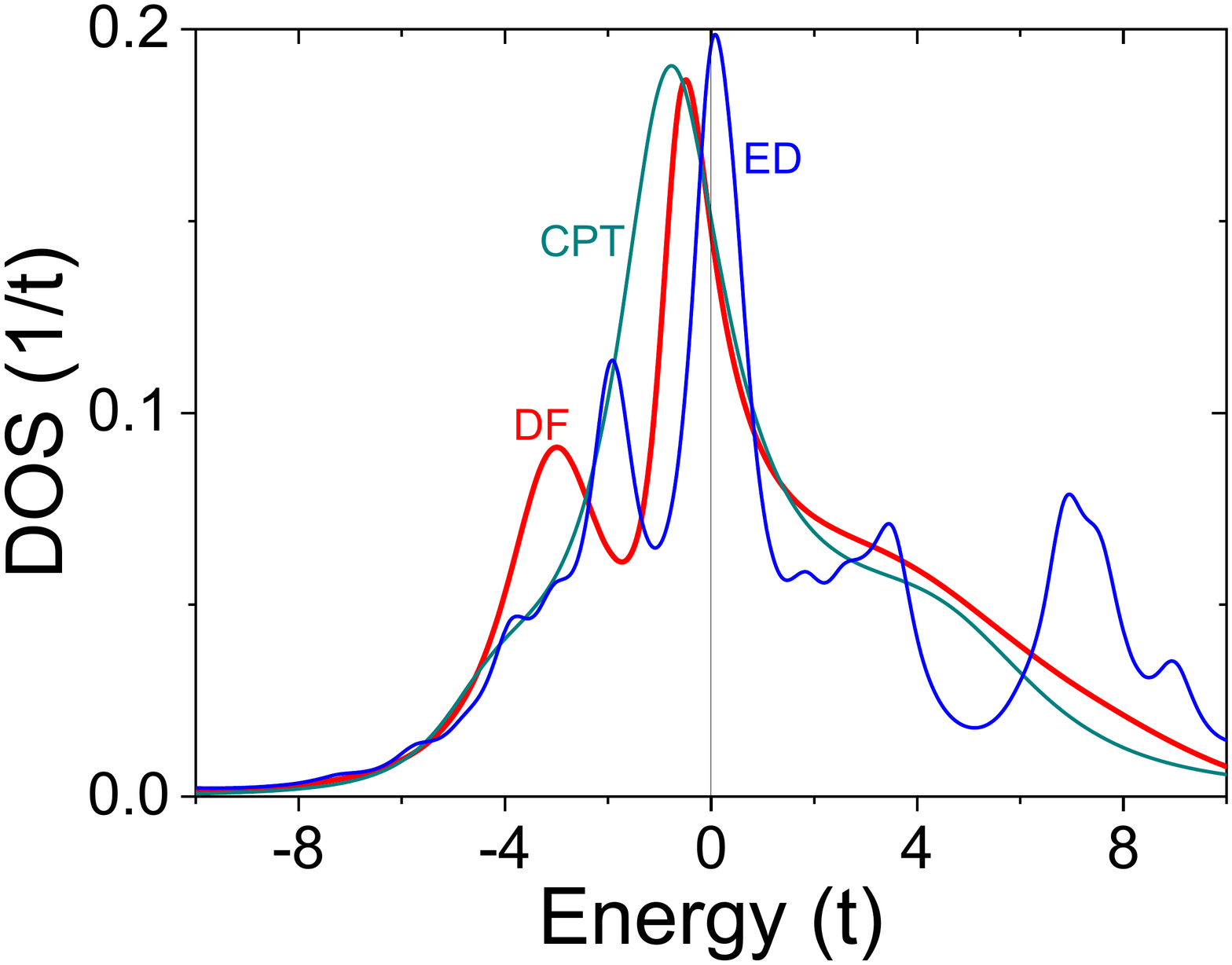}
\caption{Density of states for dual fermion plaquette second order scheme (DF)
in comparison with the cluster perturbation theory (CPT) and exact diagonalization (ED) of 2$\times$2 plaquette for $U=5.56$, $t'_0=t'=0.15$ $\mu_0=0.48$ and $\mu=1.55$, $\beta$=3.}
\label{fig:DOS_CPT}
\end{figure}

In Fig.~\ref{fig:DOS_CPT} we compare the density of states (DOS) for plaquette DF second-order perturbation (DF2) with the so-called cluster perturbation theory (CPT) which corresponds to zero dual-self energy in Eq.(\ref{DF_GF}) for quite high temperature ($\beta=3$). 
We use Pad\'e-analytical continuation from Matsubara to the real
energy axes\cite{GKKR96}. One can see that the DOS for the dual fermion theory is much more sharply peaked near the Fermi level compared to the CPT-result. For comparison we also show the ED result for the plaquette with a sharp peak exactly at Fermi level due to
six-fold degenerate ground state.
In this case there is still no signature for a pseudogap and the lattice self-energy is ``well-behaved''.

\subsection{Vertex and Bethe–Salpeter Equation}
The central idea of starting from an appropriate reference system, is that the exact solution of the latter already contains the essential correlations of the original system. These manifest themselves on the one-particle level ($g$) but especially also on the two-particle level. 
Recent studies have illustrated the value of the information encoded in vertices and susceptibilities~\cite{Rohringer12,Schafer13,Kozik15,Krien19,Harland20,Springer20,Melnick20,vanLoon20,Reitner20,Chalupa20}  even in the case of a single-orbital model. 

In the present case, we use the critical plaquette as the reference model. This plaquette has a sixfold degenerate ground state and anomalies related to transitions between these ground states manifest themselves in the two-particle correlation functions at finite temperature through the $T^{-3}$ behavior compared to the usual $T^{-1}$ one in the general case. 

The Bethe-Salpeter equation has an intertwined spin, site and frequency structure which can be simplified by looking at the different channels.  Since our main interest is superconductivity, we consider the singlet particle-particle channel. For comparison we consider also particle-hole density and magnetic channels. Regarding the frequencies, we restrict ourselves to the lowest 10 Matsubara frequencies, since the vertex function Eq.(\ref{eqn::df::gamma4}) decays strongly with $(\nu,\nu')$\cite{SM}.

\begin{figure}[t!]
\raisebox{.5cm}{\includegraphics[width=0.35\linewidth]{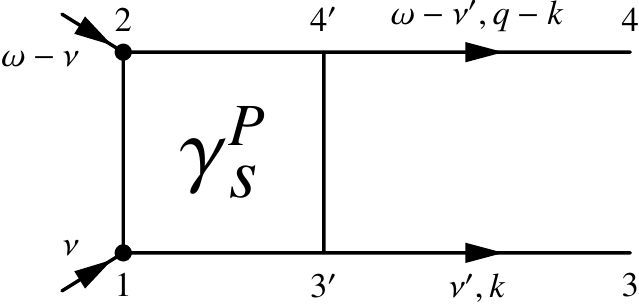}}
\includegraphics[width=0.5\linewidth]{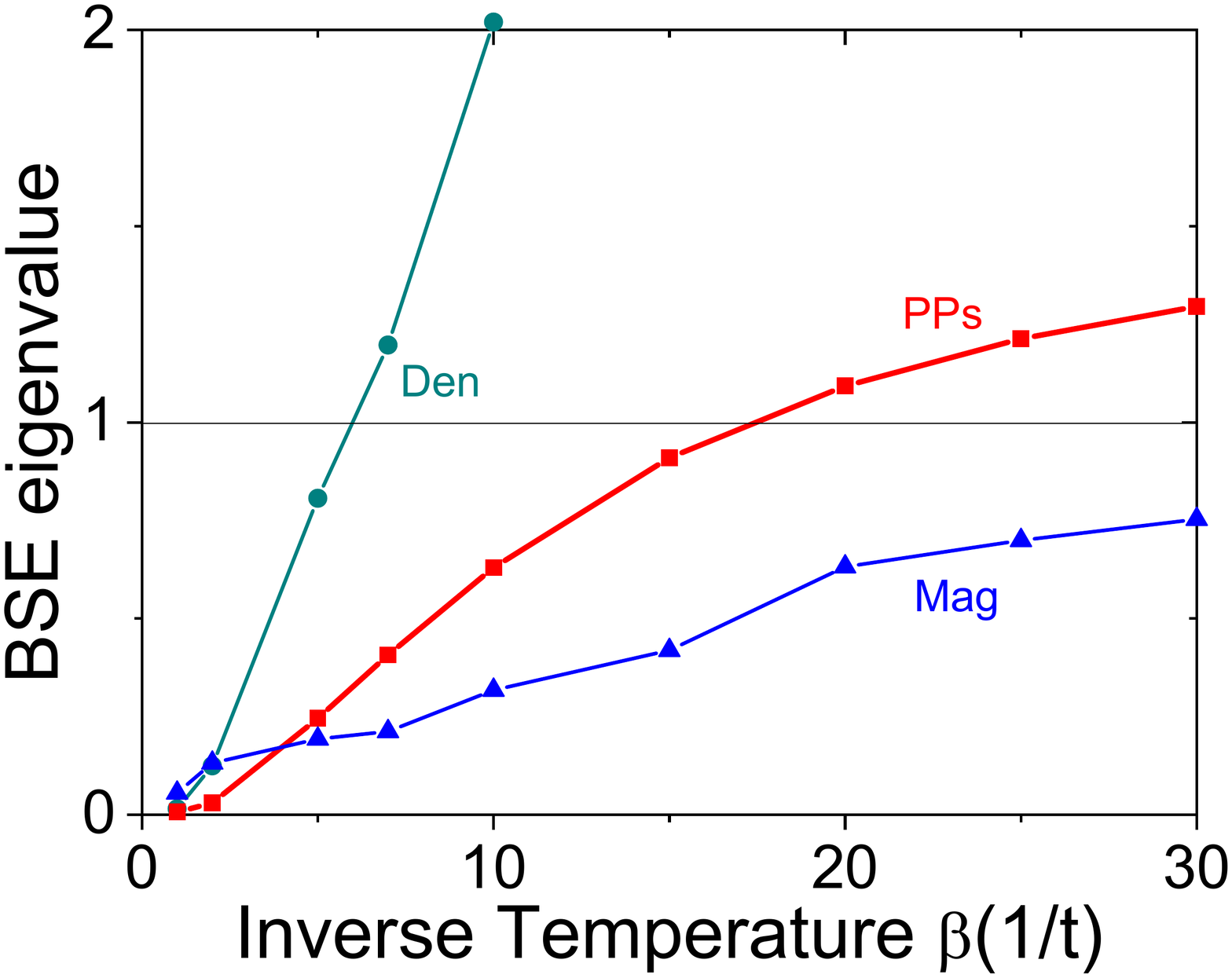}
\caption{Diagrammatic representation of the Bethe-Salpeter  kernel in the particle-particle channel (left) and its maximum eigenvalues (right) for the particle-particle singlet (PPs), density (Den) and magnetic (Mag) channel (right) for doped plaquette with U=5.56 and $t'$=-0.15t, $\mu$=1.55 and }
\label{fig:BSEpp}
\end{figure}

Within the cluster dual fermion theory, the lattice instability manifests itself by $\lambda_{max}=1$, where $\lambda_i$ are eigenvalues of the following Bethe-Salpeter matrix $\Lambda_{i,j}$  in the case of the particle-particle singlet channel:

\begin{equation}
\Lambda_{12,34}^{P\,\nu \nu'} (q,\omega)= \frac T {2 N_k} \sum_{k;\,3'4'}
\gamma_{12,3'4'}^{P\,\nu\nu'}(\omega)\tilde G^{\phantom{P}}_{4',4}(\omega-\nu',q-k)\tilde G^{\phantom{P}}_{3',3}(\nu',k) 
\end{equation}
with $i=(12,\nu)$, $j=(34,\nu')$ and $q=0$, $\omega=0$.
In this case the matrix $\gamma^P$ is Hermitian (real for $\omega=0$), while matrix $\Lambda$ is not Hermitian, but the leading eigenvalues are still found to be real for all channels (we also calculate eigenvalues of corresponding Bethe-Salpeter equations in the density and magnetic particle-hole channels). It has been shown that lattice and dual two-particle quantities have the same set of poles\cite{susceptibility_brener}. 
In the limit $T\rightarrow 0$, the plaquette vertex has several divergences ($\sim T^{-3}$), corresponding to degeneracies of the reference model, while the cluster Green's function 
has divergences ($\sim T^{-1}$) at the degenerate point.
Results for the maximum eigenvalues of the Bethe-Salpeter matrix $\Lambda$ at the critical point for $\omega=0$ and  $q=0$ are presented in the Fig.~\ref{fig:BSEpp}.

 The eigenvector corresponding to $\lambda_{max}$ for the particle-particle singlet case has $d_{x^2-y^2}$ symmetry in the plaquette space. Exactly at the plaquette degenerate point the instability (signaled by $\lambda$ crossing 1) in the density channel is very large because the $N=2,3,4$ states are degenerate. We found that this density instability is not robust against change of $\mu_0$ and as soon as we shift it towards low hole doping $\mu_0=0.8$ there is no density instability\cite{SM}. On the other hand the singlet superconducting instability is very robust and becomes the leading one for doping lower than $\delta=0.25$. The magnetic instability does not play any role for the doped case and becomes the leading one only in the half-filled case\cite{SM}.

\section{Pairing mechanism in real space, Exact diagonalization of 4$\times$4 cluster}

To understand why superconductivity occurs, it is necessary to find a pairing mechanism, i.e., an attractive interaction between pairs of fermions. So far, we have studied the eigenvalues of the Bethe-Salpeter equation to identify such a superconducting instability. Here, we will gain additional insight from a complementary real-space method. We calculated the pairing energy of two holes on the 4$\times$4 periodic cluster -- which consists of 2$\times$2 plaquettes -- through the ground state energies in the different occupation sectors,
\begin{equation}
\Delta_{2h}=\tilde{E}_{2h}-2\tilde{E}_{1h},
\label{e2h}
\end{equation}
where the energies are measured relative to the half-filled ground states $E_{0}$ with no holes, $\tilde{E}_{Nh}=E_{Nh}-E_{0}$. Note, that $\Delta_{2h}<0$ signals pairing. Calculated energies for $t'=0$ are in the perfect agreement with the standard ED results\cite{Dagotto_4x4}.

\begin{figure}[t!]
\includegraphics[width=0.7\linewidth]{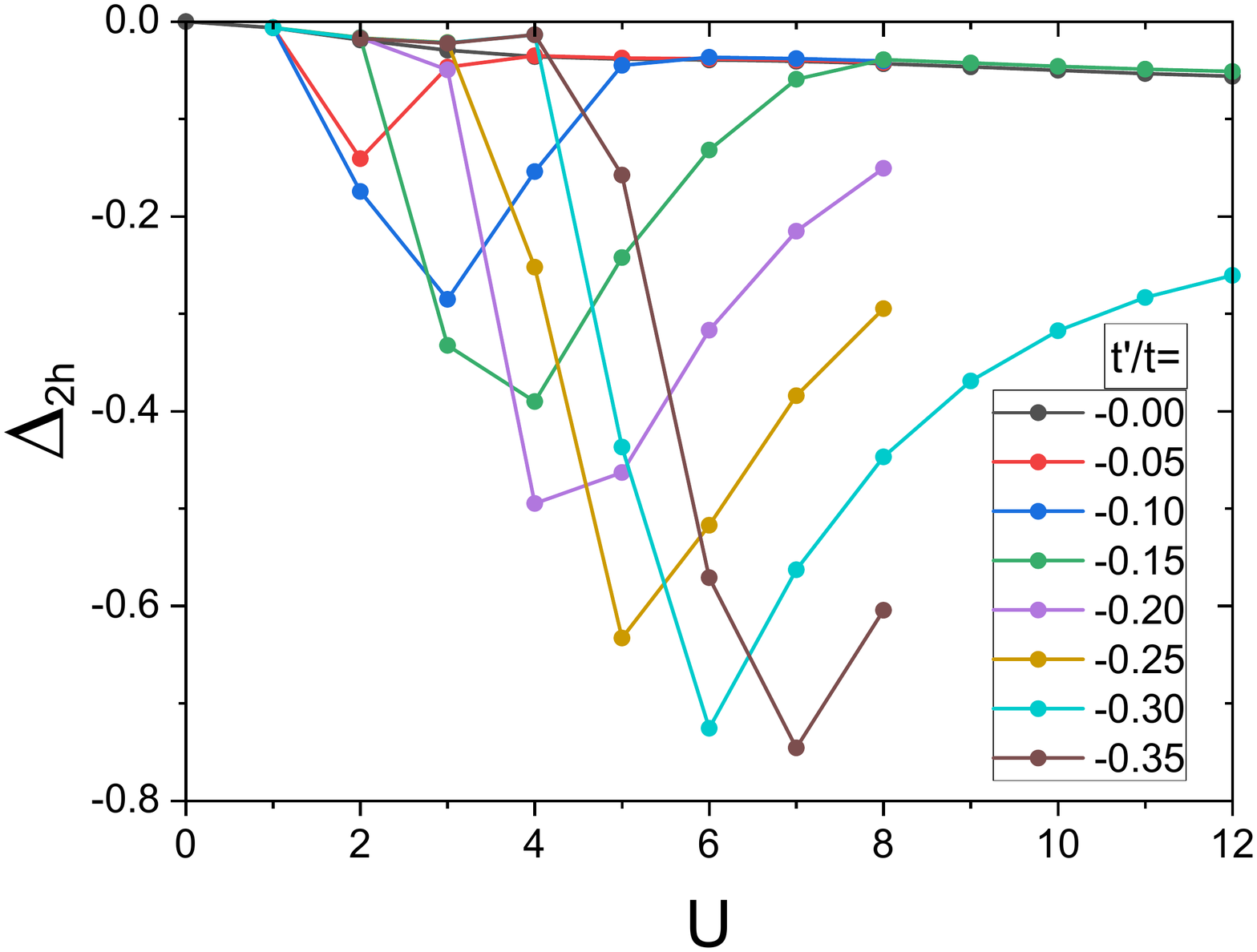}
\caption{Pairing energy $\Delta_\text{2h}$ of two holes in a 4$\times$4 cluster with periodic boundary condition as a function of $U$ and $t'$.}
\label{fig:Delta2h}
\end{figure}

Figure \ref{fig:Delta2h} shows the pair binding energy $\Delta_{2h}$ between pairs of holes for a $4\times 4$ cluster $t-t'-U$ Hubbard model with periodic boundary conditions as a function of interactions strength $U$ for different next-nearest neighbours hopping $t'$.  
There is a very strong binding of two holes around $U=6$ and $t'/t=-0.3$, which is consistent with the estimate for the cuprates~\cite{Pavarini}. The pairing energy is of the order of $\Delta_{2h}/t\approx -0.7$ which is of the order 3000 K for $t\approx0.4$ eV for generic cuprates model~\cite{Pavarini,OKA}. 
There is a clear change of behaviour of $\Delta_{2h}$ as a function of $t'$, with the vanishing of the pairing energy at small $U$. It can be attributed to the change of the ground state for the sector ($7{\uparrow},7{\downarrow}$) at $t'/t\approx 0.12$\cite{SM}.
We also observe a drastic change of the behavior of the magnetic correlations from antiferromagnetic at $t'=0$
to almost non-magnetic for $t'/t=-0.3$ in this sector\cite{SM}. 
Similar energetic of hole-binding in 4$\times$4 Hubbard cluster was found recently\cite{Orgad_ED} for a different model of inhomogeneous hopping\cite{Kivelson_inhomog}. The effects of negative $t'/t$ consider to be destructive for superconductivity in the $t-t'-J$ model\cite{Martins_ttpJ}. Our results (Figure \ref{fig:Delta2h}) show that strong effect of the hole pair binding on 4$\times$4 cluster disappear for $U\gg W= 8t$ or in $t-J$ limit.

\begin{figure}[t!]
\includegraphics[width=0.65\linewidth]{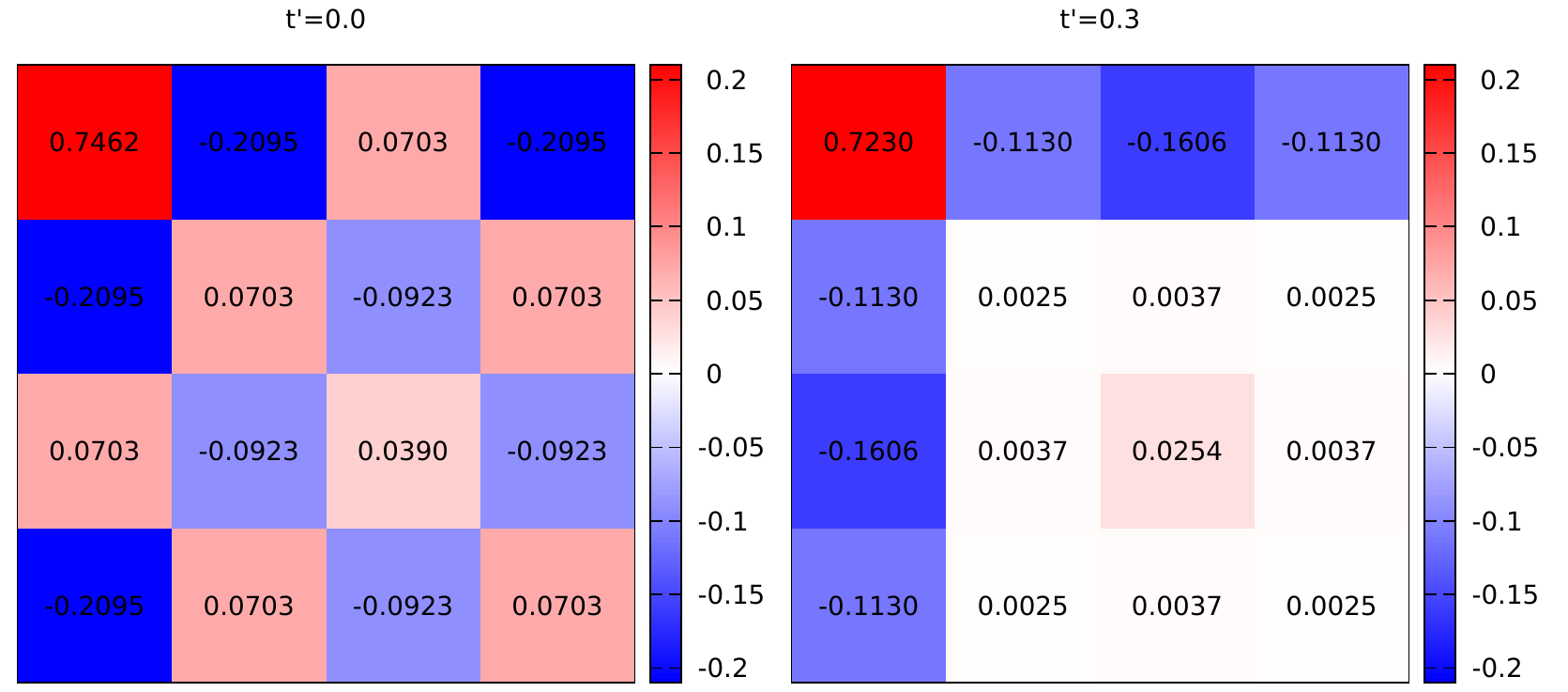}
\caption{Static spin-spin correlation function $\langle M_0 M_i \rangle$ 
obtained by exact diagonalization
for the ground state of the sector ($7{\uparrow},7{\downarrow}$) of the 4$\times$4 cluster for different t'. Whereas $t'=0$ features clear antiferromagnetic correlations, at $t'/t=-0.3$ these are replaced by stripe-like ferromagnetic correlations. The top-left corner corresponds to $i=(0,0)$.}
\label{fig:EDmag}
\end{figure}

These results clearly show the importance of $t'$, which greatly increases the pairing energy gain. At the same time, in a non-interacting systems $\Delta_{2h}=0$ by definition, so a finite value of $U$ is also necessary for the pairing. We find that the optimal $U$ increases with $t'$. A second observation is the order of magnitude of the pairing, $\Delta_{2h} \approx 0.7t \gg k_B T^{\rm exp}_c$. This tells us that bound pairs exist for temperatures far above the superconducting region. The superconducting transition should then be seen as the condensation of these pairs. Thus, the binding energy of two holes turns out to be much higher than the superconducting critical temperature which means that the pairs (``bipolarons'') should be well-defined also in non-superconducting phase, a situation dramatically different form the conventional BCS superconductivity. The difference is like the difference between purely itinerant weak ferromagnets and ferromagnets with local magnetic moments which exist until very high temperatures and only order, rather than appear, at the Curie temperature~\cite{moriya}. 

We analyzed the spin-spin correlation function in the sector $(7{\uparrow},7{\downarrow})$ with different NNN hoppings $t'$ (Fig.\ref{fig:EDmag}) and clearly see a sharp change from antiferromagnertic correlations for $t'=0$ with clear "checkerboard" structure to almost nonmagnetic case or ferromagnetic stripes in the $x$ or $y$ directions for  $t'=0.3$.
A similar reduction of AFM-correlations and existence of FM-one with $t'$ was found in a lattice QMC study\cite{Yang_DQMC_tp}.

\section{Spectral information}

\begin{figure}[t!]
\includegraphics[width=0.7\linewidth]{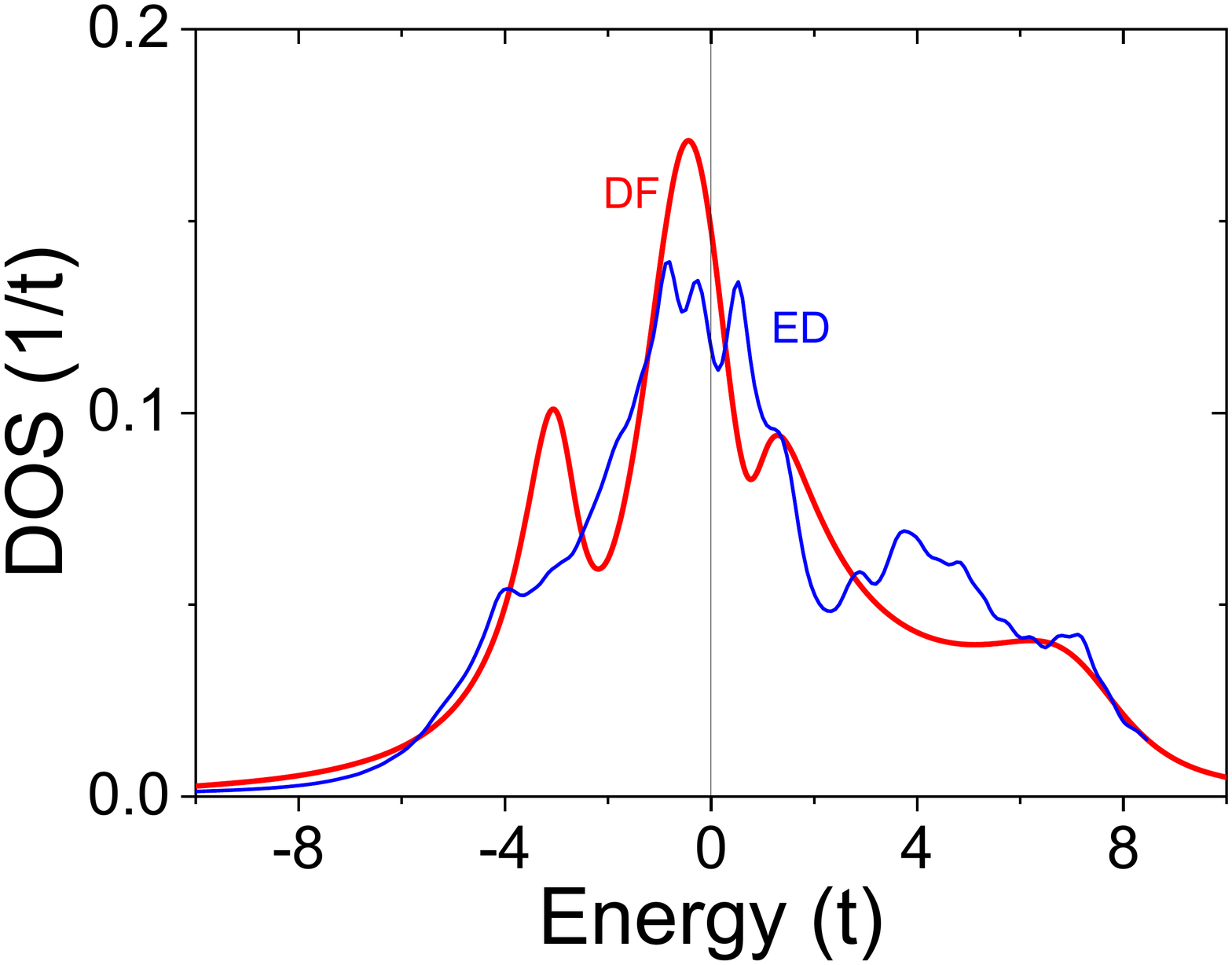}
\caption{Density of states for dual fermion plaquette perturbation (DF)
with $\beta=5$ in comparison with exact diagonalisation (ED) for $4\times4$ periodic cluster . See Fig.~\ref{fig:DOS_CPT} for other parameters. }
\label{fig:DOS_ED}
\end{figure}

Due to the degeneracy of states with different particle number, the density of states of the plaquette is large close to the Fermi level. The availability of low-energy states is the driving force behind the instabilities that occur once a lattice of plaquettes is considered. In Fig.~\ref{fig:DOS_ED} we compare the DOS for the plaquette DF perturbation theory for low temperature ($\beta=5$) with the ED results for the $4\times4$ cluster in the sector $(7\uparrow, 7\downarrow )$, which corresponds to a $2\times 2$ lattice of plaquettes. These two methods are complementary: the DF approach is perturbative in the inter-plaquette coupling and able to handle large lattices, whereas the ED is exact but limited by the cluster size. From the comparison of the two curves, we conclude that the dual fermion theory shows a tendency towards pseudogap formation which is clearly seen in the ED results. It is natural to conclude that the pseudogap in the $4\times4$ cluster is related to the coherent interactions of the large peak on the DOS in the reference plaquette or Fano-like effect of interactions with the ``soft fermion mode'' of the low-lying excitations which are encoded in the local vertex functions of the DF-approach. In this sense the pseudogap physics is not related to the magnetic fluctuations, and is more in line with the ``hidden fermion'' physics\cite{Imada_PRL,Imada_PRB} or ``destructive interference phenomena''\cite{Gunnarsson_2014}.

\section{Conclusions}
The physics of cuprate superconductors with the clear existence of a quantum critical point at $\delta_c\approx0.24$ is closely related to the degeneracy of the plaquette in the strong-coupling regime. In this sense, the plaquette and not the single site can be considered the minimal building block for cuprate physics. The renormalized dual fermion perturbation starting from the plaquette reference system with $\delta=0.25$ uncovers consequences of this degeneracy for the Green's function in the lattice and shows the basic "plaquette" mechanism of superconducting instability in the Bethe-Salpeter equation for a general cuprate model. Exact diagonalization of the 4$\times$4 cluster supports strong pair-binding related with the next-nearest hoppings $t'$. Given their large binding energy, these pairs should probably exist also at much higher temperatures than the superconducting critical temperature, remaining noncoherent. The formation of the pseudogap is related to a Fano-like effect originating from the sharply peaked DOS in the isolated plaquette embedded into the band of surrounding fermions, as was hypothesised in Ref.~\cite{Harland16}. In the overdoped regime $\delta \ge 0.25$ the strong charge fluctuations restore formation of the normal metallic phase and corresponding Bethe-Salpeter equation does not indicate any instabiliries.
For the doping  $\delta \le 0.25$ the dual perturbation theory starting from the plaquette clearly shows a low temperature $d_{x^2-y^2}$ instability. These observations can all be made by starting the perturbation theory from an isolated plaquette. For more quantitative predictions of the theoretical phase diagram, the optimal dynamical embedding of the plaquette and the implications for the resulting perturbation theory need to be studied further. 

\begin{acknowledgments}
The authors thank Alexei Rubtsov, Evgeny Stepanov, Igor Krivenko, Fedor \v{S}imkovic IV, Georg Rohringer, Andy Millis and Antoine Georges for valuable comments on the work. 
E.G.C.P.v.L. is supported by the Zentrale Forschungsf\"orderung of the Universit\"at Bremen.
This work was partially supported by the Cluster of Excellence ``Advanced Imaging of Matter'' of the Deutsche Forschungsgemeinschaft (DFG) - EXC 2056 - Project No. ID390715994,  by European Research Council via Synergy Grant 854843 - FASTCORR and by North-German Supercomputing Alliance (HLRN) under the Project No. hhp00042.
\end{acknowledgments}

\bibliography{Ref}

\begin{thebibliography}{71}%
\makeatletter
\providecommand \@ifxundefined [1]{%
 \@ifx{#1\undefined}
}%
\providecommand \@ifnum [1]{%
 \ifnum #1\expandafter \@firstoftwo
 \else \expandafter \@secondoftwo
 \fi
}%
\providecommand \@ifx [1]{%
 \ifx #1\expandafter \@firstoftwo
 \else \expandafter \@secondoftwo
 \fi
}%
\providecommand \natexlab [1]{#1}%
\providecommand \enquote  [1]{``#1''}%
\providecommand \bibnamefont  [1]{#1}%
\providecommand \bibfnamefont [1]{#1}%
\providecommand \citenamefont [1]{#1}%
\providecommand \href@noop [0]{\@secondoftwo}%
\providecommand \href [0]{\begingroup \@sanitize@url \@href}%
\providecommand \@href[1]{\@@startlink{#1}\@@href}%
\providecommand \@@href[1]{\endgroup#1\@@endlink}%
\providecommand \@sanitize@url [0]{\catcode `\\12\catcode `\$12\catcode
  `\&12\catcode `\#12\catcode `\^12\catcode `\_12\catcode `\%12\relax}%
\providecommand \@@startlink[1]{}%
\providecommand \@@endlink[0]{}%
\providecommand \url  [0]{\begingroup\@sanitize@url \@url }%
\providecommand \@url [1]{\endgroup\@href {#1}{\urlprefix }}%
\providecommand \urlprefix  [0]{URL }%
\providecommand \Eprint [0]{\href }%
\providecommand \doibase [0]{http://dx.doi.org/}%
\providecommand \selectlanguage [0]{\@gobble}%
\providecommand \bibinfo  [0]{\@secondoftwo}%
\providecommand \bibfield  [0]{\@secondoftwo}%
\providecommand \translation [1]{[#1]}%
\providecommand \BibitemOpen [0]{}%
\providecommand \bibitemStop [0]{}%
\providecommand \bibitemNoStop [0]{.\EOS\space}%
\providecommand \EOS [0]{\spacefactor3000\relax}%
\providecommand \BibitemShut  [1]{\csname bibitem#1\endcsname}%
\let\auto@bib@innerbib\@empty
\bibitem [{\citenamefont {Bednorz}\ and\ \citenamefont
  {M{\"u}ller}(1986)}]{Muller_HTSC}%
  \BibitemOpen
  \bibfield  {author} {\bibinfo {author} {\bibfnamefont {J.~G.}\ \bibnamefont
  {Bednorz}}\ and\ \bibinfo {author} {\bibfnamefont {K.~A.}\ \bibnamefont
  {M{\"u}ller}},\ }\bibfield  {title} {\enquote {\bibinfo {title} {Possible
  high {Tc} superconductivity in the {BaLaCuO} system},}\ }\href {\doibase
  10.1007/BF01303701} {\bibfield  {journal} {\bibinfo  {journal} {Zeitschrift
  f{\"u}r Physik B Condensed Matter}\ }\textbf {\bibinfo {volume} {64}},\
  \bibinfo {pages} {189--193} (\bibinfo {year} {1986})}\BibitemShut {NoStop}%
\bibitem [{\citenamefont {Zhou}\ \emph {et~al.}(2021)\citenamefont {Zhou},
  \citenamefont {Lee}, \citenamefont {Imada}, \citenamefont {Trivedi},
  \citenamefont {Phillips}, \citenamefont {Kee}, \citenamefont
  {T{\"o}rm{\"a}},\ and\ \citenamefont {Eremets}}]{HTSC_view}%
  \BibitemOpen
  \bibfield  {author} {\bibinfo {author} {\bibfnamefont {Xingjiang}\
  \bibnamefont {Zhou}}, \bibinfo {author} {\bibfnamefont {Wei-Sheng}\
  \bibnamefont {Lee}}, \bibinfo {author} {\bibfnamefont {Masatoshi}\
  \bibnamefont {Imada}}, \bibinfo {author} {\bibfnamefont {Nandini}\
  \bibnamefont {Trivedi}}, \bibinfo {author} {\bibfnamefont {Philip}\
  \bibnamefont {Phillips}}, \bibinfo {author} {\bibfnamefont {Hae-Young}\
  \bibnamefont {Kee}}, \bibinfo {author} {\bibfnamefont {P{\"a}ivi}\
  \bibnamefont {T{\"o}rm{\"a}}}, \ and\ \bibinfo {author} {\bibfnamefont
  {Mikhail}\ \bibnamefont {Eremets}},\ }\bibfield  {title} {\enquote {\bibinfo
  {title} {High-temperature superconductivity},}\ }\href {\doibase
  10.1038/s42254-021-00324-3} {\bibfield  {journal} {\bibinfo  {journal}
  {Nature Reviews Physics}\ }\textbf {\bibinfo {volume} {3}},\ \bibinfo {pages}
  {462--465} (\bibinfo {year} {2021})}\BibitemShut {NoStop}%
\bibitem [{\citenamefont {Scalapino}(2012)}]{Scalapino_RMP}%
  \BibitemOpen
  \bibfield  {author} {\bibinfo {author} {\bibfnamefont {D.~J.}\ \bibnamefont
  {Scalapino}},\ }\bibfield  {title} {\enquote {\bibinfo {title} {A common
  thread: The pairing interaction for unconventional superconductors},}\ }\href
  {\doibase 10.1103/RevModPhys.84.1383} {\bibfield  {journal} {\bibinfo
  {journal} {Rev. Mod. Phys.}\ }\textbf {\bibinfo {volume} {84}},\ \bibinfo
  {pages} {1383--1417} (\bibinfo {year} {2012})}\BibitemShut {NoStop}%
\bibitem [{\citenamefont {Esterlis}\ \emph {et~al.}(2018)\citenamefont
  {Esterlis}, \citenamefont {Kivelson},\ and\ \citenamefont
  {Scalapino}}]{Scalapino_bound}%
  \BibitemOpen
  \bibfield  {author} {\bibinfo {author} {\bibfnamefont {I.}~\bibnamefont
  {Esterlis}}, \bibinfo {author} {\bibfnamefont {S.~A.}\ \bibnamefont
  {Kivelson}}, \ and\ \bibinfo {author} {\bibfnamefont {D.~J.}\ \bibnamefont
  {Scalapino}},\ }\bibfield  {title} {\enquote {\bibinfo {title} {A bound on
  the superconducting transition temperature},}\ }\href {\doibase
  10.1038/s41535-018-0133-0} {\bibfield  {journal} {\bibinfo  {journal} {npj
  Quantum Materials}\ }\textbf {\bibinfo {volume} {3}},\ \bibinfo {pages} {59}
  (\bibinfo {year} {2018})}\BibitemShut {NoStop}%
\bibitem [{\citenamefont {Keimer}\ \emph {et~al.}(2015)\citenamefont {Keimer},
  \citenamefont {Kivelson}, \citenamefont {Norman}, \citenamefont {Uchida},\
  and\ \citenamefont {Zaanen}}]{Keimer_rev}%
  \BibitemOpen
  \bibfield  {author} {\bibinfo {author} {\bibfnamefont {B.}~\bibnamefont
  {Keimer}}, \bibinfo {author} {\bibfnamefont {S.~A.}\ \bibnamefont
  {Kivelson}}, \bibinfo {author} {\bibfnamefont {M.~R.}\ \bibnamefont
  {Norman}}, \bibinfo {author} {\bibfnamefont {S.}~\bibnamefont {Uchida}}, \
  and\ \bibinfo {author} {\bibfnamefont {J.}~\bibnamefont {Zaanen}},\
  }\bibfield  {title} {\enquote {\bibinfo {title} {From quantum matter to
  high-temperature superconductivity in copper oxides},}\ }\href {\doibase
  10.1038/nature14165} {\bibfield  {journal} {\bibinfo  {journal} {Nature}\
  }\textbf {\bibinfo {volume} {518}},\ \bibinfo {pages} {179--186} (\bibinfo
  {year} {2015})}\BibitemShut {NoStop}%
\bibitem [{\citenamefont {Jiang}\ and\ \citenamefont
  {Devereaux}(2019)}]{Deveraux_Tp}%
  \BibitemOpen
  \bibfield  {author} {\bibinfo {author} {\bibfnamefont {Hong-Chen}\
  \bibnamefont {Jiang}}\ and\ \bibinfo {author} {\bibfnamefont {Thomas~P.}\
  \bibnamefont {Devereaux}},\ }\bibfield  {title} {\enquote {\bibinfo {title}
  {Superconductivity in the doped hubbard model and its interplay with
  next-nearest hopping t'},}\ }\href {\doibase 10.1126/science.aal5304}
  {\bibfield  {journal} {\bibinfo  {journal} {Science}\ }\textbf {\bibinfo
  {volume} {365}},\ \bibinfo {pages} {1424--1428} (\bibinfo {year} {2019})},\
  \Eprint
  {http://arxiv.org/abs/https://science.sciencemag.org/content/365/6460/1424.full.pdf}
  {https://science.sciencemag.org/content/365/6460/1424.full.pdf} \BibitemShut
  {NoStop}%
\bibitem [{\citenamefont {Qin}\ \emph {et~al.}(2020)\citenamefont {Qin},
  \citenamefont {Chung}, \citenamefont {Shi}, \citenamefont {Vitali},
  \citenamefont {Hubig}, \citenamefont {Schollw\"ock}, \citenamefont {White},\
  and\ \citenamefont {Zhang}}]{Zhang_noSC}%
  \BibitemOpen
  \bibfield  {author} {\bibinfo {author} {\bibfnamefont {Mingpu}\ \bibnamefont
  {Qin}}, \bibinfo {author} {\bibfnamefont {Chia-Min}\ \bibnamefont {Chung}},
  \bibinfo {author} {\bibfnamefont {Hao}\ \bibnamefont {Shi}}, \bibinfo
  {author} {\bibfnamefont {Ettore}\ \bibnamefont {Vitali}}, \bibinfo {author}
  {\bibfnamefont {Claudius}\ \bibnamefont {Hubig}}, \bibinfo {author}
  {\bibfnamefont {Ulrich}\ \bibnamefont {Schollw\"ock}}, \bibinfo {author}
  {\bibfnamefont {Steven~R.}\ \bibnamefont {White}}, \ and\ \bibinfo {author}
  {\bibfnamefont {Shiwei}\ \bibnamefont {Zhang}} (\bibinfo {collaboration}
  {Simons Collaboration on the Many-Electron Problem}),\ }\bibfield  {title}
  {\enquote {\bibinfo {title} {Absence of superconductivity in the pure
  two-dimensional {Hubbard} model},}\ }\href {\doibase
  10.1103/PhysRevX.10.031016} {\bibfield  {journal} {\bibinfo  {journal} {Phys.
  Rev. X}\ }\textbf {\bibinfo {volume} {10}},\ \bibinfo {pages} {031016}
  (\bibinfo {year} {2020})}\BibitemShut {NoStop}%
\bibitem [{\citenamefont {Proust}\ and\ \citenamefont
  {Taillefer}(2019)}]{Taillefer_rev}%
  \BibitemOpen
  \bibfield  {author} {\bibinfo {author} {\bibfnamefont {Cyril}\ \bibnamefont
  {Proust}}\ and\ \bibinfo {author} {\bibfnamefont {Louis}\ \bibnamefont
  {Taillefer}},\ }\bibfield  {title} {\enquote {\bibinfo {title} {The
  remarkable underlying ground states of cuprate superconductors},}\ }\href
  {\doibase 10.1146/annurev-conmatphys-031218-013210} {\bibfield  {journal}
  {\bibinfo  {journal} {Annual Review of Condensed Matter Physics}\ }\textbf
  {\bibinfo {volume} {10}},\ \bibinfo {pages} {409--429} (\bibinfo {year}
  {2019})},\ \Eprint
  {http://arxiv.org/abs/https://doi.org/10.1146/annurev-conmatphys-031218-013210}
  {https://doi.org/10.1146/annurev-conmatphys-031218-013210} \BibitemShut
  {NoStop}%
\bibitem [{\citenamefont {Ayres}\ \emph {et~al.}(2021)\citenamefont {Ayres},
  \citenamefont {Berben}, \citenamefont {Culo}, \citenamefont {Hsu},
  \citenamefont {van Heumen}, \citenamefont {Huang}, \citenamefont {Zaanen},
  \citenamefont {Kondo}, \citenamefont {Takeuchi}, \citenamefont {Cooper},
  \citenamefont {Putzke}, \citenamefont {Friedemann}, \citenamefont
  {Carrington},\ and\ \citenamefont {Hussey}}]{Hussey20}%
  \BibitemOpen
  \bibfield  {author} {\bibinfo {author} {\bibfnamefont {J.}~\bibnamefont
  {Ayres}}, \bibinfo {author} {\bibfnamefont {M.}~\bibnamefont {Berben}},
  \bibinfo {author} {\bibfnamefont {M.}~\bibnamefont {Culo}}, \bibinfo {author}
  {\bibfnamefont {Y.~T.}\ \bibnamefont {Hsu}}, \bibinfo {author} {\bibfnamefont
  {E.}~\bibnamefont {van Heumen}}, \bibinfo {author} {\bibfnamefont
  {Y.}~\bibnamefont {Huang}}, \bibinfo {author} {\bibfnamefont
  {J.}~\bibnamefont {Zaanen}}, \bibinfo {author} {\bibfnamefont
  {T.}~\bibnamefont {Kondo}}, \bibinfo {author} {\bibfnamefont
  {T.}~\bibnamefont {Takeuchi}}, \bibinfo {author} {\bibfnamefont {J.~R.}\
  \bibnamefont {Cooper}}, \bibinfo {author} {\bibfnamefont {C.}~\bibnamefont
  {Putzke}}, \bibinfo {author} {\bibfnamefont {S.}~\bibnamefont {Friedemann}},
  \bibinfo {author} {\bibfnamefont {A.}~\bibnamefont {Carrington}}, \ and\
  \bibinfo {author} {\bibfnamefont {N.~E.}\ \bibnamefont {Hussey}},\ }\bibfield
   {title} {\enquote {\bibinfo {title} {{Incoherent transport across the
  strange metal regime of highly overdoped cuprates}},}\ }\href@noop {}
  {\bibfield  {journal} {\bibinfo  {journal} {Nature}\ } (\bibinfo {year}
  {2021})},\ \Eprint {http://arxiv.org/abs/2012.01208} {arXiv:2012.01208
  [cond-mat.str-el]} \BibitemShut {NoStop}%
\bibitem [{\citenamefont {Culo}\ \emph {et~al.}(2021)\citenamefont {Culo},
  \citenamefont {Duffy}, \citenamefont {Ayres}, \citenamefont {Berben},
  \citenamefont {Hsu}, \citenamefont {Hinlopen}, \citenamefont {Bernath},\ and\
  \citenamefont {Hussey}}]{Hussey21}%
  \BibitemOpen
  \bibfield  {author} {\bibinfo {author} {\bibfnamefont {M.}~\bibnamefont
  {Culo}}, \bibinfo {author} {\bibfnamefont {C.}~\bibnamefont {Duffy}},
  \bibinfo {author} {\bibfnamefont {J.}~\bibnamefont {Ayres}}, \bibinfo
  {author} {\bibfnamefont {M.}~\bibnamefont {Berben}}, \bibinfo {author}
  {\bibfnamefont {Y.-T.}\ \bibnamefont {Hsu}}, \bibinfo {author} {\bibfnamefont
  {R.~D.~H.}\ \bibnamefont {Hinlopen}}, \bibinfo {author} {\bibfnamefont
  {B.}~\bibnamefont {Bernath}}, \ and\ \bibinfo {author} {\bibfnamefont
  {N.~E.}\ \bibnamefont {Hussey}},\ }\bibfield  {title} {\enquote {\bibinfo
  {title} {{Possible superconductivity from incoherent carriers in overdoped
  cuprates}},}\ }\href {\doibase 10.21468/SciPostPhys.11.1.012} {\bibfield
  {journal} {\bibinfo  {journal} {SciPost Phys.}\ }\textbf {\bibinfo {volume}
  {11}},\ \bibinfo {pages} {12} (\bibinfo {year} {2021})}\BibitemShut {NoStop}%
\bibitem [{\citenamefont {Andersen}\ \emph {et~al.}(1994)\citenamefont
  {Andersen}, \citenamefont {Jepsen}, \citenamefont {Liechtenstein},\ and\
  \citenamefont {Mazin}}]{OKA}%
  \BibitemOpen
  \bibfield  {author} {\bibinfo {author} {\bibfnamefont {O.~K.}\ \bibnamefont
  {Andersen}}, \bibinfo {author} {\bibfnamefont {O.}~\bibnamefont {Jepsen}},
  \bibinfo {author} {\bibfnamefont {A.~I.}\ \bibnamefont {Liechtenstein}}, \
  and\ \bibinfo {author} {\bibfnamefont {I.~I.}\ \bibnamefont {Mazin}},\
  }\bibfield  {title} {\enquote {\bibinfo {title} {Plane dimpling and
  saddle-point bifurcation in the band structures of optimally doped
  high-temperature superconductors: A tight-binding model},}\ }\href {\doibase
  10.1103/PhysRevB.49.4145} {\bibfield  {journal} {\bibinfo  {journal} {Phys.
  Rev. B}\ }\textbf {\bibinfo {volume} {49}},\ \bibinfo {pages} {4145--4157}
  (\bibinfo {year} {1994})}\BibitemShut {NoStop}%
\bibitem [{\citenamefont {Collignon}\ \emph {et~al.}(2017)\citenamefont
  {Collignon}, \citenamefont {Badoux}, \citenamefont {Afshar}, \citenamefont
  {Michon}, \citenamefont {Lalibert\'e}, \citenamefont {Cyr-Choini\`ere},
  \citenamefont {Zhou}, \citenamefont {Licciardello}, \citenamefont {Wiedmann},
  \citenamefont {Doiron-Leyraud},\ and\ \citenamefont
  {Taillefer}}]{Taillefer_FS}%
  \BibitemOpen
  \bibfield  {author} {\bibinfo {author} {\bibfnamefont {C.}~\bibnamefont
  {Collignon}}, \bibinfo {author} {\bibfnamefont {S.}~\bibnamefont {Badoux}},
  \bibinfo {author} {\bibfnamefont {S.~A.~A.}\ \bibnamefont {Afshar}}, \bibinfo
  {author} {\bibfnamefont {B.}~\bibnamefont {Michon}}, \bibinfo {author}
  {\bibfnamefont {F.}~\bibnamefont {Lalibert\'e}}, \bibinfo {author}
  {\bibfnamefont {O.}~\bibnamefont {Cyr-Choini\`ere}}, \bibinfo {author}
  {\bibfnamefont {J.-S.}\ \bibnamefont {Zhou}}, \bibinfo {author}
  {\bibfnamefont {S.}~\bibnamefont {Licciardello}}, \bibinfo {author}
  {\bibfnamefont {S.}~\bibnamefont {Wiedmann}}, \bibinfo {author}
  {\bibfnamefont {N.}~\bibnamefont {Doiron-Leyraud}}, \ and\ \bibinfo {author}
  {\bibfnamefont {Louis}\ \bibnamefont {Taillefer}},\ }\bibfield  {title}
  {\enquote {\bibinfo {title} {Fermi-surface transformation across the
  pseudogap critical point of the cuprate superconductor
  {${\mathrm{La}}_{1.6\ensuremath{-}x}{\mathrm{Nd}}_{0.4}{\mathrm{Sr}}_{x}{\mathrm{CuO}}_{4}$}},}\
  }\href {\doibase 10.1103/PhysRevB.95.224517} {\bibfield  {journal} {\bibinfo
  {journal} {Phys. Rev. B}\ }\textbf {\bibinfo {volume} {95}},\ \bibinfo
  {pages} {224517} (\bibinfo {year} {2017})}\BibitemShut {NoStop}%
\bibitem [{\citenamefont {Rohringer}\ \emph {et~al.}(2018)\citenamefont
  {Rohringer}, \citenamefont {Hafermann}, \citenamefont {Toschi}, \citenamefont
  {Katanin}, \citenamefont {Antipov}, \citenamefont {Katsnelson}, \citenamefont
  {Lichtenstein}, \citenamefont {Rubtsov},\ and\ \citenamefont
  {Held}}]{DF_Rev}%
  \BibitemOpen
  \bibfield  {author} {\bibinfo {author} {\bibfnamefont {G.}~\bibnamefont
  {Rohringer}}, \bibinfo {author} {\bibfnamefont {H.}~\bibnamefont
  {Hafermann}}, \bibinfo {author} {\bibfnamefont {A.}~\bibnamefont {Toschi}},
  \bibinfo {author} {\bibfnamefont {A.~A.}\ \bibnamefont {Katanin}}, \bibinfo
  {author} {\bibfnamefont {A.~E.}\ \bibnamefont {Antipov}}, \bibinfo {author}
  {\bibfnamefont {M.~I.}\ \bibnamefont {Katsnelson}}, \bibinfo {author}
  {\bibfnamefont {A.~I.}\ \bibnamefont {Lichtenstein}}, \bibinfo {author}
  {\bibfnamefont {A.~N.}\ \bibnamefont {Rubtsov}}, \ and\ \bibinfo {author}
  {\bibfnamefont {K.}~\bibnamefont {Held}},\ }\bibfield  {title} {\enquote
  {\bibinfo {title} {Diagrammatic routes to nonlocal correlations beyond
  dynamical mean field theory},}\ }\href {\doibase
  10.1103/RevModPhys.90.025003} {\bibfield  {journal} {\bibinfo  {journal}
  {Rev. Mod. Phys.}\ }\textbf {\bibinfo {volume} {90}},\ \bibinfo {pages}
  {025003} (\bibinfo {year} {2018})}\BibitemShut {NoStop}%
\bibitem [{\citenamefont {Hafermann}\ \emph {et~al.}(2008)\citenamefont
  {Hafermann}, \citenamefont {Brener}, \citenamefont {Rubtsov}, \citenamefont
  {Katsnelson},\ and\ \citenamefont {Lichtenstein}}]{HH_CDF}%
  \BibitemOpen
  \bibfield  {author} {\bibinfo {author} {\bibfnamefont {H.}~\bibnamefont
  {Hafermann}}, \bibinfo {author} {\bibfnamefont {S.}~\bibnamefont {Brener}},
  \bibinfo {author} {\bibfnamefont {A.~N.}\ \bibnamefont {Rubtsov}}, \bibinfo
  {author} {\bibfnamefont {M.~I.}\ \bibnamefont {Katsnelson}}, \ and\ \bibinfo
  {author} {\bibfnamefont {A.~I.}\ \bibnamefont {Lichtenstein}},\ }\bibfield
  {title} {\enquote {\bibinfo {title} {Cluster dual fermion approach to
  nonlocal correlations},}\ }\href {\doibase 10.1134/S0021364007220134}
  {\bibfield  {journal} {\bibinfo  {journal} {JETP Letters}\ }\textbf {\bibinfo
  {volume} {86}},\ \bibinfo {pages} {677--682} (\bibinfo {year}
  {2008})}\BibitemShut {NoStop}%
\bibitem [{\citenamefont {Pavarini}\ \emph {et~al.}(2001)\citenamefont
  {Pavarini}, \citenamefont {Dasgupta}, \citenamefont {Saha-Dasgupta},
  \citenamefont {Jepsen},\ and\ \citenamefont {Andersen}}]{Pavarini}%
  \BibitemOpen
  \bibfield  {author} {\bibinfo {author} {\bibfnamefont {E.}~\bibnamefont
  {Pavarini}}, \bibinfo {author} {\bibfnamefont {I.}~\bibnamefont {Dasgupta}},
  \bibinfo {author} {\bibfnamefont {T.}~\bibnamefont {Saha-Dasgupta}}, \bibinfo
  {author} {\bibfnamefont {O.}~\bibnamefont {Jepsen}}, \ and\ \bibinfo {author}
  {\bibfnamefont {O.~K.}\ \bibnamefont {Andersen}},\ }\bibfield  {title}
  {\enquote {\bibinfo {title} {Band-structure trend in hole-doped cuprates and
  correlation with ${\mathit{t}}_{\mathit{c}\mathrm{max}}$},}\ }\href {\doibase
  10.1103/PhysRevLett.87.047003} {\bibfield  {journal} {\bibinfo  {journal}
  {Phys. Rev. Lett.}\ }\textbf {\bibinfo {volume} {87}},\ \bibinfo {pages}
  {047003} (\bibinfo {year} {2001})}\BibitemShut {NoStop}%
\bibitem [{\citenamefont {Harland}\ \emph {et~al.}(2016)\citenamefont
  {Harland}, \citenamefont {Katsnelson},\ and\ \citenamefont
  {Lichtenstein}}]{Harland16}%
  \BibitemOpen
  \bibfield  {author} {\bibinfo {author} {\bibfnamefont {Malte}\ \bibnamefont
  {Harland}}, \bibinfo {author} {\bibfnamefont {Mikhail~I.}\ \bibnamefont
  {Katsnelson}}, \ and\ \bibinfo {author} {\bibfnamefont {Alexander~I.}\
  \bibnamefont {Lichtenstein}},\ }\bibfield  {title} {\enquote {\bibinfo
  {title} {Plaquette valence bond theory of high-temperature
  superconductivity},}\ }\href {\doibase 10.1103/PhysRevB.94.125133} {\bibfield
   {journal} {\bibinfo  {journal} {Phys. Rev. B}\ }\textbf {\bibinfo {volume}
  {94}},\ \bibinfo {pages} {125133} (\bibinfo {year} {2016})}\BibitemShut
  {NoStop}%
\bibitem [{\citenamefont {Lichtenstein}\ and\ \citenamefont
  {Katsnelson}(2000)}]{dwCDMFT}%
  \BibitemOpen
  \bibfield  {author} {\bibinfo {author} {\bibfnamefont {A.~I.}\ \bibnamefont
  {Lichtenstein}}\ and\ \bibinfo {author} {\bibfnamefont {M.~I.}\ \bibnamefont
  {Katsnelson}},\ }\bibfield  {title} {\enquote {\bibinfo {title}
  {Antiferromagnetism and d-wave superconductivity in cuprates: A cluster
  dynamical mean-field theory},}\ }\href {\doibase 10.1103/PhysRevB.62.R9283}
  {\bibfield  {journal} {\bibinfo  {journal} {Phys. Rev. B}\ }\textbf {\bibinfo
  {volume} {62}},\ \bibinfo {pages} {R9283--R9286} (\bibinfo {year}
  {2000})}\BibitemShut {NoStop}%
\bibitem [{\citenamefont {Altman}\ and\ \citenamefont
  {Auerbach}(2002)}]{Auerbach}%
  \BibitemOpen
  \bibfield  {author} {\bibinfo {author} {\bibfnamefont {Ehud}\ \bibnamefont
  {Altman}}\ and\ \bibinfo {author} {\bibfnamefont {Assa}\ \bibnamefont
  {Auerbach}},\ }\bibfield  {title} {\enquote {\bibinfo {title} {Plaquette
  boson-fermion model of cuprates},}\ }\href {\doibase
  10.1103/PhysRevB.65.104508} {\bibfield  {journal} {\bibinfo  {journal} {Phys.
  Rev. B}\ }\textbf {\bibinfo {volume} {65}},\ \bibinfo {pages} {104508}
  (\bibinfo {year} {2002})}\BibitemShut {NoStop}%
\bibitem [{\citenamefont {Hewson}(1993)}]{hewson}%
  \BibitemOpen
  \bibfield  {author} {\bibinfo {author} {\bibfnamefont {Alexander~Cyril}\
  \bibnamefont {Hewson}},\ }\href {\doibase 10.1017/CBO9780511470752} {\emph
  {\bibinfo {title} {{The Kondo Problem to Heavy Fermions}}}}\ (\bibinfo
  {publisher} {Cambridge University Press},\ \bibinfo {address} {Cambridge},\
  \bibinfo {year} {1993})\BibitemShut {NoStop}%
\bibitem [{\citenamefont {Krivenko}\ \emph {et~al.}(2010)\citenamefont
  {Krivenko}, \citenamefont {Rubtsov}, \citenamefont {Katsnelson},\ and\
  \citenamefont {Lichtenstein}}]{Krivenko}%
  \BibitemOpen
  \bibfield  {author} {\bibinfo {author} {\bibfnamefont {I.~S.}\ \bibnamefont
  {Krivenko}}, \bibinfo {author} {\bibfnamefont {A.~N.}\ \bibnamefont
  {Rubtsov}}, \bibinfo {author} {\bibfnamefont {M.~I.}\ \bibnamefont
  {Katsnelson}}, \ and\ \bibinfo {author} {\bibfnamefont {A.~I.}\ \bibnamefont
  {Lichtenstein}},\ }\bibfield  {title} {\enquote {\bibinfo {title} {Analytical
  approximation for single-impurity anderson model},}\ }\href {\doibase
  10.1134/S0021364010060123} {\bibfield  {journal} {\bibinfo  {journal} {JETP
  Letters}\ }\textbf {\bibinfo {volume} {91}},\ \bibinfo {pages} {319--325}
  (\bibinfo {year} {2010})}\BibitemShut {NoStop}%
\bibitem [{\citenamefont {Wu}\ \emph {et~al.}(2017)\citenamefont {Wu},
  \citenamefont {Ferrero}, \citenamefont {Georges},\ and\ \citenamefont
  {Kozik}}]{Wei_point}%
  \BibitemOpen
  \bibfield  {author} {\bibinfo {author} {\bibfnamefont {Wei}\ \bibnamefont
  {Wu}}, \bibinfo {author} {\bibfnamefont {Michel}\ \bibnamefont {Ferrero}},
  \bibinfo {author} {\bibfnamefont {Antoine}\ \bibnamefont {Georges}}, \ and\
  \bibinfo {author} {\bibfnamefont {Evgeny}\ \bibnamefont {Kozik}},\ }\bibfield
   {title} {\enquote {\bibinfo {title} {Controlling feynman diagrammatic
  expansions: Physical nature of the pseudogap in the two-dimensional hubbard
  model},}\ }\href {\doibase 10.1103/PhysRevB.96.041105} {\bibfield  {journal}
  {\bibinfo  {journal} {Phys. Rev. B}\ }\textbf {\bibinfo {volume} {96}},\
  \bibinfo {pages} {041105} (\bibinfo {year} {2017})}\BibitemShut {NoStop}%
\bibitem [{\citenamefont {Harland}\ \emph {et~al.}(2020)\citenamefont
  {Harland}, \citenamefont {Brener}, \citenamefont {Katsnelson},\ and\
  \citenamefont {Lichtenstein}}]{Harland20}%
  \BibitemOpen
  \bibfield  {author} {\bibinfo {author} {\bibfnamefont {Malte}\ \bibnamefont
  {Harland}}, \bibinfo {author} {\bibfnamefont {Sergey}\ \bibnamefont
  {Brener}}, \bibinfo {author} {\bibfnamefont {Mikhail~I.}\ \bibnamefont
  {Katsnelson}}, \ and\ \bibinfo {author} {\bibfnamefont {Alexander~I.}\
  \bibnamefont {Lichtenstein}},\ }\bibfield  {title} {\enquote {\bibinfo
  {title} {Exactly solvable model of strongly correlated $d$-wave
  superconductivity},}\ }\href {\doibase 10.1103/PhysRevB.101.045119}
  {\bibfield  {journal} {\bibinfo  {journal} {Phys. Rev. B}\ }\textbf {\bibinfo
  {volume} {101}},\ \bibinfo {pages} {045119} (\bibinfo {year}
  {2020})}\BibitemShut {NoStop}%
\bibitem [{\citenamefont {Rubtsov}\ \emph {et~al.}(2008)\citenamefont
  {Rubtsov}, \citenamefont {Katsnelson},\ and\ \citenamefont
  {Lichtenstein}}]{RKL08}%
  \BibitemOpen
  \bibfield  {author} {\bibinfo {author} {\bibfnamefont {A.~N.}\ \bibnamefont
  {Rubtsov}}, \bibinfo {author} {\bibfnamefont {M.~I.}\ \bibnamefont
  {Katsnelson}}, \ and\ \bibinfo {author} {\bibfnamefont {A.~I.}\ \bibnamefont
  {Lichtenstein}},\ }\bibfield  {title} {\enquote {\bibinfo {title} {Dual
  fermion approach to nonlocal correlations in the hubbard model},}\ }\href
  {\doibase 10.1103/PhysRevB.77.033101} {\bibfield  {journal} {\bibinfo
  {journal} {Phys. Rev. B}\ }\textbf {\bibinfo {volume} {77}},\ \bibinfo
  {pages} {033101} (\bibinfo {year} {2008})}\BibitemShut {NoStop}%
\bibitem [{\citenamefont {Brener}\ \emph {et~al.}(2020)\citenamefont {Brener},
  \citenamefont {Stepanov}, \citenamefont {Rubtsov}, \citenamefont
  {Katsnelson},\ and\ \citenamefont {Lichtenstein}}]{Brener_refsys}%
  \BibitemOpen
  \bibfield  {author} {\bibinfo {author} {\bibfnamefont {Sergey}\ \bibnamefont
  {Brener}}, \bibinfo {author} {\bibfnamefont {Evgeny~A.}\ \bibnamefont
  {Stepanov}}, \bibinfo {author} {\bibfnamefont {Alexey~N.}\ \bibnamefont
  {Rubtsov}}, \bibinfo {author} {\bibfnamefont {Mikhail~I.}\ \bibnamefont
  {Katsnelson}}, \ and\ \bibinfo {author} {\bibfnamefont {Alexander~I.}\
  \bibnamefont {Lichtenstein}},\ }\bibfield  {title} {\enquote {\bibinfo
  {title} {{Dual fermion method as a prototype of generic reference-system
  approach for correlated fermions}},}\ }\href {\doibase
  https://doi.org/10.1016/j.aop.2020.168310} {\bibfield  {journal} {\bibinfo
  {journal} {Annals of Physics}\ }\textbf {\bibinfo {volume} {422}},\ \bibinfo
  {pages} {168310} (\bibinfo {year} {2020})}\BibitemShut {NoStop}%
\bibitem [{\citenamefont {Bagrov}\ \emph {et~al.}(2020)\citenamefont {Bagrov},
  \citenamefont {Danilov}, \citenamefont {Brener}, \citenamefont {Harland},
  \citenamefont {Lichtenstein},\ and\ \citenamefont
  {Katsnelson}}]{Bagrov_4x4Network}%
  \BibitemOpen
  \bibfield  {author} {\bibinfo {author} {\bibfnamefont {Andrey~A.}\
  \bibnamefont {Bagrov}}, \bibinfo {author} {\bibfnamefont {Mikhail}\
  \bibnamefont {Danilov}}, \bibinfo {author} {\bibfnamefont {Sergey}\
  \bibnamefont {Brener}}, \bibinfo {author} {\bibfnamefont {Malte}\
  \bibnamefont {Harland}}, \bibinfo {author} {\bibfnamefont {Alexander~I.}\
  \bibnamefont {Lichtenstein}}, \ and\ \bibinfo {author} {\bibfnamefont
  {Mikhail~I.}\ \bibnamefont {Katsnelson}},\ }\bibfield  {title} {\enquote
  {\bibinfo {title} {Detecting quantum critical points in the $t-t'$
  {Fermi-Hubbard} model via complex network theory},}\ }\href {\doibase
  10.1038/s41598-020-77513-0} {\bibfield  {journal} {\bibinfo  {journal}
  {Scientific Reports}\ }\textbf {\bibinfo {volume} {10}},\ \bibinfo {pages}
  {20470} (\bibinfo {year} {2020})}\BibitemShut {NoStop}%
\bibitem [{\citenamefont {Hafermann}\ \emph {et~al.}(2012)\citenamefont
  {Hafermann}, \citenamefont {Lechermann}, \citenamefont {Rubtsov},
  \citenamefont {Katsnelson}, \citenamefont {Georges},\ and\ \citenamefont
  {Lichtenstein}}]{leshouches}%
  \BibitemOpen
  \bibfield  {author} {\bibinfo {author} {\bibfnamefont {Hartmut}\ \bibnamefont
  {Hafermann}}, \bibinfo {author} {\bibfnamefont {Frank}\ \bibnamefont
  {Lechermann}}, \bibinfo {author} {\bibfnamefont {Alexey~N.}\ \bibnamefont
  {Rubtsov}}, \bibinfo {author} {\bibfnamefont {Mikhail~I.}\ \bibnamefont
  {Katsnelson}}, \bibinfo {author} {\bibfnamefont {Antoine}\ \bibnamefont
  {Georges}}, \ and\ \bibinfo {author} {\bibfnamefont {Alexander~I.}\
  \bibnamefont {Lichtenstein}},\ }\bibfield  {title} {\enquote {\bibinfo
  {title} {Strong electronic correlations: Dynamical mean-field theory and
  beyond},}\ }in\ \href@noop {} {\emph {\bibinfo {booktitle} {Modern theories
  of many-particle systems in condensed matter physics}}},\ Vol.\ \bibinfo
  {volume} {843},\ \bibinfo {editor} {edited by\ \bibinfo {editor}
  {\bibfnamefont {Daniel~C}\ \bibnamefont {Cabra}}, \bibinfo {editor}
  {\bibfnamefont {Andreas}\ \bibnamefont {Honecker}}, \ and\ \bibinfo {editor}
  {\bibfnamefont {Pierre}\ \bibnamefont {Pujol}}}\ (\bibinfo  {publisher}
  {Springer Science \& Business Media},\ \bibinfo {year} {2012})\
  Chap.~\bibinfo {chapter} {4}\BibitemShut {NoStop}%
\bibitem [{\citenamefont {Takemori}\ \emph {et~al.}(2018)\citenamefont
  {Takemori}, \citenamefont {Koga},\ and\ \citenamefont
  {Hafermann}}]{Takemori18}%
  \BibitemOpen
  \bibfield  {author} {\bibinfo {author} {\bibfnamefont {Nayuta}\ \bibnamefont
  {Takemori}}, \bibinfo {author} {\bibfnamefont {Akihisa}\ \bibnamefont
  {Koga}}, \ and\ \bibinfo {author} {\bibfnamefont {Hartmut}\ \bibnamefont
  {Hafermann}},\ }\href@noop {} {\enquote {\bibinfo {title} {Intersite electron
  correlations on inhomogeneous lattices: a real-space dual fermion
  approach},}\ } (\bibinfo {year} {2018}),\ \Eprint
  {http://arxiv.org/abs/1801.02441} {arXiv:1801.02441 [cond-mat.str-el]}
  \BibitemShut {NoStop}%
\bibitem [{\citenamefont {Gull}\ \emph {et~al.}(2011)\citenamefont {Gull},
  \citenamefont {Millis}, \citenamefont {Lichtenstein}, \citenamefont
  {Rubtsov}, \citenamefont {Troyer},\ and\ \citenamefont {Werner}}]{CTQMC}%
  \BibitemOpen
  \bibfield  {author} {\bibinfo {author} {\bibfnamefont {Emanuel}\ \bibnamefont
  {Gull}}, \bibinfo {author} {\bibfnamefont {Andrew~J.}\ \bibnamefont
  {Millis}}, \bibinfo {author} {\bibfnamefont {Alexander~I.}\ \bibnamefont
  {Lichtenstein}}, \bibinfo {author} {\bibfnamefont {Alexey~N.}\ \bibnamefont
  {Rubtsov}}, \bibinfo {author} {\bibfnamefont {Matthias}\ \bibnamefont
  {Troyer}}, \ and\ \bibinfo {author} {\bibfnamefont {Philipp}\ \bibnamefont
  {Werner}},\ }\bibfield  {title} {\enquote {\bibinfo {title} {Continuous-time
  monte carlo methods for quantum impurity models},}\ }\href {\doibase
  10.1103/RevModPhys.83.349} {\bibfield  {journal} {\bibinfo  {journal} {Rev.
  Mod. Phys.}\ }\textbf {\bibinfo {volume} {83}},\ \bibinfo {pages} {349--404}
  (\bibinfo {year} {2011})}\BibitemShut {NoStop}%
\bibitem [{\citenamefont {Stratonovich}(1957)}]{Stratonovich_HS}%
  \BibitemOpen
  \bibfield  {author} {\bibinfo {author} {\bibfnamefont {R.~L.}\ \bibnamefont
  {Stratonovich}},\ }\bibfield  {title} {\enquote {\bibinfo {title} {On a
  method of calculating quantum distribution functions},}\ }in\ \href@noop {}
  {\emph {\bibinfo {booktitle} {Soviet Physics Doklady}}},\ Vol.~\bibinfo
  {volume} {2}\ (\bibinfo {year} {1957})\ p.\ \bibinfo {pages}
  {416}\BibitemShut {NoStop}%
\bibitem [{\citenamefont {Hubbard}(1959)}]{Hubbard_HS}%
  \BibitemOpen
  \bibfield  {author} {\bibinfo {author} {\bibfnamefont {J.}~\bibnamefont
  {Hubbard}},\ }\bibfield  {title} {\enquote {\bibinfo {title} {Calculation of
  partition functions},}\ }\href {\doibase 10.1103/PhysRevLett.3.77} {\bibfield
   {journal} {\bibinfo  {journal} {Phys. Rev. Lett.}\ }\textbf {\bibinfo
  {volume} {3}},\ \bibinfo {pages} {77--78} (\bibinfo {year}
  {1959})}\BibitemShut {NoStop}%
\bibitem [{\citenamefont {van Loon}(2021)}]{DF2}%
  \BibitemOpen
  \bibfield  {author} {\bibinfo {author} {\bibfnamefont {Erik G C~P}\
  \bibnamefont {van Loon}},\ }\bibfield  {title} {\enquote {\bibinfo {title}
  {Second-order dual fermion for multi-orbital systems},}\ }\href {\doibase
  10.1088/1361-648x/abd9ed} {\bibfield  {journal} {\bibinfo  {journal} {Journal
  of Physics: Condensed Matter}\ }\textbf {\bibinfo {volume} {33}},\ \bibinfo
  {pages} {135601} (\bibinfo {year} {2021})}\BibitemShut {NoStop}%
\bibitem [{\citenamefont {Sarker}(1988)}]{Sarker_1988}%
  \BibitemOpen
  \bibfield  {author} {\bibinfo {author} {\bibfnamefont {S~K}\ \bibnamefont
  {Sarker}},\ }\bibfield  {title} {\enquote {\bibinfo {title} {A new functional
  integral formalism for strongly correlated fermi systems},}\ }\href {\doibase
  10.1088/0022-3719/21/18/002} {\bibfield  {journal} {\bibinfo  {journal}
  {Journal of Physics C: Solid State Physics}\ }\textbf {\bibinfo {volume}
  {21}},\ \bibinfo {pages} {L667--L672} (\bibinfo {year} {1988})}\BibitemShut
  {NoStop}%
\bibitem [{\citenamefont {Pairault}\ \emph {et~al.}(1998)\citenamefont
  {Pairault}, \citenamefont {S\'en\'echal},\ and\ \citenamefont
  {Tremblay}}]{Pairault_PRL}%
  \BibitemOpen
  \bibfield  {author} {\bibinfo {author} {\bibfnamefont {St\'ephane}\
  \bibnamefont {Pairault}}, \bibinfo {author} {\bibfnamefont {David}\
  \bibnamefont {S\'en\'echal}}, \ and\ \bibinfo {author} {\bibfnamefont
  {A.-M.~S.}\ \bibnamefont {Tremblay}},\ }\bibfield  {title} {\enquote
  {\bibinfo {title} {Strong-coupling expansion for the {Hubbard} model},}\
  }\href {\doibase 10.1103/PhysRevLett.80.5389} {\bibfield  {journal} {\bibinfo
   {journal} {Phys. Rev. Lett.}\ }\textbf {\bibinfo {volume} {80}},\ \bibinfo
  {pages} {5389--5392} (\bibinfo {year} {1998})}\BibitemShut {NoStop}%
\bibitem [{\citenamefont {{Pairault, S.}}\ \emph {et~al.}(2000)\citenamefont
  {{Pairault, S.}}, \citenamefont {{S\'en\'echal, D.}},\ and\ \citenamefont
  {{A.-M.S. Tremblay}}}]{Pairault_EPJB}%
  \BibitemOpen
  \bibfield  {author} {\bibinfo {author} {\bibnamefont {{Pairault, S.}}},
  \bibinfo {author} {\bibnamefont {{S\'en\'echal, D.}}}, \ and\ \bibinfo
  {author} {\bibnamefont {{A.-M.S. Tremblay}}},\ }\bibfield  {title} {\enquote
  {\bibinfo {title} {Strong-coupling perturbation theory of the {Hubbard}
  model},}\ }\href {\doibase 10.1007/s100510070253} {\bibfield  {journal}
  {\bibinfo  {journal} {Eur. Phys. J. B}\ }\textbf {\bibinfo {volume} {16}},\
  \bibinfo {pages} {85--105} (\bibinfo {year} {2000})}\BibitemShut {NoStop}%
\bibitem [{\citenamefont {Dupuis}\ and\ \citenamefont
  {Pairault}(2000)}]{Dupuis_Pairault}%
  \BibitemOpen
  \bibfield  {author} {\bibinfo {author} {\bibfnamefont {N.}~\bibnamefont
  {Dupuis}}\ and\ \bibinfo {author} {\bibfnamefont {S.}~\bibnamefont
  {Pairault}},\ }\bibfield  {title} {\enquote {\bibinfo {title} {A
  strong-coupling expansion for the {Hubbard} model},}\ }\href {\doibase
  10.1142/S0217979200002430} {\bibfield  {journal} {\bibinfo  {journal}
  {International Journal of Modern Physics B}\ }\textbf {\bibinfo {volume}
  {14}},\ \bibinfo {pages} {2529--2560} (\bibinfo {year} {2000})},\ \Eprint
  {http://arxiv.org/abs/https://doi.org/10.1142/S0217979200002430}
  {https://doi.org/10.1142/S0217979200002430} \BibitemShut {NoStop}%
\bibitem [{\citenamefont {Dupuis}(2001)}]{Dupuis}%
  \BibitemOpen
  \bibfield  {author} {\bibinfo {author} {\bibfnamefont {N.}~\bibnamefont
  {Dupuis}},\ }\bibfield  {title} {\enquote {\bibinfo {title} {A new approach
  to strongly correlated fermion systems: the spin–particle–hole
  coherent-state path integral},}\ }\href {\doibase
  https://doi.org/10.1016/S0550-3213(01)00465-5} {\bibfield  {journal}
  {\bibinfo  {journal} {Nuclear Physics B}\ }\textbf {\bibinfo {volume}
  {618}},\ \bibinfo {pages} {617 -- 649} (\bibinfo {year} {2001})}\BibitemShut
  {NoStop}%
\bibitem [{\citenamefont {Gros}\ and\ \citenamefont
  {Valent\'{i}}(1993)}]{Valenti_CPT}%
  \BibitemOpen
  \bibfield  {author} {\bibinfo {author} {\bibfnamefont {Claudius}\
  \bibnamefont {Gros}}\ and\ \bibinfo {author} {\bibfnamefont {Roser}\
  \bibnamefont {Valent\'{i}}},\ }\bibfield  {title} {\enquote {\bibinfo {title}
  {Cluster expansion for the self-energy: A simple many-body method for
  interpreting the photoemission spectra of correlated fermi systems},}\ }\href
  {\doibase 10.1103/PhysRevB.48.418} {\bibfield  {journal} {\bibinfo  {journal}
  {Phys. Rev. B}\ }\textbf {\bibinfo {volume} {48}},\ \bibinfo {pages}
  {418--425} (\bibinfo {year} {1993})}\BibitemShut {NoStop}%
\bibitem [{\citenamefont {Hafermann}(2009)}]{hafermann_thesis}%
  \BibitemOpen
  \bibfield  {author} {\bibinfo {author} {\bibfnamefont {H.}~\bibnamefont
  {Hafermann}},\ }\emph {\bibinfo {title} {Numerical Approaches to Spatial
  Correlations in Strongly Interacting Fermion Systems}},\ \href@noop {} {Ph.D.
  thesis},\ \bibinfo  {school} {Universität Hamburg} (\bibinfo {year}
  {2009})\BibitemShut {NoStop}%
\bibitem [{\citenamefont {Hafermann}\ \emph
  {et~al.}(2009{\natexlab{a}})\citenamefont {Hafermann}, \citenamefont {Li},
  \citenamefont {Rubtsov}, \citenamefont {Katsnelson}, \citenamefont
  {Lichtenstein},\ and\ \citenamefont {Monien}}]{LDFA}%
  \BibitemOpen
  \bibfield  {author} {\bibinfo {author} {\bibfnamefont {H.}~\bibnamefont
  {Hafermann}}, \bibinfo {author} {\bibfnamefont {G.}~\bibnamefont {Li}},
  \bibinfo {author} {\bibfnamefont {A.~N.}\ \bibnamefont {Rubtsov}}, \bibinfo
  {author} {\bibfnamefont {M.~I.}\ \bibnamefont {Katsnelson}}, \bibinfo
  {author} {\bibfnamefont {A.~I.}\ \bibnamefont {Lichtenstein}}, \ and\
  \bibinfo {author} {\bibfnamefont {H.}~\bibnamefont {Monien}},\ }\bibfield
  {title} {\enquote {\bibinfo {title} {Efficient perturbation theory for
  quantum lattice models},}\ }\href {\doibase 10.1103/PhysRevLett.102.206401}
  {\bibfield  {journal} {\bibinfo  {journal} {Phys. Rev. Lett.}\ }\textbf
  {\bibinfo {volume} {102}},\ \bibinfo {pages} {206401} (\bibinfo {year}
  {2009}{\natexlab{a}})}\BibitemShut {NoStop}%
\bibitem [{\citenamefont {Krien}\ \emph {et~al.}(2020)\citenamefont {Krien},
  \citenamefont {Valli}, \citenamefont {Chalupa}, \citenamefont {Capone},
  \citenamefont {Lichtenstein},\ and\ \citenamefont {Toschi}}]{DF_Parq}%
  \BibitemOpen
  \bibfield  {author} {\bibinfo {author} {\bibfnamefont {Friedrich}\
  \bibnamefont {Krien}}, \bibinfo {author} {\bibfnamefont {Angelo}\
  \bibnamefont {Valli}}, \bibinfo {author} {\bibfnamefont {Patrick}\
  \bibnamefont {Chalupa}}, \bibinfo {author} {\bibfnamefont {Massimo}\
  \bibnamefont {Capone}}, \bibinfo {author} {\bibfnamefont {Alexander~I.}\
  \bibnamefont {Lichtenstein}}, \ and\ \bibinfo {author} {\bibfnamefont
  {Alessandro}\ \bibnamefont {Toschi}},\ }\bibfield  {title} {\enquote
  {\bibinfo {title} {Boson-exchange parquet solver for dual fermions},}\ }\href
  {\doibase 10.1103/PhysRevB.102.195131} {\bibfield  {journal} {\bibinfo
  {journal} {Phys. Rev. B}\ }\textbf {\bibinfo {volume} {102}},\ \bibinfo
  {pages} {195131} (\bibinfo {year} {2020})}\BibitemShut {NoStop}%
\bibitem [{\citenamefont {Iskakov}\ \emph {et~al.}(2016)\citenamefont
  {Iskakov}, \citenamefont {Antipov},\ and\ \citenamefont {Gull}}]{DiagDFQMC1}%
  \BibitemOpen
  \bibfield  {author} {\bibinfo {author} {\bibfnamefont {Sergei}\ \bibnamefont
  {Iskakov}}, \bibinfo {author} {\bibfnamefont {Andrey~E.}\ \bibnamefont
  {Antipov}}, \ and\ \bibinfo {author} {\bibfnamefont {Emanuel}\ \bibnamefont
  {Gull}},\ }\bibfield  {title} {\enquote {\bibinfo {title} {Diagrammatic monte
  carlo for dual fermions},}\ }\href {\doibase 10.1103/PhysRevB.94.035102}
  {\bibfield  {journal} {\bibinfo  {journal} {Phys. Rev. B}\ }\textbf {\bibinfo
  {volume} {94}},\ \bibinfo {pages} {035102} (\bibinfo {year}
  {2016})}\BibitemShut {NoStop}%
\bibitem [{\citenamefont {Gukelberger}\ \emph {et~al.}(2017)\citenamefont
  {Gukelberger}, \citenamefont {Kozik},\ and\ \citenamefont
  {Hafermann}}]{DiagDFQMC2}%
  \BibitemOpen
  \bibfield  {author} {\bibinfo {author} {\bibfnamefont {Jan}\ \bibnamefont
  {Gukelberger}}, \bibinfo {author} {\bibfnamefont {Evgeny}\ \bibnamefont
  {Kozik}}, \ and\ \bibinfo {author} {\bibfnamefont {Hartmut}\ \bibnamefont
  {Hafermann}},\ }\bibfield  {title} {\enquote {\bibinfo {title} {Diagrammatic
  monte carlo approach for diagrammatic extensions of dynamical mean-field
  theory: Convergence analysis of the dual fermion technique},}\ }\href
  {\doibase 10.1103/PhysRevB.96.035152} {\bibfield  {journal} {\bibinfo
  {journal} {Phys. Rev. B}\ }\textbf {\bibinfo {volume} {96}},\ \bibinfo
  {pages} {035152} (\bibinfo {year} {2017})}\BibitemShut {NoStop}%
\bibitem [{\citenamefont {Vandelli}\ \emph {et~al.}(2020)\citenamefont
  {Vandelli}, \citenamefont {Harkov}, \citenamefont {Stepanov}, \citenamefont
  {Gukelberger}, \citenamefont {Kozik}, \citenamefont {Rubio},\ and\
  \citenamefont {Lichtenstein}}]{DiagDBQMC}%
  \BibitemOpen
  \bibfield  {author} {\bibinfo {author} {\bibfnamefont {M.}~\bibnamefont
  {Vandelli}}, \bibinfo {author} {\bibfnamefont {V.}~\bibnamefont {Harkov}},
  \bibinfo {author} {\bibfnamefont {E.~A.}\ \bibnamefont {Stepanov}}, \bibinfo
  {author} {\bibfnamefont {J.}~\bibnamefont {Gukelberger}}, \bibinfo {author}
  {\bibfnamefont {E.}~\bibnamefont {Kozik}}, \bibinfo {author} {\bibfnamefont
  {A.}~\bibnamefont {Rubio}}, \ and\ \bibinfo {author} {\bibfnamefont {A.~I.}\
  \bibnamefont {Lichtenstein}},\ }\bibfield  {title} {\enquote {\bibinfo
  {title} {Dual boson diagrammatic monte carlo approach applied to the extended
  hubbard model},}\ }\href {\doibase 10.1103/PhysRevB.102.195109} {\bibfield
  {journal} {\bibinfo  {journal} {Phys. Rev. B}\ }\textbf {\bibinfo {volume}
  {102}},\ \bibinfo {pages} {195109} (\bibinfo {year} {2020})}\BibitemShut
  {NoStop}%
\bibitem [{\citenamefont {van Loon}(2020)}]{DF2github}%
  \BibitemOpen
  \bibfield  {author} {\bibinfo {author} {\bibfnamefont {E.G.C.P.}\
  \bibnamefont {van Loon}},\ }\href@noop {} {\enquote {\bibinfo {title}
  {dualfermion},}\ }\bibinfo {howpublished}
  {\url{https://github.com/egcpvanloon/dualfermion/}} (\bibinfo {year}
  {2020})\BibitemShut {NoStop}%
\bibitem [{\citenamefont {Parcollet}\ \emph {et~al.}(2015)\citenamefont
  {Parcollet}, \citenamefont {Ferrero}, \citenamefont {Ayral}, \citenamefont
  {Hafermann}, \citenamefont {Krivenko}, \citenamefont {Messio},\ and\
  \citenamefont {Seth}}]{triqs}%
  \BibitemOpen
  \bibfield  {author} {\bibinfo {author} {\bibfnamefont {Olivier}\ \bibnamefont
  {Parcollet}}, \bibinfo {author} {\bibfnamefont {Michel}\ \bibnamefont
  {Ferrero}}, \bibinfo {author} {\bibfnamefont {Thomas}\ \bibnamefont {Ayral}},
  \bibinfo {author} {\bibfnamefont {Hartmut}\ \bibnamefont {Hafermann}},
  \bibinfo {author} {\bibfnamefont {Igor}\ \bibnamefont {Krivenko}}, \bibinfo
  {author} {\bibfnamefont {Laura}\ \bibnamefont {Messio}}, \ and\ \bibinfo
  {author} {\bibfnamefont {Priyanka}\ \bibnamefont {Seth}},\ }\bibfield
  {title} {\enquote {\bibinfo {title} {Triqs: A toolbox for research on
  interacting quantum systems},}\ }\href {\doibase
  http://dx.doi.org/10.1016/j.cpc.2015.04.023} {\bibfield  {journal} {\bibinfo
  {journal} {Computer Physics Communications}\ }\textbf {\bibinfo {volume}
  {196}},\ \bibinfo {pages} {398 -- 415} (\bibinfo {year} {2015})}\BibitemShut
  {NoStop}%
\bibitem [{\citenamefont {Antipov}\ \emph {et~al.}(2017)\citenamefont
  {Antipov}, \citenamefont {Krivenko},\ and\ \citenamefont
  {Iskakov}}]{pomerol}%
  \BibitemOpen
  \bibfield  {author} {\bibinfo {author} {\bibfnamefont {Andrey~E.}\
  \bibnamefont {Antipov}}, \bibinfo {author} {\bibfnamefont {Igor}\
  \bibnamefont {Krivenko}}, \ and\ \bibinfo {author} {\bibfnamefont {Sergei}\
  \bibnamefont {Iskakov}},\ }\href {\doibase 10.5281/zenodo.825870} {\enquote
  {\bibinfo {title} {aeantipov/pomerol: 1.2},}\ } (\bibinfo {year}
  {2017})\BibitemShut {NoStop}%
\bibitem [{\citenamefont {Iskakov}\ \emph {et~al.}(2018)\citenamefont
  {Iskakov}, \citenamefont {Terletska},\ and\ \citenamefont {Gull}}]{Iskakov1}%
  \BibitemOpen
  \bibfield  {author} {\bibinfo {author} {\bibfnamefont {Sergei}\ \bibnamefont
  {Iskakov}}, \bibinfo {author} {\bibfnamefont {Hanna}\ \bibnamefont
  {Terletska}}, \ and\ \bibinfo {author} {\bibfnamefont {Emanuel}\ \bibnamefont
  {Gull}},\ }\bibfield  {title} {\enquote {\bibinfo {title} {Momentum-space
  cluster dual-fermion method},}\ }\href {\doibase 10.1103/PhysRevB.97.125114}
  {\bibfield  {journal} {\bibinfo  {journal} {Phys. Rev. B}\ }\textbf {\bibinfo
  {volume} {97}},\ \bibinfo {pages} {125114} (\bibinfo {year}
  {2018})}\BibitemShut {NoStop}%
\bibitem [{\citenamefont {Hafermann}\ \emph
  {et~al.}(2009{\natexlab{b}})\citenamefont {Hafermann}, \citenamefont {Jung},
  \citenamefont {Brener}, \citenamefont {Katsnelson}, \citenamefont {Rubtsov},\
  and\ \citenamefont {Lichtenstein}}]{Superpert}%
  \BibitemOpen
  \bibfield  {author} {\bibinfo {author} {\bibfnamefont {H.}~\bibnamefont
  {Hafermann}}, \bibinfo {author} {\bibfnamefont {C.}~\bibnamefont {Jung}},
  \bibinfo {author} {\bibfnamefont {S.}~\bibnamefont {Brener}}, \bibinfo
  {author} {\bibfnamefont {M.~I.}\ \bibnamefont {Katsnelson}}, \bibinfo
  {author} {\bibfnamefont {A.~N.}\ \bibnamefont {Rubtsov}}, \ and\ \bibinfo
  {author} {\bibfnamefont {A.~I.}\ \bibnamefont {Lichtenstein}},\ }\bibfield
  {title} {\enquote {\bibinfo {title} {Superperturbation solver for quantum
  impurity models},}\ }\href {\doibase 10.1209/0295-5075/85/27007} {\bibfield
  {journal} {\bibinfo  {journal} {{EPL} (Europhysics Letters)}\ }\textbf
  {\bibinfo {volume} {85}},\ \bibinfo {pages} {27007} (\bibinfo {year}
  {2009}{\natexlab{b}})}\BibitemShut {NoStop}%
\bibitem [{\citenamefont {Georges}\ \emph {et~al.}(1996)\citenamefont
  {Georges}, \citenamefont {Kotliar}, \citenamefont {Krauth},\ and\
  \citenamefont {Rozenberg}}]{GKKR96}%
  \BibitemOpen
  \bibfield  {author} {\bibinfo {author} {\bibfnamefont {Antoine}\ \bibnamefont
  {Georges}}, \bibinfo {author} {\bibfnamefont {Gabriel}\ \bibnamefont
  {Kotliar}}, \bibinfo {author} {\bibfnamefont {Werner}\ \bibnamefont
  {Krauth}}, \ and\ \bibinfo {author} {\bibfnamefont {Marcelo~J.}\ \bibnamefont
  {Rozenberg}},\ }\bibfield  {title} {\enquote {\bibinfo {title} {Dynamical
  mean-field theory of strongly correlated fermion systems and the limit of
  infinite dimensions},}\ }\href {\doibase 10.1103/RevModPhys.68.13} {\bibfield
   {journal} {\bibinfo  {journal} {Rev. Mod. Phys.}\ }\textbf {\bibinfo
  {volume} {68}},\ \bibinfo {pages} {13--125} (\bibinfo {year}
  {1996})}\BibitemShut {NoStop}%
\bibitem [{\citenamefont {Rohringer}\ \emph {et~al.}(2012)\citenamefont
  {Rohringer}, \citenamefont {Valli},\ and\ \citenamefont
  {Toschi}}]{Rohringer12}%
  \BibitemOpen
  \bibfield  {author} {\bibinfo {author} {\bibfnamefont {G.}~\bibnamefont
  {Rohringer}}, \bibinfo {author} {\bibfnamefont {A.}~\bibnamefont {Valli}}, \
  and\ \bibinfo {author} {\bibfnamefont {A.}~\bibnamefont {Toschi}},\
  }\bibfield  {title} {\enquote {\bibinfo {title} {Local electronic correlation
  at the two-particle level},}\ }\href {\doibase 10.1103/PhysRevB.86.125114}
  {\bibfield  {journal} {\bibinfo  {journal} {Phys. Rev. B}\ }\textbf {\bibinfo
  {volume} {86}},\ \bibinfo {pages} {125114} (\bibinfo {year}
  {2012})}\BibitemShut {NoStop}%
\bibitem [{\citenamefont {Sch\"afer}\ \emph {et~al.}(2013)\citenamefont
  {Sch\"afer}, \citenamefont {Rohringer}, \citenamefont {Gunnarsson},
  \citenamefont {Ciuchi}, \citenamefont {Sangiovanni},\ and\ \citenamefont
  {Toschi}}]{Schafer13}%
  \BibitemOpen
  \bibfield  {author} {\bibinfo {author} {\bibfnamefont {T.}~\bibnamefont
  {Sch\"afer}}, \bibinfo {author} {\bibfnamefont {G.}~\bibnamefont
  {Rohringer}}, \bibinfo {author} {\bibfnamefont {O.}~\bibnamefont
  {Gunnarsson}}, \bibinfo {author} {\bibfnamefont {S.}~\bibnamefont {Ciuchi}},
  \bibinfo {author} {\bibfnamefont {G.}~\bibnamefont {Sangiovanni}}, \ and\
  \bibinfo {author} {\bibfnamefont {A.}~\bibnamefont {Toschi}},\ }\bibfield
  {title} {\enquote {\bibinfo {title} {Divergent precursors of the
  {Mott-Hubbard} transition at the two-particle level},}\ }\href {\doibase
  10.1103/PhysRevLett.110.246405} {\bibfield  {journal} {\bibinfo  {journal}
  {Phys. Rev. Lett.}\ }\textbf {\bibinfo {volume} {110}},\ \bibinfo {pages}
  {246405} (\bibinfo {year} {2013})}\BibitemShut {NoStop}%
\bibitem [{\citenamefont {Kozik}\ \emph {et~al.}(2015)\citenamefont {Kozik},
  \citenamefont {Ferrero},\ and\ \citenamefont {Georges}}]{Kozik15}%
  \BibitemOpen
  \bibfield  {author} {\bibinfo {author} {\bibfnamefont {Evgeny}\ \bibnamefont
  {Kozik}}, \bibinfo {author} {\bibfnamefont {Michel}\ \bibnamefont {Ferrero}},
  \ and\ \bibinfo {author} {\bibfnamefont {Antoine}\ \bibnamefont {Georges}},\
  }\bibfield  {title} {\enquote {\bibinfo {title} {Nonexistence of the
  {Luttinger-Ward} functional and misleading convergence of skeleton
  diagrammatic series for {H}ubbard-like models},}\ }\href {\doibase
  10.1103/PhysRevLett.114.156402} {\bibfield  {journal} {\bibinfo  {journal}
  {Phys. Rev. Lett.}\ }\textbf {\bibinfo {volume} {114}},\ \bibinfo {pages}
  {156402} (\bibinfo {year} {2015})}\BibitemShut {NoStop}%
\bibitem [{\citenamefont {Krien}\ \emph {et~al.}(2019)\citenamefont {Krien},
  \citenamefont {van Loon}, \citenamefont {Katsnelson}, \citenamefont
  {Lichtenstein},\ and\ \citenamefont {Capone}}]{Krien19}%
  \BibitemOpen
  \bibfield  {author} {\bibinfo {author} {\bibfnamefont {Friedrich}\
  \bibnamefont {Krien}}, \bibinfo {author} {\bibfnamefont {Erik G. C.~P.}\
  \bibnamefont {van Loon}}, \bibinfo {author} {\bibfnamefont {Mikhail~I.}\
  \bibnamefont {Katsnelson}}, \bibinfo {author} {\bibfnamefont {Alexander~I.}\
  \bibnamefont {Lichtenstein}}, \ and\ \bibinfo {author} {\bibfnamefont
  {Massimo}\ \bibnamefont {Capone}},\ }\bibfield  {title} {\enquote {\bibinfo
  {title} {Two-particle fermi liquid parameters at the mott transition: Vertex
  divergences, landau parameters, and incoherent response in dynamical
  mean-field theory},}\ }\href {\doibase 10.1103/PhysRevB.99.245128} {\bibfield
   {journal} {\bibinfo  {journal} {Phys. Rev. B}\ }\textbf {\bibinfo {volume}
  {99}},\ \bibinfo {pages} {245128} (\bibinfo {year} {2019})}\BibitemShut
  {NoStop}%
\bibitem [{\citenamefont {Springer}\ \emph {et~al.}(2020)\citenamefont
  {Springer}, \citenamefont {Chalupa}, \citenamefont {Ciuchi}, \citenamefont
  {Sangiovanni},\ and\ \citenamefont {Toschi}}]{Springer20}%
  \BibitemOpen
  \bibfield  {author} {\bibinfo {author} {\bibfnamefont {D.}~\bibnamefont
  {Springer}}, \bibinfo {author} {\bibfnamefont {P.}~\bibnamefont {Chalupa}},
  \bibinfo {author} {\bibfnamefont {S.}~\bibnamefont {Ciuchi}}, \bibinfo
  {author} {\bibfnamefont {G.}~\bibnamefont {Sangiovanni}}, \ and\ \bibinfo
  {author} {\bibfnamefont {A.}~\bibnamefont {Toschi}},\ }\bibfield  {title}
  {\enquote {\bibinfo {title} {Interplay between local response and vertex
  divergences in many-fermion systems with on-site attraction},}\ }\href
  {\doibase 10.1103/PhysRevB.101.155148} {\bibfield  {journal} {\bibinfo
  {journal} {Phys. Rev. B}\ }\textbf {\bibinfo {volume} {101}},\ \bibinfo
  {pages} {155148} (\bibinfo {year} {2020})}\BibitemShut {NoStop}%
\bibitem [{\citenamefont {Melnick}\ and\ \citenamefont
  {Kotliar}(2020)}]{Melnick20}%
  \BibitemOpen
  \bibfield  {author} {\bibinfo {author} {\bibfnamefont {Corey}\ \bibnamefont
  {Melnick}}\ and\ \bibinfo {author} {\bibfnamefont {Gabriel}\ \bibnamefont
  {Kotliar}},\ }\bibfield  {title} {\enquote {\bibinfo {title} {Fermi-liquid
  theory and divergences of the two-particle irreducible vertex in the periodic
  anderson lattice},}\ }\href {\doibase 10.1103/PhysRevB.101.165105} {\bibfield
   {journal} {\bibinfo  {journal} {Phys. Rev. B}\ }\textbf {\bibinfo {volume}
  {101}},\ \bibinfo {pages} {165105} (\bibinfo {year} {2020})}\BibitemShut
  {NoStop}%
\bibitem [{\citenamefont {van Loon}\ \emph {et~al.}(2020)\citenamefont {van
  Loon}, \citenamefont {Krien},\ and\ \citenamefont {Katanin}}]{vanLoon20}%
  \BibitemOpen
  \bibfield  {author} {\bibinfo {author} {\bibfnamefont {Erik G. C.~P.}\
  \bibnamefont {van Loon}}, \bibinfo {author} {\bibfnamefont {Friedrich}\
  \bibnamefont {Krien}}, \ and\ \bibinfo {author} {\bibfnamefont {Andrey~A.}\
  \bibnamefont {Katanin}},\ }\bibfield  {title} {\enquote {\bibinfo {title}
  {Bethe-salpeter equation at the critical end point of the mott transition},}\
  }\href {\doibase 10.1103/PhysRevLett.125.136402} {\bibfield  {journal}
  {\bibinfo  {journal} {Phys. Rev. Lett.}\ }\textbf {\bibinfo {volume} {125}},\
  \bibinfo {pages} {136402} (\bibinfo {year} {2020})}\BibitemShut {NoStop}%
\bibitem [{\citenamefont {Reitner}\ \emph {et~al.}(2020)\citenamefont
  {Reitner}, \citenamefont {Chalupa}, \citenamefont {Del~Re}, \citenamefont
  {Springer}, \citenamefont {Ciuchi}, \citenamefont {Sangiovanni},\ and\
  \citenamefont {Toschi}}]{Reitner20}%
  \BibitemOpen
  \bibfield  {author} {\bibinfo {author} {\bibfnamefont {M.}~\bibnamefont
  {Reitner}}, \bibinfo {author} {\bibfnamefont {P.}~\bibnamefont {Chalupa}},
  \bibinfo {author} {\bibfnamefont {L.}~\bibnamefont {Del~Re}}, \bibinfo
  {author} {\bibfnamefont {D.}~\bibnamefont {Springer}}, \bibinfo {author}
  {\bibfnamefont {S.}~\bibnamefont {Ciuchi}}, \bibinfo {author} {\bibfnamefont
  {G.}~\bibnamefont {Sangiovanni}}, \ and\ \bibinfo {author} {\bibfnamefont
  {A.}~\bibnamefont {Toschi}},\ }\bibfield  {title} {\enquote {\bibinfo {title}
  {Attractive effect of a strong electronic repulsion: The physics of vertex
  divergences},}\ }\href {\doibase 10.1103/PhysRevLett.125.196403} {\bibfield
  {journal} {\bibinfo  {journal} {Phys. Rev. Lett.}\ }\textbf {\bibinfo
  {volume} {125}},\ \bibinfo {pages} {196403} (\bibinfo {year}
  {2020})}\BibitemShut {NoStop}%
\bibitem [{\citenamefont {Chalupa}\ \emph {et~al.}(2021)\citenamefont
  {Chalupa}, \citenamefont {Sch\"afer}, \citenamefont {Reitner}, \citenamefont
  {Springer}, \citenamefont {Andergassen},\ and\ \citenamefont
  {Toschi}}]{Chalupa20}%
  \BibitemOpen
  \bibfield  {author} {\bibinfo {author} {\bibfnamefont {P.}~\bibnamefont
  {Chalupa}}, \bibinfo {author} {\bibfnamefont {T.}~\bibnamefont {Sch\"afer}},
  \bibinfo {author} {\bibfnamefont {M.}~\bibnamefont {Reitner}}, \bibinfo
  {author} {\bibfnamefont {D.}~\bibnamefont {Springer}}, \bibinfo {author}
  {\bibfnamefont {S.}~\bibnamefont {Andergassen}}, \ and\ \bibinfo {author}
  {\bibfnamefont {A.}~\bibnamefont {Toschi}},\ }\bibfield  {title} {\enquote
  {\bibinfo {title} {Fingerprints of the local moment formation and its kondo
  screening in the generalized susceptibilities of many-electron problems},}\
  }\href {\doibase 10.1103/PhysRevLett.126.056403} {\bibfield  {journal}
  {\bibinfo  {journal} {Phys. Rev. Lett.}\ }\textbf {\bibinfo {volume} {126}},\
  \bibinfo {pages} {056403} (\bibinfo {year} {2021})}\BibitemShut {NoStop}%
\bibitem [{\citenamefont {et. al.}(2021)}]{SM}%
  \BibitemOpen
  \bibfield  {author} {\bibinfo {author} {\bibfnamefont {Michael~Danilov}\
  \bibnamefont {et. al.}},\ }\bibfield  {title} {\enquote {\bibinfo {title}
  {Supplemental materials},}\ }\href@noop {} {\  (\bibinfo {year}
  {2021})}\BibitemShut {NoStop}%
\bibitem [{\citenamefont {Brener}\ \emph {et~al.}(2008)\citenamefont {Brener},
  \citenamefont {Hafermann}, \citenamefont {Rubtsov}, \citenamefont
  {Katsnelson},\ and\ \citenamefont {Lichtenstein}}]{susceptibility_brener}%
  \BibitemOpen
  \bibfield  {author} {\bibinfo {author} {\bibfnamefont {S.}~\bibnamefont
  {Brener}}, \bibinfo {author} {\bibfnamefont {H.}~\bibnamefont {Hafermann}},
  \bibinfo {author} {\bibfnamefont {A.~N.}\ \bibnamefont {Rubtsov}}, \bibinfo
  {author} {\bibfnamefont {M.~I.}\ \bibnamefont {Katsnelson}}, \ and\ \bibinfo
  {author} {\bibfnamefont {A.~I.}\ \bibnamefont {Lichtenstein}},\ }\bibfield
  {title} {\enquote {\bibinfo {title} {Dual fermion approach to susceptibility
  of correlated lattice fermions},}\ }\href {\doibase
  10.1103/PhysRevB.77.195105} {\bibfield  {journal} {\bibinfo  {journal} {Phys.
  Rev. B}\ }\textbf {\bibinfo {volume} {77}},\ \bibinfo {pages} {195105}
  (\bibinfo {year} {2008})}\BibitemShut {NoStop}%
\bibitem [{\citenamefont {Dagotto}\ \emph {et~al.}(1992)\citenamefont
  {Dagotto}, \citenamefont {Moreo}, \citenamefont {Ortolani}, \citenamefont
  {Poilblanc},\ and\ \citenamefont {Riera}}]{Dagotto_4x4}%
  \BibitemOpen
  \bibfield  {author} {\bibinfo {author} {\bibfnamefont {E.}~\bibnamefont
  {Dagotto}}, \bibinfo {author} {\bibfnamefont {A.}~\bibnamefont {Moreo}},
  \bibinfo {author} {\bibfnamefont {F.}~\bibnamefont {Ortolani}}, \bibinfo
  {author} {\bibfnamefont {D.}~\bibnamefont {Poilblanc}}, \ and\ \bibinfo
  {author} {\bibfnamefont {J.}~\bibnamefont {Riera}},\ }\bibfield  {title}
  {\enquote {\bibinfo {title} {Static and dynamical properties of doped hubbard
  clusters},}\ }\href {\doibase 10.1103/PhysRevB.45.10741} {\bibfield
  {journal} {\bibinfo  {journal} {Phys. Rev. B}\ }\textbf {\bibinfo {volume}
  {45}},\ \bibinfo {pages} {10741--10760} (\bibinfo {year} {1992})}\BibitemShut
  {NoStop}%
\bibitem [{\citenamefont {Wachtel}\ \emph {et~al.}(2017)\citenamefont
  {Wachtel}, \citenamefont {Baruch},\ and\ \citenamefont {Orgad}}]{Orgad_ED}%
  \BibitemOpen
  \bibfield  {author} {\bibinfo {author} {\bibfnamefont {Gideon}\ \bibnamefont
  {Wachtel}}, \bibinfo {author} {\bibfnamefont {Shirit}\ \bibnamefont
  {Baruch}}, \ and\ \bibinfo {author} {\bibfnamefont {Dror}\ \bibnamefont
  {Orgad}},\ }\bibfield  {title} {\enquote {\bibinfo {title} {Optimal
  inhomogeneity for pairing in hubbard systems with next-nearest-neighbor
  hopping},}\ }\href {\doibase 10.1103/PhysRevB.96.064527} {\bibfield
  {journal} {\bibinfo  {journal} {Phys. Rev. B}\ }\textbf {\bibinfo {volume}
  {96}},\ \bibinfo {pages} {064527} (\bibinfo {year} {2017})}\BibitemShut
  {NoStop}%
\bibitem [{\citenamefont {Tsai}\ \emph {et~al.}(2008)\citenamefont {Tsai},
  \citenamefont {Yao}, \citenamefont {L\"auchli},\ and\ \citenamefont
  {Kivelson}}]{Kivelson_inhomog}%
  \BibitemOpen
  \bibfield  {author} {\bibinfo {author} {\bibfnamefont {Wei-Feng}\
  \bibnamefont {Tsai}}, \bibinfo {author} {\bibfnamefont {Hong}\ \bibnamefont
  {Yao}}, \bibinfo {author} {\bibfnamefont {Andreas}\ \bibnamefont
  {L\"auchli}}, \ and\ \bibinfo {author} {\bibfnamefont {Steven~A.}\
  \bibnamefont {Kivelson}},\ }\bibfield  {title} {\enquote {\bibinfo {title}
  {Optimal inhomogeneity for superconductivity: Finite-size studies},}\ }\href
  {\doibase 10.1103/PhysRevB.77.214502} {\bibfield  {journal} {\bibinfo
  {journal} {Phys. Rev. B}\ }\textbf {\bibinfo {volume} {77}},\ \bibinfo
  {pages} {214502} (\bibinfo {year} {2008})}\BibitemShut {NoStop}%
\bibitem [{\citenamefont {Martins}\ \emph {et~al.}(2001)\citenamefont
  {Martins}, \citenamefont {Xavier}, \citenamefont {Arrachea},\ and\
  \citenamefont {Dagotto}}]{Martins_ttpJ}%
  \BibitemOpen
  \bibfield  {author} {\bibinfo {author} {\bibfnamefont {G.~B.}\ \bibnamefont
  {Martins}}, \bibinfo {author} {\bibfnamefont {J.~C.}\ \bibnamefont {Xavier}},
  \bibinfo {author} {\bibfnamefont {L.}~\bibnamefont {Arrachea}}, \ and\
  \bibinfo {author} {\bibfnamefont {E.}~\bibnamefont {Dagotto}},\ }\bibfield
  {title} {\enquote {\bibinfo {title} {Qualitative understanding of the sign of
  ${t}^{\ensuremath{'}}$ asymmetry in the extended $t\ensuremath{-}j$ model and
  relevance for pairing properties},}\ }\href {\doibase
  10.1103/PhysRevB.64.180513} {\bibfield  {journal} {\bibinfo  {journal} {Phys.
  Rev. B}\ }\textbf {\bibinfo {volume} {64}},\ \bibinfo {pages} {180513}
  (\bibinfo {year} {2001})}\BibitemShut {NoStop}%
\bibitem [{\citenamefont {Moriya}(1985)}]{moriya}%
  \BibitemOpen
  \bibfield  {author} {\bibinfo {author} {\bibfnamefont {Toru}\ \bibnamefont
  {Moriya}},\ }\href@noop {} {\emph {\bibinfo {title} {Spin Fluctuations in
  Itinerant Electron Magnetism}}}\ (\bibinfo  {publisher} {Springer Verlag},\
  \bibinfo {address} {New York},\ \bibinfo {year} {1985})\BibitemShut {NoStop}%
\bibitem [{\citenamefont {Yang}\ \emph {et~al.}(2020)\citenamefont {Yang},
  \citenamefont {Ying}, \citenamefont {Li}, \citenamefont {Yang}, \citenamefont
  {Sun},\ and\ \citenamefont {Li}}]{Yang_DQMC_tp}%
  \BibitemOpen
  \bibfield  {author} {\bibinfo {author} {\bibfnamefont {Shuhui}\ \bibnamefont
  {Yang}}, \bibinfo {author} {\bibfnamefont {Tao}\ \bibnamefont {Ying}},
  \bibinfo {author} {\bibfnamefont {Weiqi}\ \bibnamefont {Li}}, \bibinfo
  {author} {\bibfnamefont {Jianqun}\ \bibnamefont {Yang}}, \bibinfo {author}
  {\bibfnamefont {Xiudong}\ \bibnamefont {Sun}}, \ and\ \bibinfo {author}
  {\bibfnamefont {Xingji}\ \bibnamefont {Li}},\ }\bibfield  {title} {\enquote
  {\bibinfo {title} {Quantum monte carlo study of the hubbard model with
  next-nearest-neighbor hopping $t'$: pairing and magnetism},}\ }\href
  {\doibase 10.1088/1361-648x/abd33a} {\bibfield  {journal} {\bibinfo
  {journal} {Journal of Physics: Condensed Matter}\ }\textbf {\bibinfo {volume}
  {33}},\ \bibinfo {pages} {115601} (\bibinfo {year} {2020})}\BibitemShut
  {NoStop}%
\bibitem [{\citenamefont {Sakai}\ \emph
  {et~al.}(2016{\natexlab{a}})\citenamefont {Sakai}, \citenamefont {Civelli},\
  and\ \citenamefont {Imada}}]{Imada_PRL}%
  \BibitemOpen
  \bibfield  {author} {\bibinfo {author} {\bibfnamefont {Shiro}\ \bibnamefont
  {Sakai}}, \bibinfo {author} {\bibfnamefont {Marcello}\ \bibnamefont
  {Civelli}}, \ and\ \bibinfo {author} {\bibfnamefont {Masatoshi}\ \bibnamefont
  {Imada}},\ }\bibfield  {title} {\enquote {\bibinfo {title} {Hidden fermionic
  excitation boosting high-temperature superconductivity in cuprates},}\ }\href
  {\doibase 10.1103/PhysRevLett.116.057003} {\bibfield  {journal} {\bibinfo
  {journal} {Phys. Rev. Lett.}\ }\textbf {\bibinfo {volume} {116}},\ \bibinfo
  {pages} {057003} (\bibinfo {year} {2016}{\natexlab{a}})}\BibitemShut
  {NoStop}%
\bibitem [{\citenamefont {Sakai}\ \emph
  {et~al.}(2016{\natexlab{b}})\citenamefont {Sakai}, \citenamefont {Civelli},\
  and\ \citenamefont {Imada}}]{Imada_PRB}%
  \BibitemOpen
  \bibfield  {author} {\bibinfo {author} {\bibfnamefont {Shiro}\ \bibnamefont
  {Sakai}}, \bibinfo {author} {\bibfnamefont {Marcello}\ \bibnamefont
  {Civelli}}, \ and\ \bibinfo {author} {\bibfnamefont {Masatoshi}\ \bibnamefont
  {Imada}},\ }\bibfield  {title} {\enquote {\bibinfo {title} {Hidden-fermion
  representation of self-energy in pseudogap and superconducting states of the
  two-dimensional hubbard model},}\ }\href {\doibase
  10.1103/PhysRevB.94.115130} {\bibfield  {journal} {\bibinfo  {journal} {Phys.
  Rev. B}\ }\textbf {\bibinfo {volume} {94}},\ \bibinfo {pages} {115130}
  (\bibinfo {year} {2016}{\natexlab{b}})}\BibitemShut {NoStop}%
\bibitem [{\citenamefont {Merino}\ and\ \citenamefont
  {Gunnarsson}(2014)}]{Gunnarsson_2014}%
  \BibitemOpen
  \bibfield  {author} {\bibinfo {author} {\bibfnamefont {J.}~\bibnamefont
  {Merino}}\ and\ \bibinfo {author} {\bibfnamefont {O.}~\bibnamefont
  {Gunnarsson}},\ }\bibfield  {title} {\enquote {\bibinfo {title} {Pseudogap
  and singlet formation in organic and cuprate superconductors},}\ }\href
  {\doibase 10.1103/PhysRevB.89.245130} {\bibfield  {journal} {\bibinfo
  {journal} {Phys. Rev. B}\ }\textbf {\bibinfo {volume} {89}},\ \bibinfo
  {pages} {245130} (\bibinfo {year} {2014})}\BibitemShut {NoStop}%
\bibitem [{\citenamefont {Rost}\ \emph {et~al.}(2012)\citenamefont {Rost},
  \citenamefont {Gorelik}, \citenamefont {Assaad},\ and\ \citenamefont
  {Bl\"umer}}]{Rost_QMC}%
  \BibitemOpen
  \bibfield  {author} {\bibinfo {author} {\bibfnamefont {D.}~\bibnamefont
  {Rost}}, \bibinfo {author} {\bibfnamefont {E.~V.}\ \bibnamefont {Gorelik}},
  \bibinfo {author} {\bibfnamefont {F.}~\bibnamefont {Assaad}}, \ and\ \bibinfo
  {author} {\bibfnamefont {N.}~\bibnamefont {Bl\"umer}},\ }\bibfield  {title}
  {\enquote {\bibinfo {title} {Momentum-dependent pseudogaps in the half-filled
  two-dimensional hubbard model},}\ }\href {\doibase
  10.1103/PhysRevB.86.155109} {\bibfield  {journal} {\bibinfo  {journal} {Phys.
  Rev. B}\ }\textbf {\bibinfo {volume} {86}},\ \bibinfo {pages} {155109}
  (\bibinfo {year} {2012})}\BibitemShut {NoStop}%
\bibitem [{\citenamefont {Dagotto}\ \emph {et~al.}(1990)\citenamefont
  {Dagotto}, \citenamefont {Joynt}, \citenamefont {Moreo}, \citenamefont
  {Bacci},\ and\ \citenamefont {Gagliano}}]{Dagotto_2222}%
  \BibitemOpen
  \bibfield  {author} {\bibinfo {author} {\bibfnamefont {Elbio}\ \bibnamefont
  {Dagotto}}, \bibinfo {author} {\bibfnamefont {Robert}\ \bibnamefont {Joynt}},
  \bibinfo {author} {\bibfnamefont {Adriana}\ \bibnamefont {Moreo}}, \bibinfo
  {author} {\bibfnamefont {Silvia}\ \bibnamefont {Bacci}}, \ and\ \bibinfo
  {author} {\bibfnamefont {Eduardo}\ \bibnamefont {Gagliano}},\ }\bibfield
  {title} {\enquote {\bibinfo {title} {Strongly correlated electronic systems
  with one hole: Dynamical properties},}\ }\href {\doibase
  10.1103/PhysRevB.41.9049} {\bibfield  {journal} {\bibinfo  {journal} {Phys.
  Rev. B}\ }\textbf {\bibinfo {volume} {41}},\ \bibinfo {pages} {9049--9073}
  (\bibinfo {year} {1990})}\BibitemShut {NoStop}%
\end{thebibliography}%


\appendix
\section{Exact relations for Green's function}
\label{App_Exact}

After appropriate diagrammatic results for the dual self-energy and the dual Green function has been obtained, it has to be transformed back to the corresponding physical quantities in terms of real lattice fermions. The fact that dual fermions are introduced through the exact Hubbard-Stratonovich transformation  Eq.~\eqref{hs_transfo} allows to establish exact identities between dual and lattice Greens function \cite{RKL08,hafermann_thesis}.

The relations between the $n$-particle cumulants of the dual and lattice fermions can be established using the cumulant (linked cluster) technique. To this end, one may consider two different, equivalent representations of the following generating functional:
\begin{align}
e^{-F[J^*J,L^*L]} = \int\mathcal{D}[c^*c,d^*d] e^{ -S[c^*c,d^*,d]  +J^*_1 c_1 + c^*_2 J_2 + L^*_1 d_1 + d^*_2 L_2  }
\label{df_functional0}
\end{align}
Integrating out the lattice fermions from this functional similar to (\ref{eqn::df::def_V}) (this can be done with the sources $J$ and $J^*$ set to zero) yields
\begin{equation}
e^{-F[L^*,L]} =  \int \mathcal{D}[d^*,d] e^{-\tilde{S}[d^*,d] + L^*_1 d_1 + d^*_2 L_2 }
\label{eqn::df::functional2}
\end{equation}
 The dual Green function and the two-particle correlator related to the non-local susceptibilities are obtained from (\ref{eqn::df::functional2}) by suitable functional derivatives, e.g.
\begin{align}
\left. \tilde{G}_{12} = \frac{\delta^2 F}{\delta L_2 \delta L^*_1}\right|_{L^*=L=0}
\nonumber\\
\label{eqn::df::funcderiv}
\end{align}
Integrating out the dual fermions from Eq.(\ref{df_functional0}) using the HST, one obtains an alternative representation, which more clearly reveals a connection of the functional derivatives with respect to the sources $J$,$J^*$ and $L$, $L^*$.
The result is
\begin{align}
F[J^*J,L^*L] = & L^*_1 (t-\Delta)_{12} L_2 - \ln \int\mathcal{D}[c^*,c] \exp\Big(-S[c^*,c]\ + \nonumber\\
&+ J_1^*c_1 + c_2^* J_2 - L^*_1 (t-\Delta)_{12} c_2 - c^*_1 (t-\Delta)_{12}L_2 \Big).
\label{eqn::df::functional3}
\end{align}
In analogy to (\ref{eqn::df::funcderiv}), the cumulants in terms of lattice fermions are obviously obtained by functional derivative with respect to the sources $J$ and $J^*$ with $L$ and $L^*$ set to zero. Applying the derivatives with respect to $L$, $L^*$ to (\ref{eqn::df::functional3}) with $J=J^*=0$ and comparing to (\ref{eqn::df::funcderiv}), e.g. yields the following identity:
\begin{equation}
G_{12} = -(t-\Delta)^{-1}_{12} + (t-\Delta)^{-1}_{11'} \tilde{G}_{1'2'} (t-\Delta)^{-1}_{2'2}.
\label{app::exactgf}
\end{equation}

\section{Plaquette periodization}
\label{App_period}

By breaking up the original lattice into plaquettes, the translational symmetry is broken: bonds within a plaquette are treated differently from bonds between plaquettes. 
We need to restore the full translational symmetry and then all quantities can be written in terms of the momentum $\kv$ in the Brillouin Zone of the original lattice.

\begin{figure}[t!]
\includegraphics[width=0.3\linewidth]{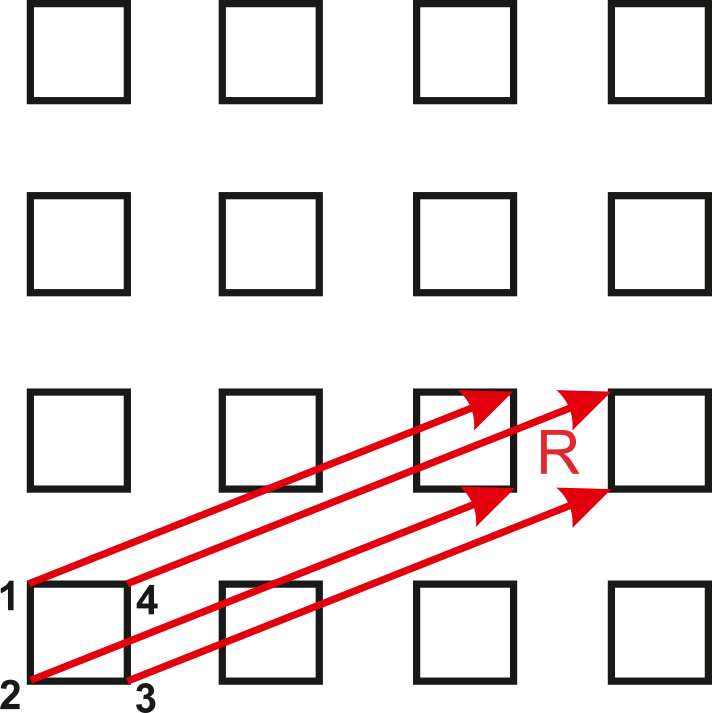}
\caption{Scheme for the real-space periodization of the $2\times 2$ plaquette lattice.}
\label{fig:period}
\end{figure}

Let us discuss a periodization of plaquette self-energy $\Sigma_{ij}({{\mathbf{r}},\nu})$   Eq.~\eqref{DF_SE} where $\mathbf{r}\equiv(r_x,r_y)$ is the supercell translation and $i,j$ are cluster sites (see Fig.({\ref{fig:period}})). The latter can be alternatively described by intra-plaquette translation vectors $\mathbf{i},\mathbf{j}$ taking values $[(0,0),(0,1),(1,1),(1,0)]$ for the site indices 0 to 3 respectively. We would like to get a lattice periodic self-energy $\Sigma({{\mathbf{R}},\nu})$
where $\mathbf{R}\equiv(R_x,R_y)$ is the original square lattice translations. By construction $\Sigma$ is periodic in $\mathbf{r}$, but not in $\mathbf{R}$. The natural periodization procedure would be taking all four possible values of $\mathbf{i}\equiv(i_x,i_y)$ and average over them for a given value of $\mathbf{R}$. This is done straightforwardly with a minor technical challenge of determining the supercell translation $\mathbf{r}$ and final site index $\mathbf{j}$ that correspond to a given value of lattice translation $\mathbf{R}$ and initial site index $\mathbf{i}$. By recasting
\begin{equation}
    \mathbf{i}+\mathbf{R}\equiv(i_x+R_x,i_y+R_y)=(2[(i_x+R_x)/2]+2\{(i_x+R_x)/2\},2[(i_y+R_y)/2]+2\{(i_y+R_y)/2\}),
\end{equation}
and noticing that $\mathbf{i}+\mathbf{R}=\mathbf{j}+2\mathbf{r}$, we immediately find $\mathbf{r}(\mathbf{i},\mathbf{R})=([(i_x+R_x)/2],[(i_y+R_y)/2])$ and $\mathbf{j}(\mathbf{i},\mathbf{R})=(2\{(i_x+R_x)/2\},2\{(i_y+R_y)/2\})$. Here $[x]$ and $\{x\}$ are the integer and fractional parts of $x$ respectively. Finally we take for the periodized self-energy:
\begin{equation}
    \Sigma(\mathbf{R},\nu)=\frac14\sum_{\mathbf{i}}\Sigma_{ij(\mathbf{i},\mathbf{R})}(\mathbf{r}(\mathbf{i},\mathbf{R}),\nu),
\end{equation}
with the sum being taken over four cluster sites.

\section{The non-interacting system}

\begin{figure}[t!]
\includegraphics[width=0.42\linewidth]{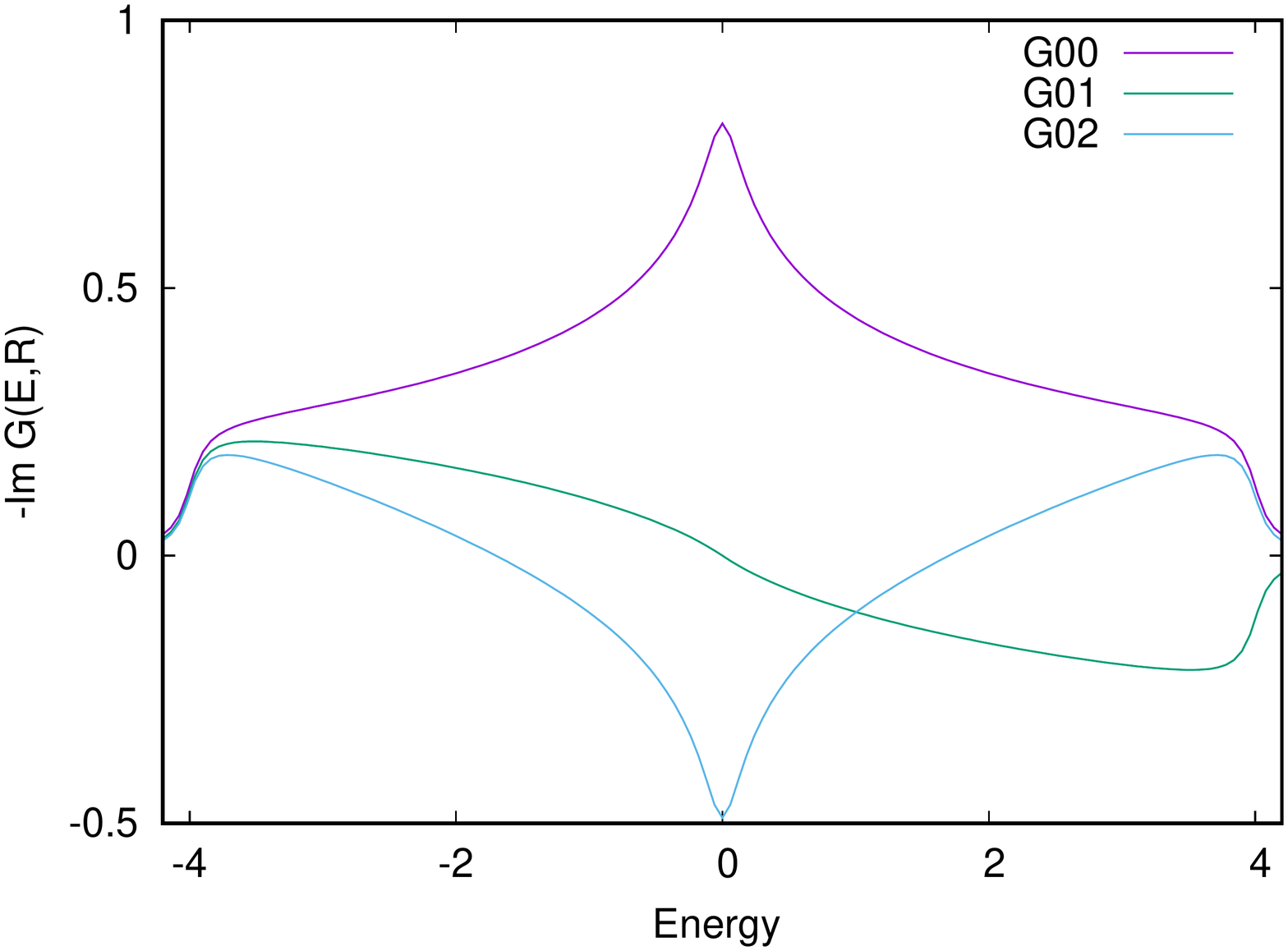}
\includegraphics[width=0.42\linewidth]{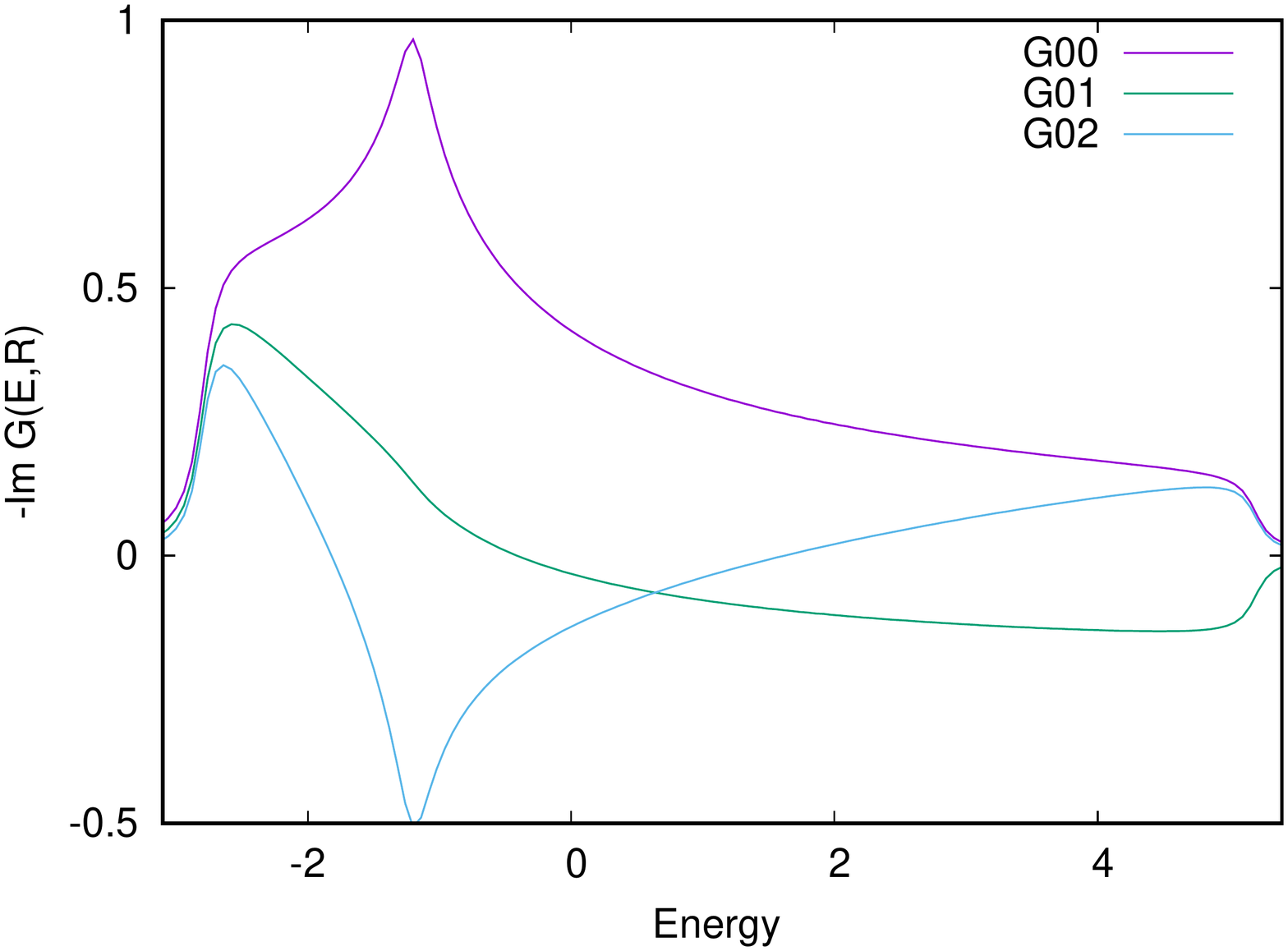}
\caption{Plaquette supercell Green's function $-\Imag G(E,r)$  for $U=0$, $t'/t=0$ (left) and $t'/t=-0.3$ (right) with $r=(0,0)$, $r=(0,1)$ and $r=(1,1)$.
}
\label{fig:TBdos}
\end{figure}

At $U=0$, the Hubbard model becomes a tight-binding model that is diagonalized by going to the momentum basis. Figure~\ref{fig:TBdos} presents the resulting bare Green's function. 
The local part ($G_{11}$, only the imaginary part is shown) is related to the local DOS and displays the familiar Van Hove singularity at $E=0$ for $t'=0$. For $t'/t=-0.3$, the Van Hove singularity shifts down to $E=-1.2=4t'$, and becomes more extended.
More interesting is that the next-nearest-neighbor part $G_{13}$ (again, we show the imaginary part) is also very large in magnitude and close to the Van Hove singularity it has the same order of magnitude as the local part. The next-nearest-neighbor part corresponds to the same displacement as the hopping parameter $t'$.

In the DF perturbation theory based on the plaquette, we calculate the self-energy corrections to this bare Green's function and we find that this self-energy effectively increases $t'$ and makes the extended Van Hove bands almost flat with large density of states near the Fermi level. This boosts the tendency towards instabilities such as superconductivity. 

In CDMFT, it is necessary to perform CT-QMC calculations to solve the impurity problem, which is computationally expensive and always introduces (substantial) numerical uncertainty. Furthermore, hybridization with the bath will wash out the degenerate point of the plaquette. Since we believe that the degenerate point contains the essential physical ingredients for the cuprate problem, here we decided to use the plaquette at the degenerate point as the dual fermion reference model for all calculations and discuss the related fluctuations and instabilities.

\section{Dual perturbation theory at half-filling}
\label{App_Plaq0}

\begin{figure}[t!]
\includegraphics[width=0.5\linewidth]{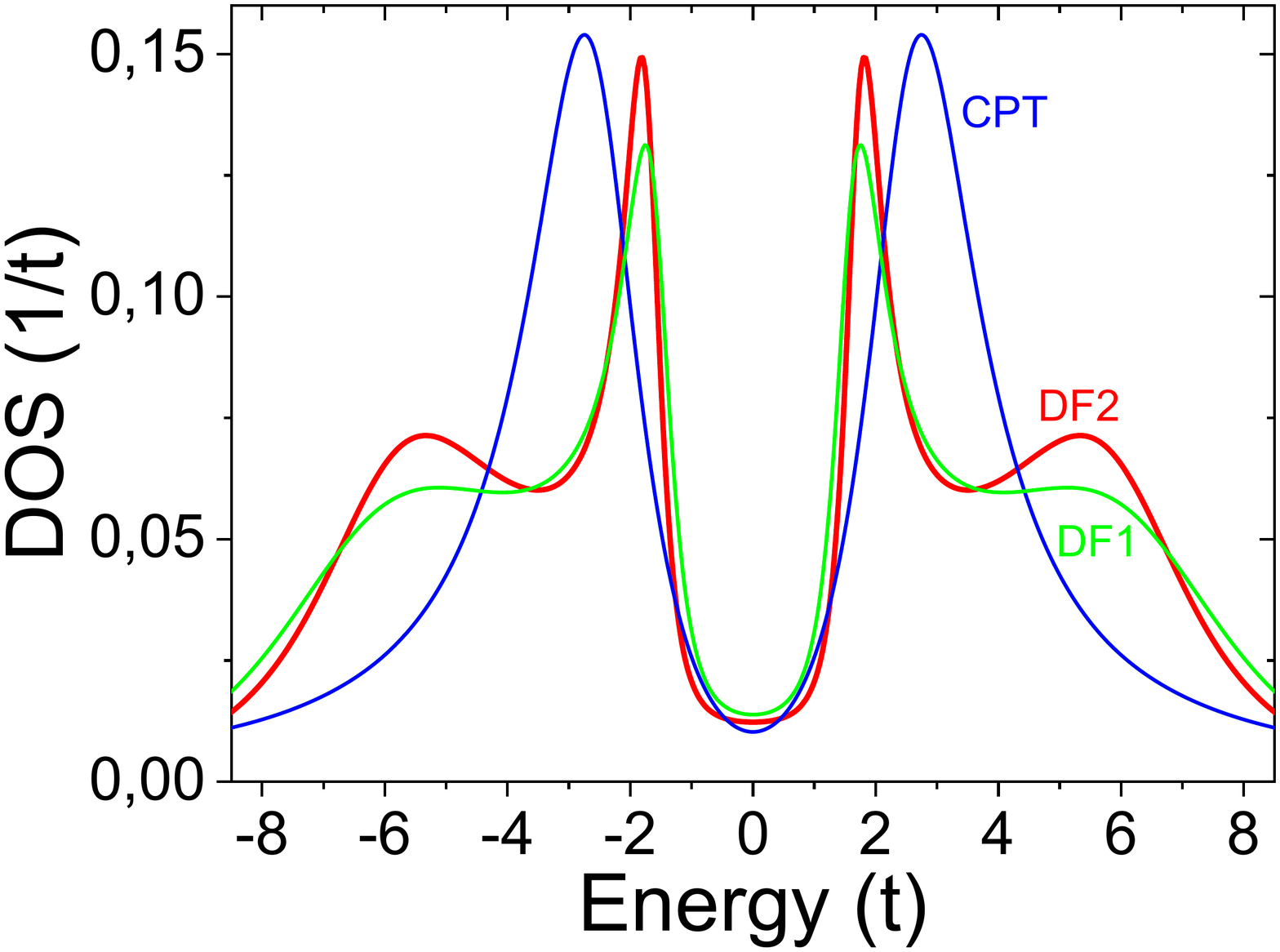}
\caption{Density of states for the half-filled case with $t'=0$ and $U=W=8$ in the zeroth-order DF approximation (CPT), the first-order (DF1) and the second order DF-theory (DF2)}
\label{fig:DOSplaq0}
\end{figure}

We also did calculations for the half-filled  square lattice Hubbard model and compared the dual fermion self-energy with the doped case.
The parameters were chosen as following: $t=1$, $t'=0$, $U=8$ (equal to the bandwidth $W=8t$) and the temperature $\beta=5$. Similar calculations have previously been done for higher temperature~\cite{Brener_refsys}. The density of states for the different approximations is presented in Fig.~\ref{fig:DOSplaq0}. The zeroth order dual approximation ($\tilde{\Sigma}=0$) corresponds to the cluster perturbation theory \cite{Valenti_CPT}, where only the cluster self-energy was taken into account in the lattice model. The DOS for the first- and second order dual fermion plaquette perturbations are quite similar and are very different from the CPT approximation. The DF theory reproduces the so-called four-peak structure of the half-filled Hubbard model, which is also obtained in lattice QMC calculations \cite{Rost_QMC}.

\begin{figure}[t!]
\includegraphics[width=0.33\linewidth]{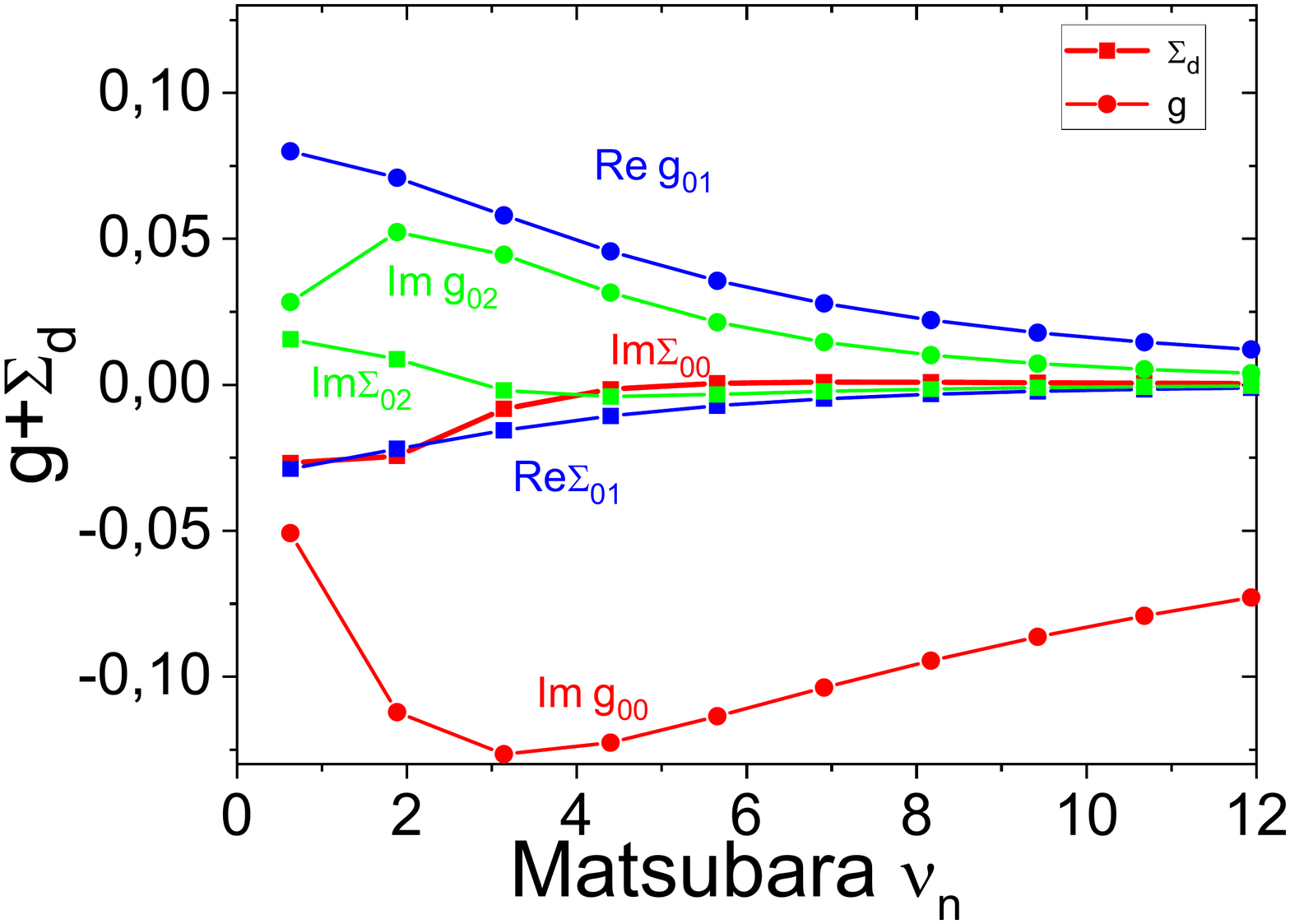}
\includegraphics[width=0.33\linewidth]{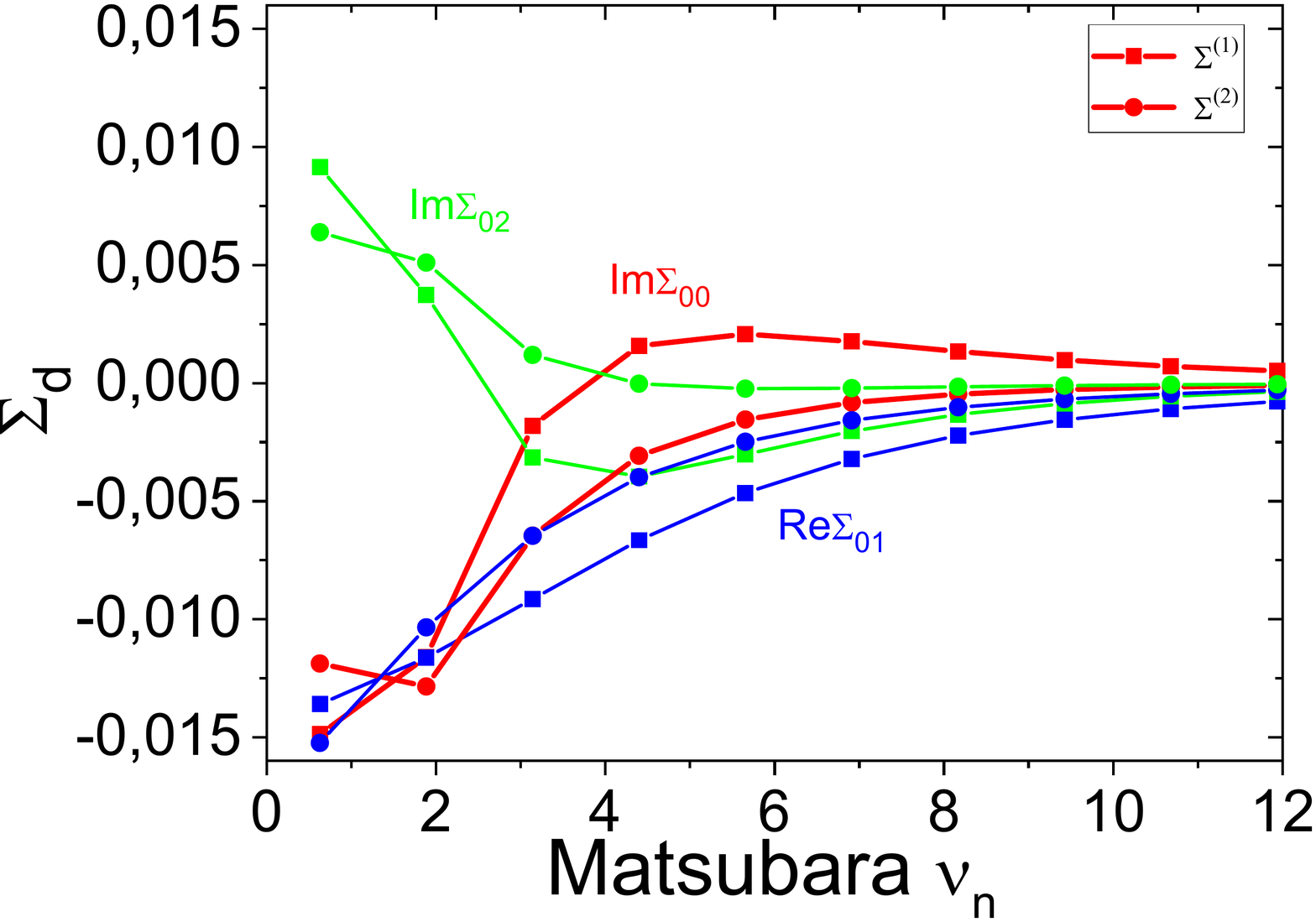}
\includegraphics[width=0.33\linewidth]{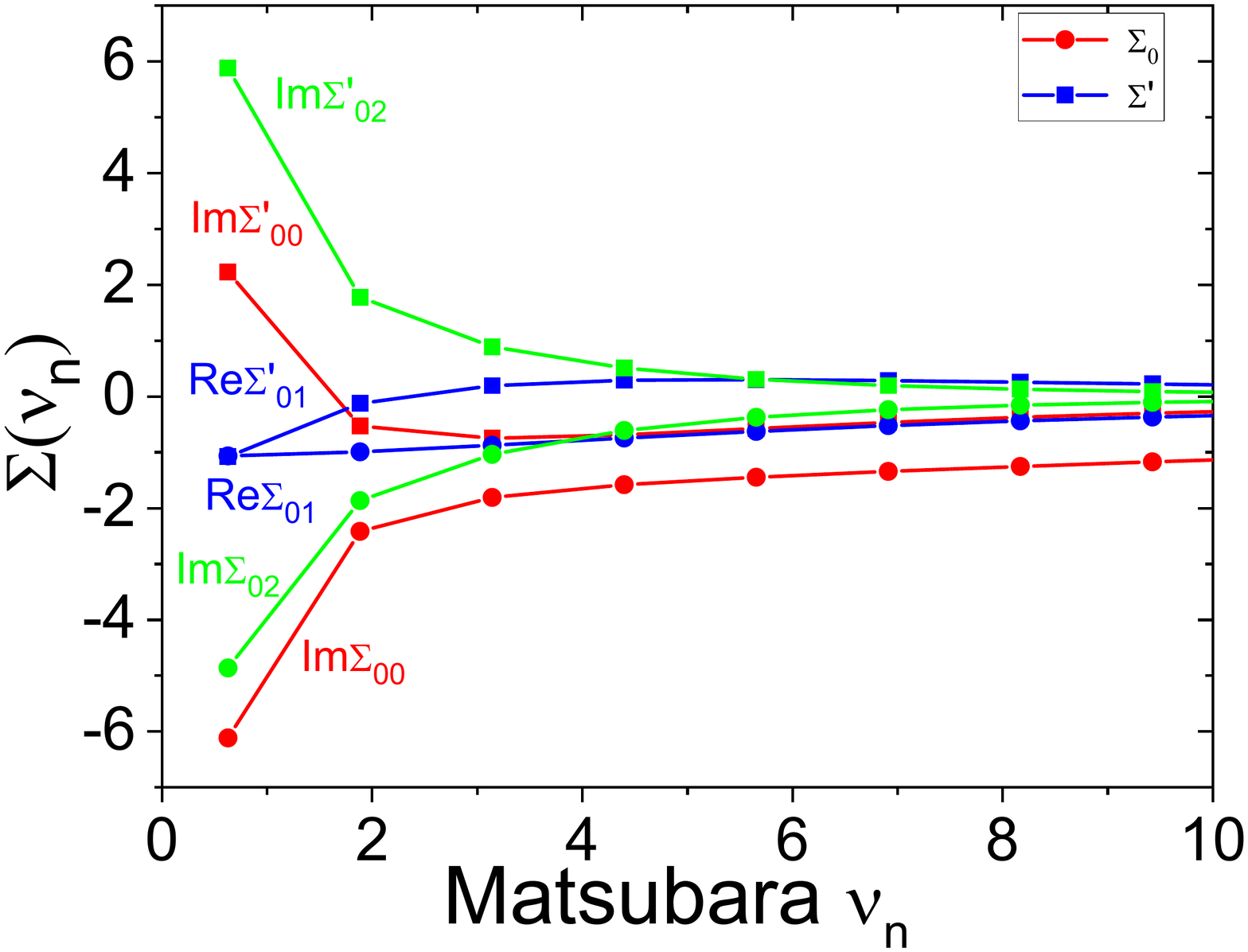}
\caption{Green's function of the half-filled reference plaquette (g) and local plaquette part of the dual self-energy  ($\tilde{\Sigma}$) of the second order DF perturbation (left). 
Dual self-energy for the 1-st ($\tilde{\Sigma}^{(1)}$) and 2-nd ($\tilde{\Sigma}^{(2)}$) order DF perturbation (middle).
Self-energy for the reference system and the  non-local DF part ($\Sigma^\prime$ (right)}
\label{fig:Gsigd0}
\end{figure}

Let us discuss, how such a strong renormalization can be seen in the dual-self energy.
We start from the definition of the lattice Green's function in the dual fermion theory in Eq.~\eqref{DF_GF}, where the calculated self-energy enters in the following combination
$\left( g_\nu+{\widetilde{\Sigma}_{{\mathbf{k}}\nu} }\right)$, which shows that $\tilde{\Sigma}$ has the meaning of a T-matrix-like quantity which adds to the ``bare'' plaquette Green's function $g$.
Therefore we show these two summands in Fig.~\ref{fig:Gsigd0}, as a function of the Matsubara frequency. We show only the plaquette-local ({\it i.e.} summed over $\mathbf{k}$ but keeping the plaquette site indices) part of the dual self-energy. We use the anticlockwise numbering of plaquette sites from 0 to 3, similar to Fig.~\ref{fig:period}. There are only three non-equivalent elements of the plaquette-local self-energy: the diagonal part $\tilde{\Sigma}_{00}$, nearest neighbour (along hopping $t$)
$\tilde{\Sigma}_{01}$ and next-nearest neighbour (along hopping $t'$) $\tilde{\Sigma}_{02}$. The same holds for the plaquette Green's function $g_{ij}$. Moreover, due to the particle-hole symmetry for the half-filled case with $t'=0$ the only non-zero elements of these complex functions in Matsubara space are the imaginary parts of $\tilde{\Sigma}_{00}$ and $\tilde{\Sigma}_{02}$ and the real part of $\tilde{\Sigma}_{01}$. One can see from the left panel os the Fig.(\ref{fig:Gsigd0}) that $\operatorname{Im}\tilde{\Sigma}_{00}$ effectively makes $\operatorname{Im} g_{00}$ less insulating, i.e., it reduces the gap. There is also a appreciable reduction of $\operatorname{Re} g_{01}$
due to the opposite sign of $\operatorname{Re}\tilde{\Sigma}_{01}$, which corresponds to the effective reduction of the nearest-neighbour hopping $t$ due to correlation effects. The effect of $\operatorname{Im}\tilde{\Sigma}_{02}$ is quite small compare to $\operatorname{Im} g_{02}$ .

In the middle panel of the Fig.~\ref{fig:Gsigd0} we show separate contributions to the dual self-energy from the first-order diagram, Eq.~\eqref{df:1order}, labelled with dots and a plaquette-local part of the second-order diagram,  Eq.~\ref{eqn::sigma2}, labelled with squares. Both contributions are of the same order and have the same sign, which is quite different from the doped case where we observed a significant cancellation effect by the second-order contributions. 

We can also transform self-energies back to the the real fermionic representations and compare the plaquette self-energy $\Sigma_0$ with the plaquette-local part of the additional DF-contribution $\Sigma'$ (Fig.~\ref{fig:Gsigd0}, right panel). There is strong compensation of the local part $\operatorname{Im} \Sigma_{00}$  as well as the second-neighbour part $\operatorname{Im} \Sigma_{02}$,
while the nearest-neighbour self-energies $\operatorname{Re} \Sigma_{01}$ are small and have the same sign.

\section{Plaquette generalized susceptibility and the degenerate point}
For all DF-calculations we used 44 fermionic Masubara frequencies for the dual Green's function and for the vertex we used 22/21 fermion/boson frequencies.
We checked few calculations with up to 160 Matsubara frequencies and results are
not very sensitive and well converged due to fast decay of the four point correlation functions for $\beta \le 10$. 

The generalized susceptibility $\chi^{P}_{ijkl}(\nu,\nu',\omega)$ is an important quantity that describes the two-particle correlations in a given channel. For the particle-particle case it merely coincides with the two-particle Green's function $\kappa^{P}_{ijkl}(\nu,\nu',\omega)$, while in the particle-hole channel $\chi$ differs from $\kappa$ by the disconnected contribution $g_{ij}(\nu)g_{kl}(nu')\delta_{\omega,0}$. In this appendix $\omega$ will be set to $0$ everywhere and we omit it in the notation.

Let us look at the structure of the particle-particle generalized susceptibility in more detail. We define the superconducting nonlocal pairing operator as $\Delta_{ij}(\nu)=c_{i,\nu}^{\uparrow}c_{j,-\nu}^{\downarrow}$ (in order not to overload the formulas we consider only the $\omega=0$ situation). The singlet pairing corresponds to $(\uparrow\downarrow-\downarrow\uparrow)/\sqrt{2}$ combination, or equivalently the singlet pairing operator is given by $\Delta^{s}_{ij}(\nu)=(\Delta_{ij}(\nu)+\Delta_{ji}(-\nu))/\sqrt{2}$. The generalized particle-particle singlet susceptibility is defined as 
\begin{equation}
\chi^{s}_{ijkl}(\nu,\nu')=\langle\Delta^{s}_{ij}(\nu)\Delta^{s\dagger}_{kl}(\nu')\rangle=\chi_{ijkl}^{P\uparrow\downarrow\uparrow\downarrow}(\nu,\nu')-\chi_{ijkl}^{P\uparrow\downarrow\downarrow\uparrow}(\nu,\nu')=\chi_{ijkl}^{P\uparrow\downarrow\uparrow\downarrow}(\nu,\nu')+\chi_{ijlk}^{P\uparrow\downarrow\uparrow\downarrow}(\nu,-\nu')
\end{equation}.

\begin{figure}[t!]
\includegraphics[width=.5\linewidth]{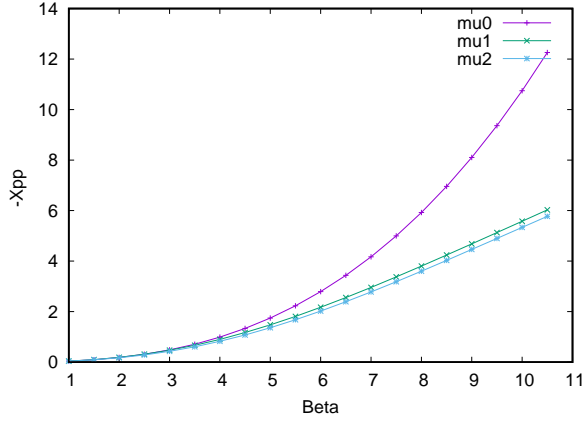}
\caption{The main non-local component of particle-particle generalized susceptibility $\chi_{0110}$ for plaquette with optimal parameters for degenerate point $\mu_0=0.48$ (mu0) and shifted $\mu_1=0.8$  (mu1) and $\mu_2=0.0$  (mu2) as function of inverse temperature $\beta$.}
\label{fig:Xi1221}
\end{figure}

To understand the special role of the plaquette degenerate point, we consider a single component of the particle-particle singlet susceptibility of the plaquette, namely $\chi^{s}_{0110}$ at $\nu=\nu'=\pi/\beta$.
Figure~\ref{fig:Xi1221} shows this objects as a function of the inverse temperature for three different chemical potentials $\mu_i$. At high temperature, the three different chemical potentials give essentially the same result, since $\Delta \mu \times \beta \ll 1$. However, at low temperature, we find both quantitative and qualitative changes. The degenerate point corresponds to $\mu_0\approx 0.48$ and at low temperature we find the scaling scaling $-\chi_{0110}\sim \beta^3$. For slightly shifted chemical potential: $\mu_1=0.8$ and  $\mu_2=0.0$ 
(see Fig.\ref{fig::2x2phase}) we obtain the standard linear behaviour as a function of $\beta$. We conclude that the degeneracy strongly enhances the corresponding components of the susceptibility at low temperature. This $\chi_{0110}$ can be considered as a plaquette-local precursor of the d-wave superconducting instability in the plaquette-lattice.

Now we proceed to constructing the $d$-wave pairing operator and consequently the $d$-wave generalized susceptibility. The pairing operator is given by (up to a normailzation factor):
\begin{equation}
    \Delta^{d}_{i}(\nu)=\sum_ja_{ij}\Delta_{ij}^{s}(\nu),
\end{equation}
where $a_{ij}=1$ when $ij$ is a horizontal bond, $a_{ij}=-1$ when it is a vertical bond, and $a_{ij}=0$ otherwise. Thus for example $\Delta^d_0=\Delta^s_{03}-\Delta_{01}^s$. The $d$-wave generalized susceptibility is then given by $\chi^d_{ik}(\nu,\nu')=\langle\Delta^d_i(\nu)\Delta^d_k(\nu')\rangle$. We will be interested only in the homogeneous part of it given by:
\begin{equation}
    \chi^d(\nu,\nu')=\sum_k\chi^d_{0k}(\nu,\nu').
\end{equation}

Breaking this expression into single contributions and keeping only the non-equivalent terms (in other words fixing the first pair of indices to $01$) we get:

\begin{equation}
\chi^{s}=\chi^s_{0110}+\chi^s_{0101}-\chi^s_{0121}-\chi^s_{0112}+\chi^s_{0132}+\chi^s_{0123}-\chi^s_{0103}-\chi^s_{0130}.
\end{equation}

\begin{figure}[t!]
\includegraphics[width=.4\linewidth]{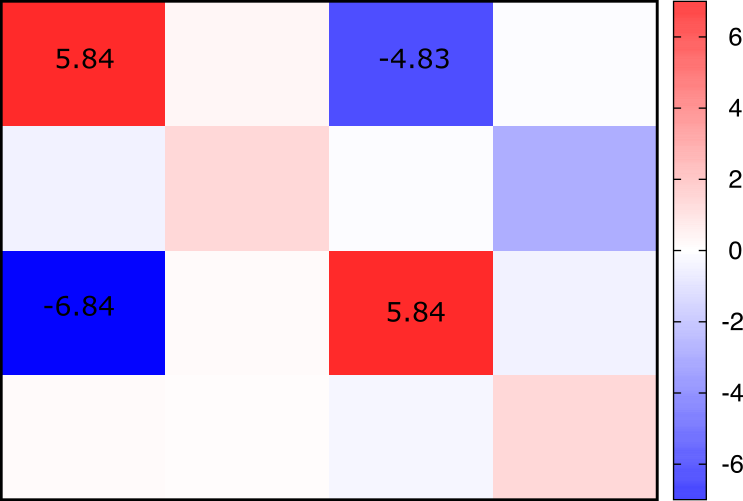}
\caption{ Elements of particle-particle susceptibility matrix $\chi_{ijkl}$ with the fixed $(i,j)=(0,1)$, $k$ going from 0 to 3 from left to right and $l$ from 0 to 3 from bottom to top. }
\label{fig:Xi01kl}
\end{figure}

Fig.~\ref{fig:Xi01kl} shows the elements $\chi^s_{01kl}(\nu=\pi/\beta,\nu'=\pi/\beta,\omega=0)$ for different $k=0\div 3$ and $l=0 \div 3$.
We can see that all non-negligible elements coherently increase each other according to the $d$-wave symmetry, therefore there is no cancellation and no frustration.   

\begin{figure}[t!]
\includegraphics[width=0.5\linewidth]{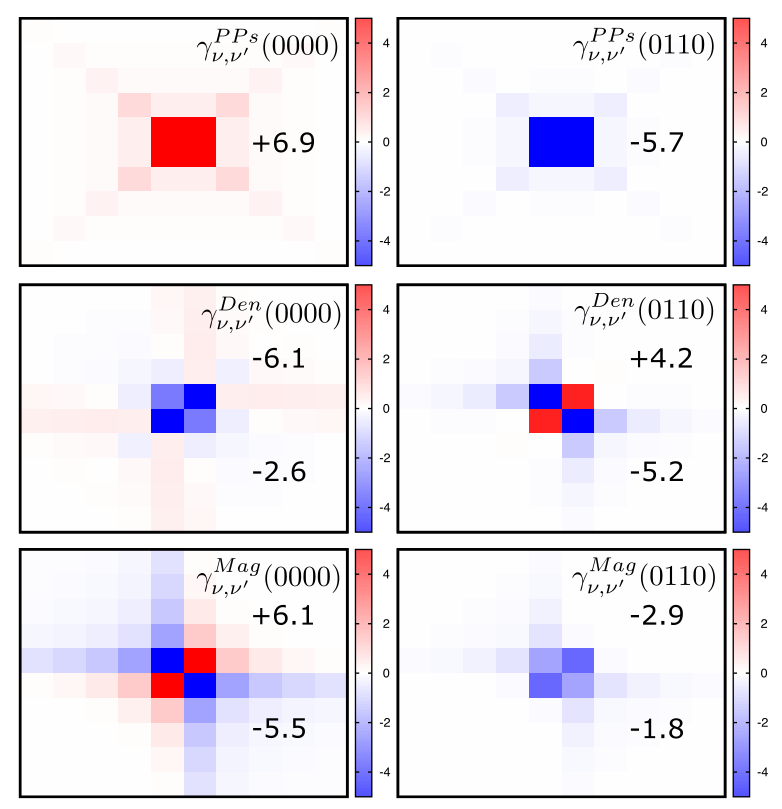}
\caption{Real part of Vertex function $\gamma^{PP/PH}_{\nu,\nu',\omega=0}$(ijkl) for different plaquette indices with U=5.56 and $t'/$=-0.3t, $\mu_{c}$=0.479 $\beta$=10. Maximum positive and negative values are shown}
\label{fig:vertex}
\end{figure}

Fig.~\ref{fig:vertex} shows the vertex (in our case just connected correlator ) $\gamma(\nu,\nu',\omega=0)$ of the reference model as a function of inverse temperature. 

\begin{figure}[t!]
\includegraphics[width=0.33\linewidth]{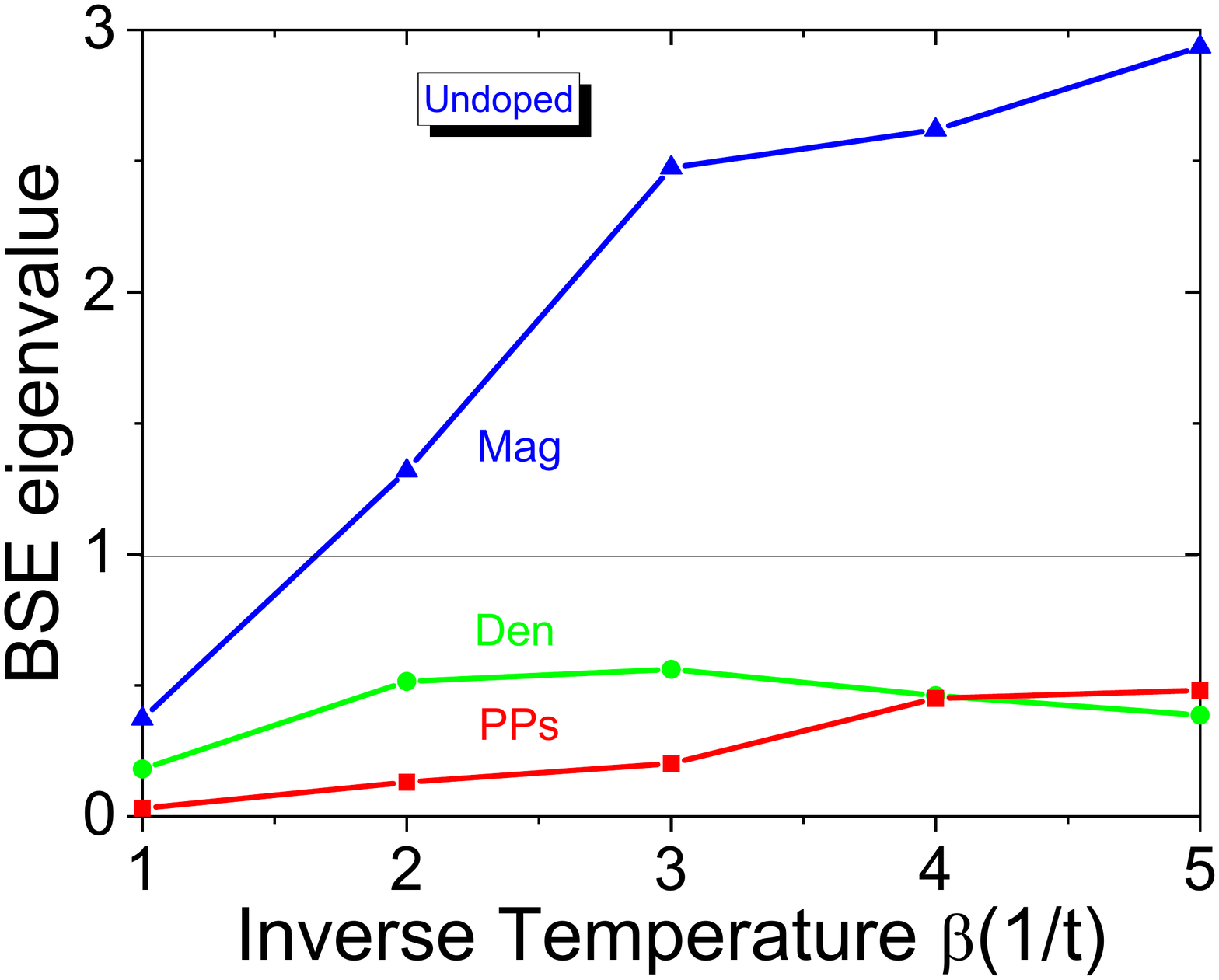}
\includegraphics[width=0.33\linewidth]{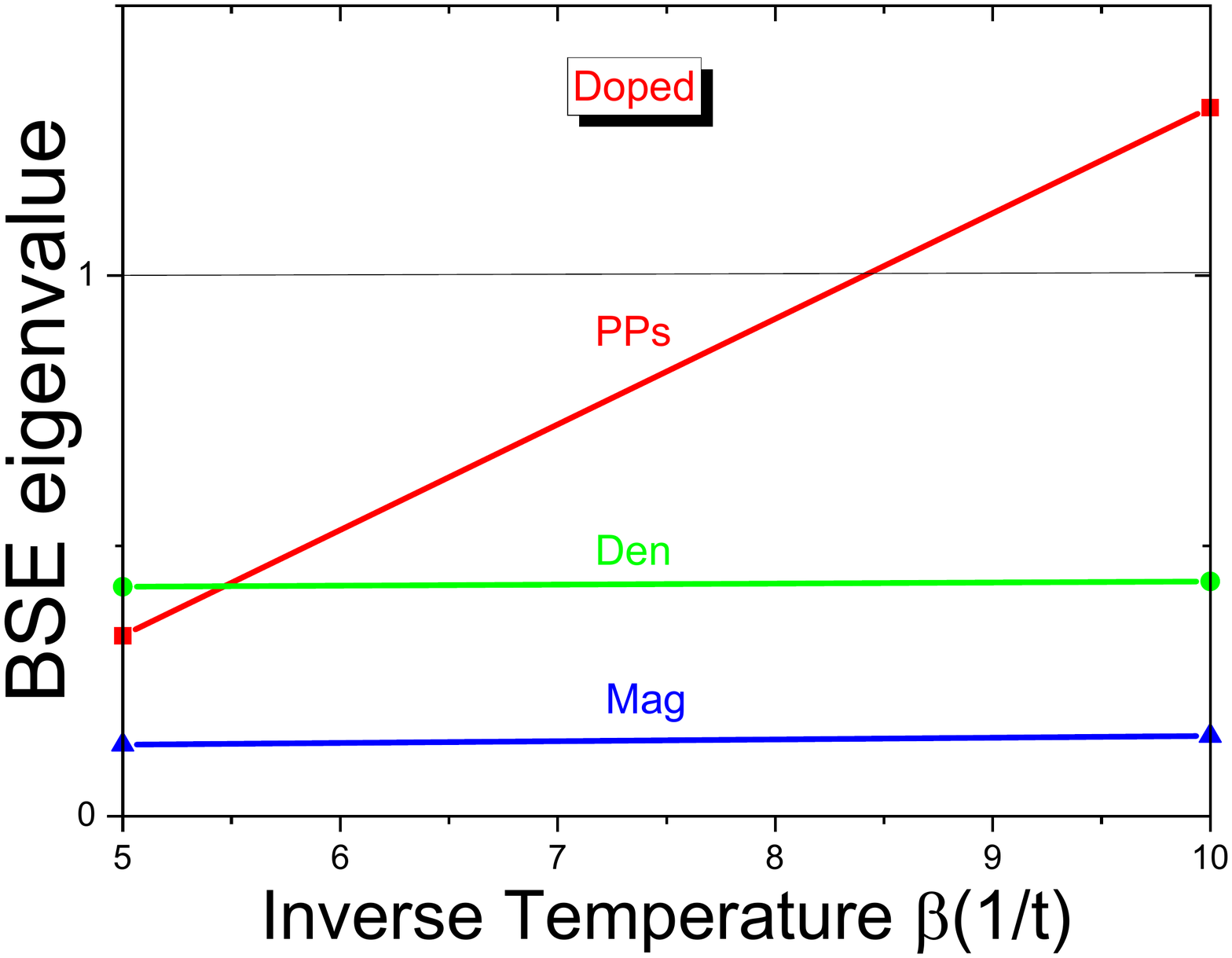}
\includegraphics[width=0.33\linewidth]{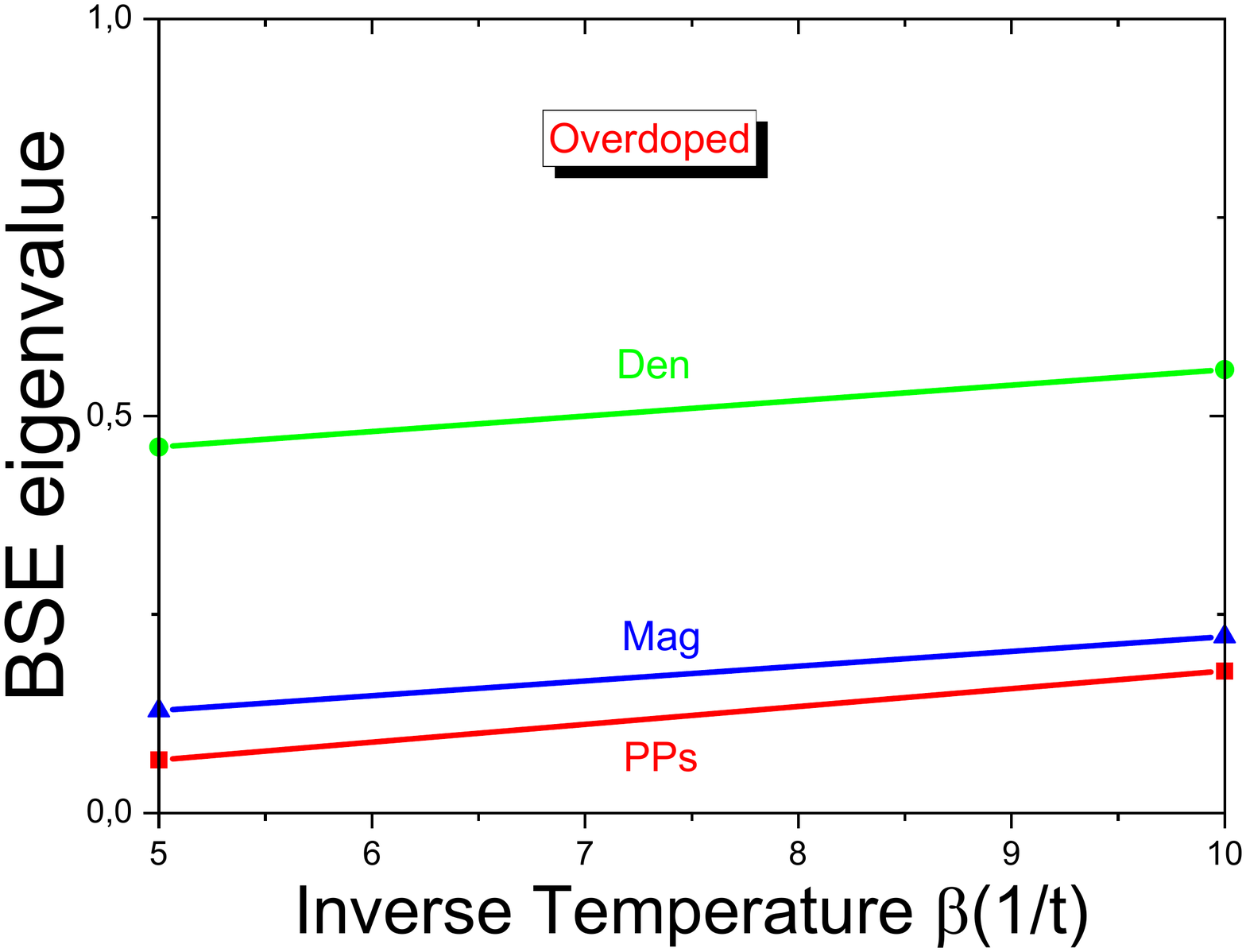}
\caption{Maximum eigenvalues of BSE for half-filled plaquette with U=8 and t'=0 (left),
for doped plaquette with U=5.56 and $t'$=-0.3t, $\mu_{0}$=0.8 $\mu$=1.55 (middle), and
for overdoped plaquette with U=5.56 and $t'$=-0.3t, $\mu_{0}$=0.0 $\mu$=1.5 (right)
}
\label{fig:BSEdoping}
\end{figure}

Results for the maximum eigenvalues of the Bethe-Salpeter matrix $\Lambda_{ij}$ for different hole doping are presented in the Fig.~\ref{fig:BSEdoping}. 
In the half-field case with $t'=0$ and $\mu_0=U/2$ the main instability is related to the particle-hole magnetic channel with the eigenvector corresponding to an
antiferromagnetic checkerboard structure. In this case there is no density or superconducting instability.
For the optimally doped case ($\mu_0=0.8$) the largest instability is related to the particle-particle singlet $d_{x^2-y^2}$ superconductivity. The density instability is not robust again small change in ($\mu_0$) for the plaquette.
Finally for the overdoped case ($\mu_0=0.0$) there is no instability until $\beta=10$ which may indicate a
formation of a normal metallic phase.

\begin{figure}[t!]
\includegraphics[width=0.48\linewidth]{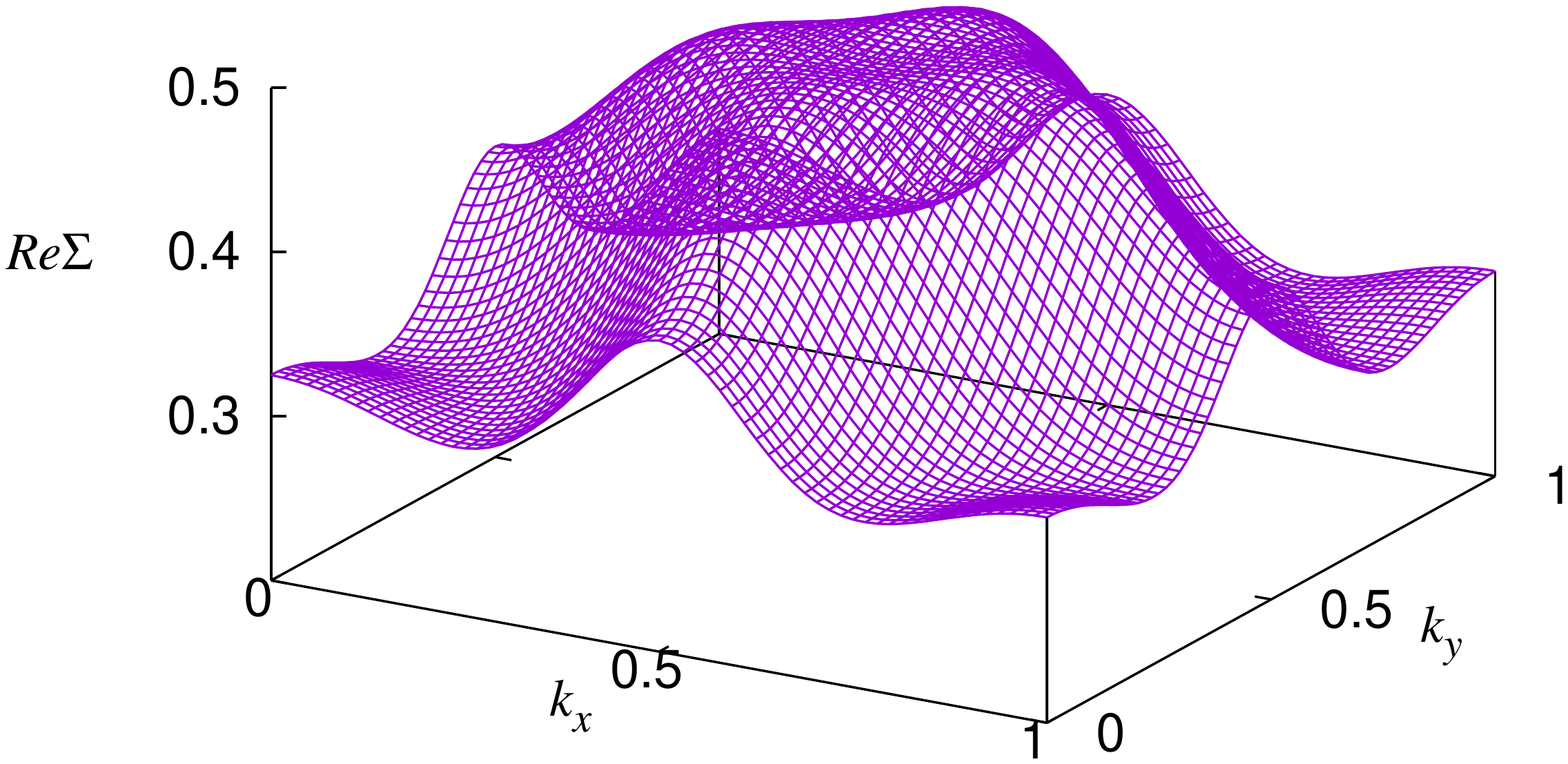}
\includegraphics[width=0.48\linewidth]{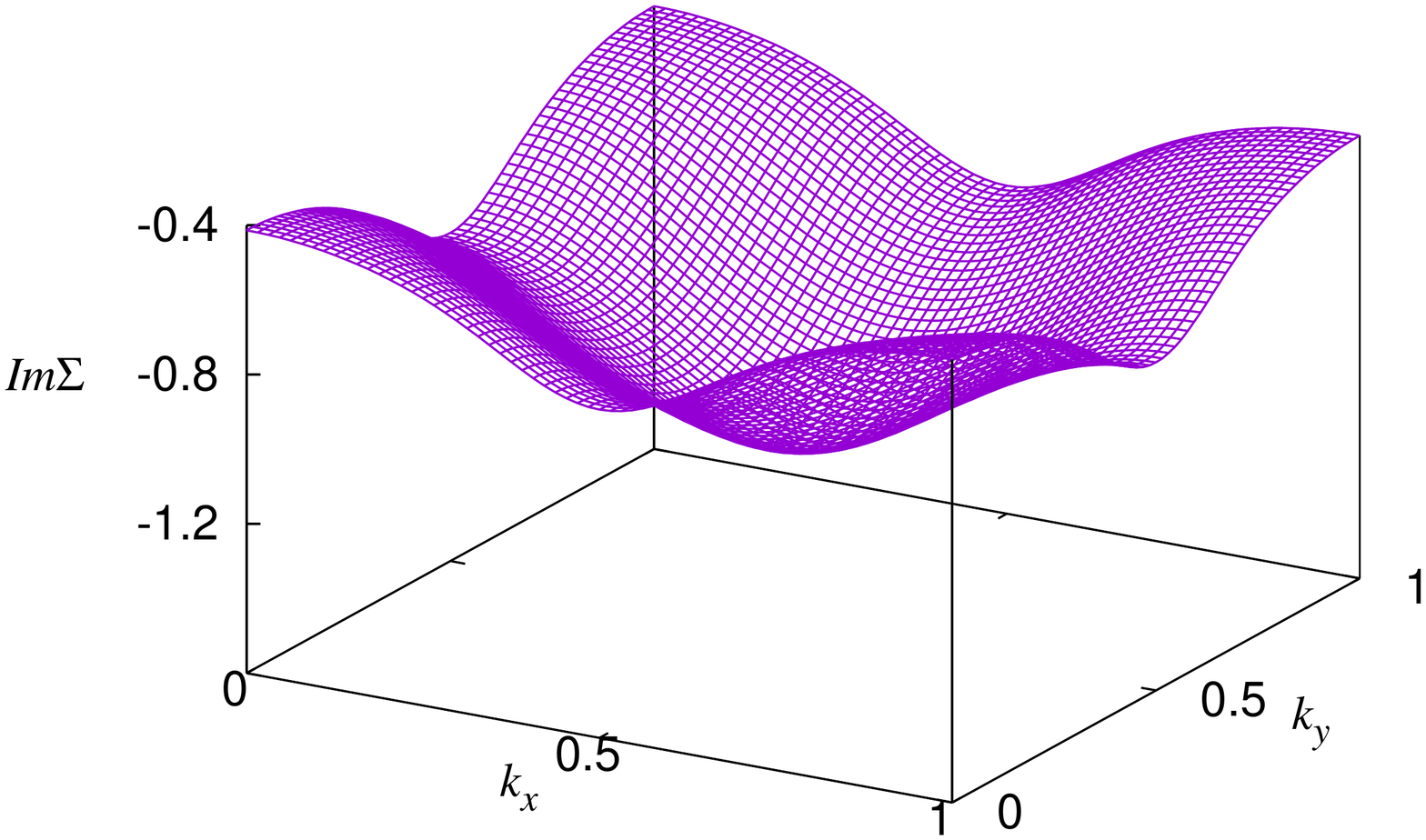}
\caption{Real (left) and imaginary (right) part of lattice $\Sigma_{{\mathbf{k}},\nu=\pi T}$ for dual fermion plaquette theory for $U$=5.56, $t'/t$=-0.15$, \mu_0=0.48$, $\mu$=1.55 and $\beta=3$.}
\label{fig:SigReIm}
\end{figure}
The Fig.\ref{fig:SigReIm} shows the periodized lattice DF plaquette self-energy according to Eq.~\eqref{DF_SE} in the full Brillouin zone $64\times64$ $\bf k$-mesh for the lowest Matsubara frequency. The real part of the self-energy already has an anomalously sharp feature near  $X$ point in the Brillouin zone.

\begin{figure}[t!]
\includegraphics[width=0.5\linewidth]{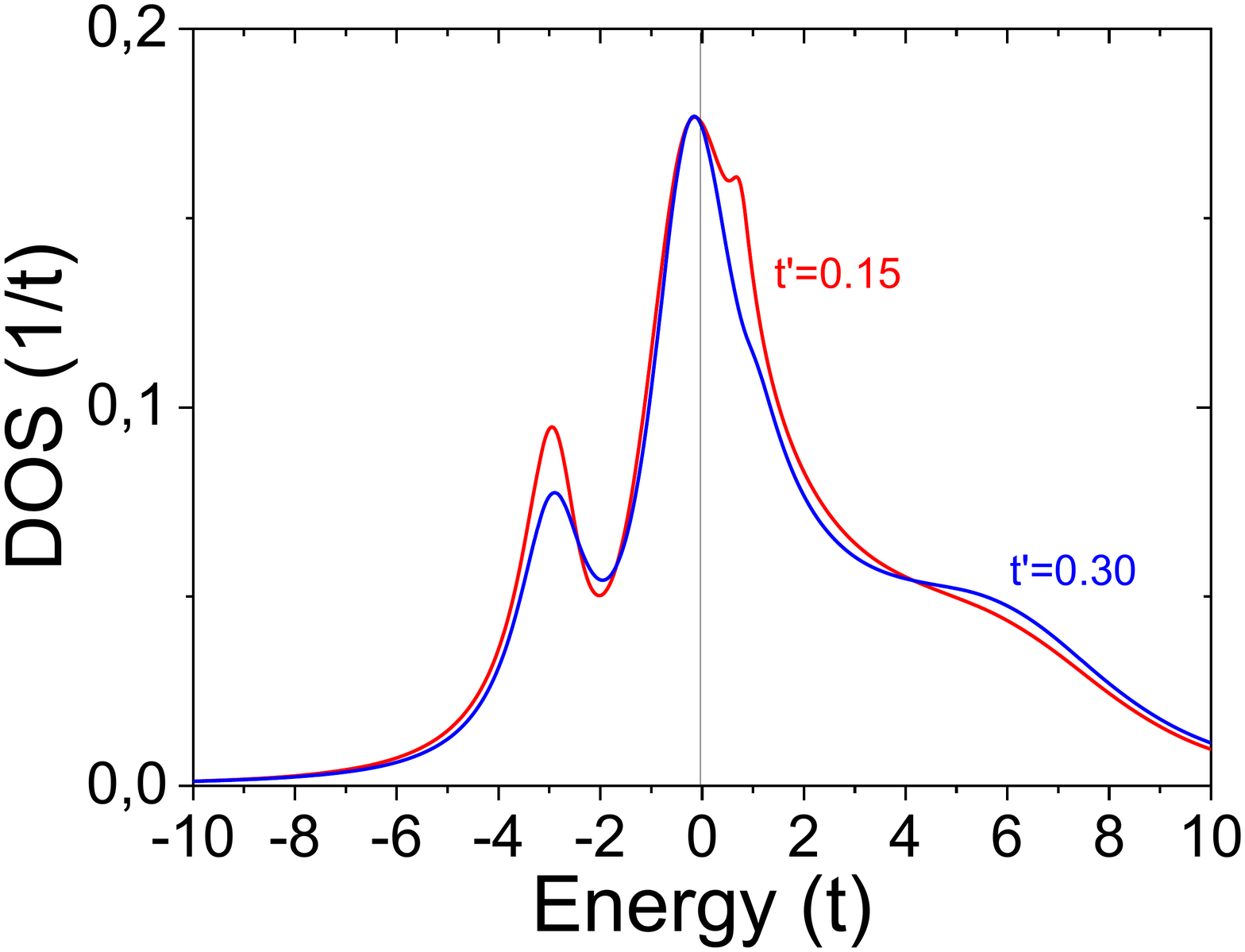}
\caption{Density of states for dual fermion plaquette superperturbation (DF) for for $U$=5.56, $t_0'/t$=-0.15$, \mu_0=0.48$
in two different approximations: $t'/t$=-0.15, $\mu$=1.0 and $t'/t$=-0.3,  $\mu$=0.7}
\label{fig:DOS}
\end{figure}

In the Fig.\ref{fig:DOS} we compare the density of states (DOS) for plaquette DF perturbation for two different approximations: the lable $t'=0.15$ corresponds to $t'_0=t'=0.15$ and $\mu=1.0$, while the case $t'=0.3$ corresponds to $t'=0.3$ with reference hopping $t'_0=0.15$ and $\mu=0.7$. Both calculations correspond approximately to hole doping $\delta=0.15$.

\begin{figure}[t!]
\includegraphics[width=0.4\linewidth]{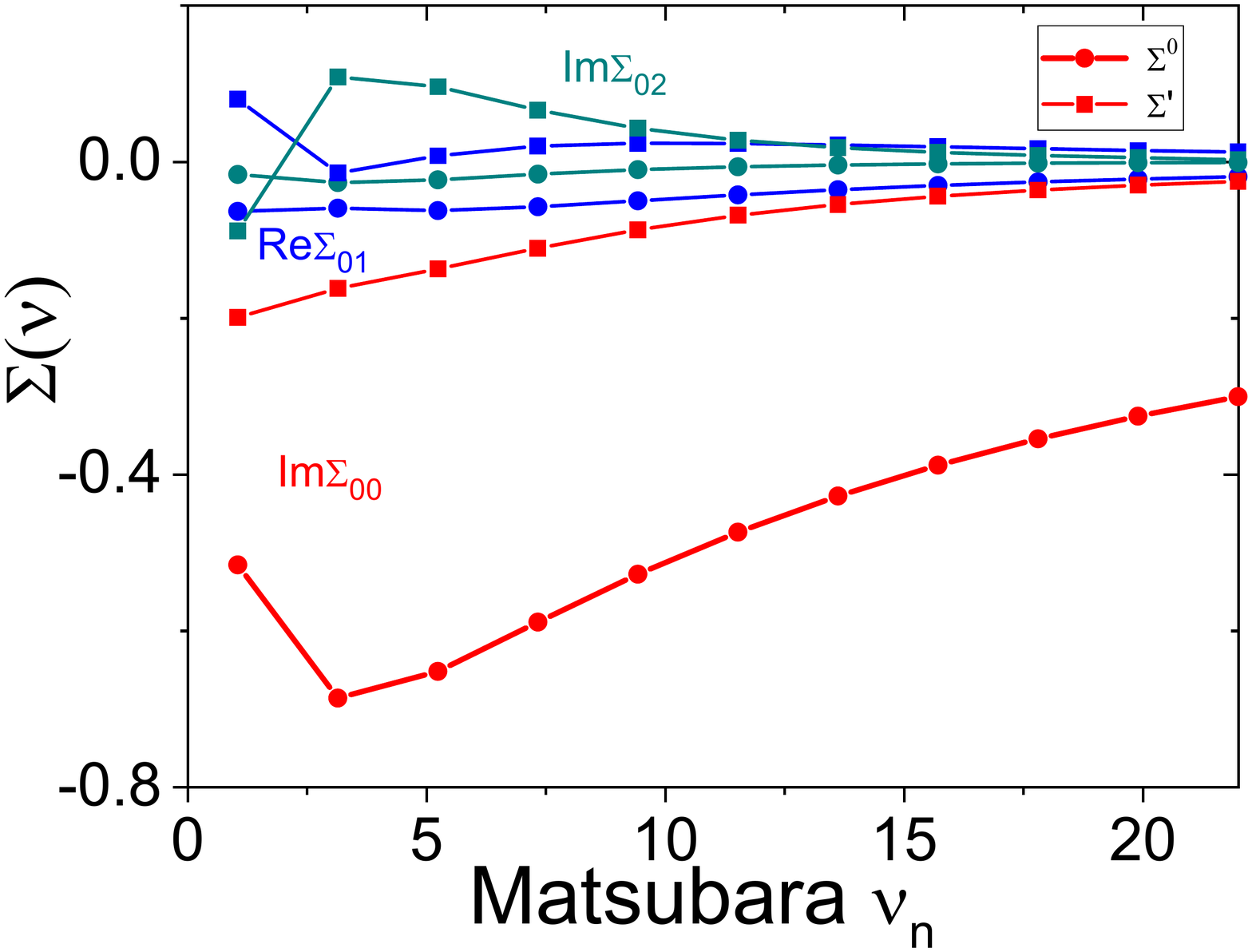}
\includegraphics[width=0.4\linewidth]{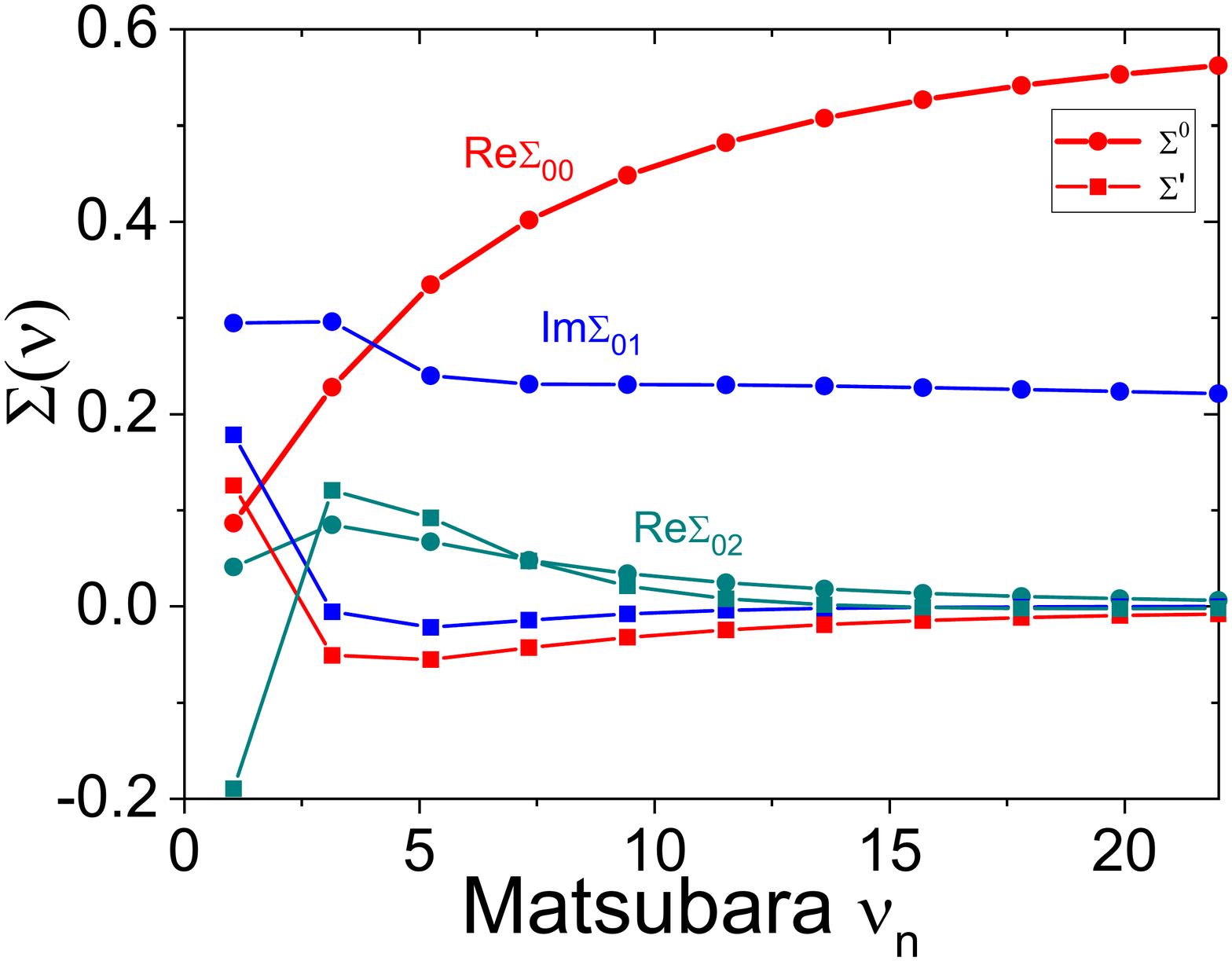}
\caption{Self-energy for the optimally doped case for the reference system ($\Sigma_0$, circle) and the non-local DF part ($\Sigma^\prime$, square) for $\beta=3$, $\mu=1.55$ from the plaquette DF-perturbation at the degenerate point.}
\label{fig:SigB3Mu155}
\end{figure}

\begin{figure}[t!]
\includegraphics[width=0.4\linewidth]{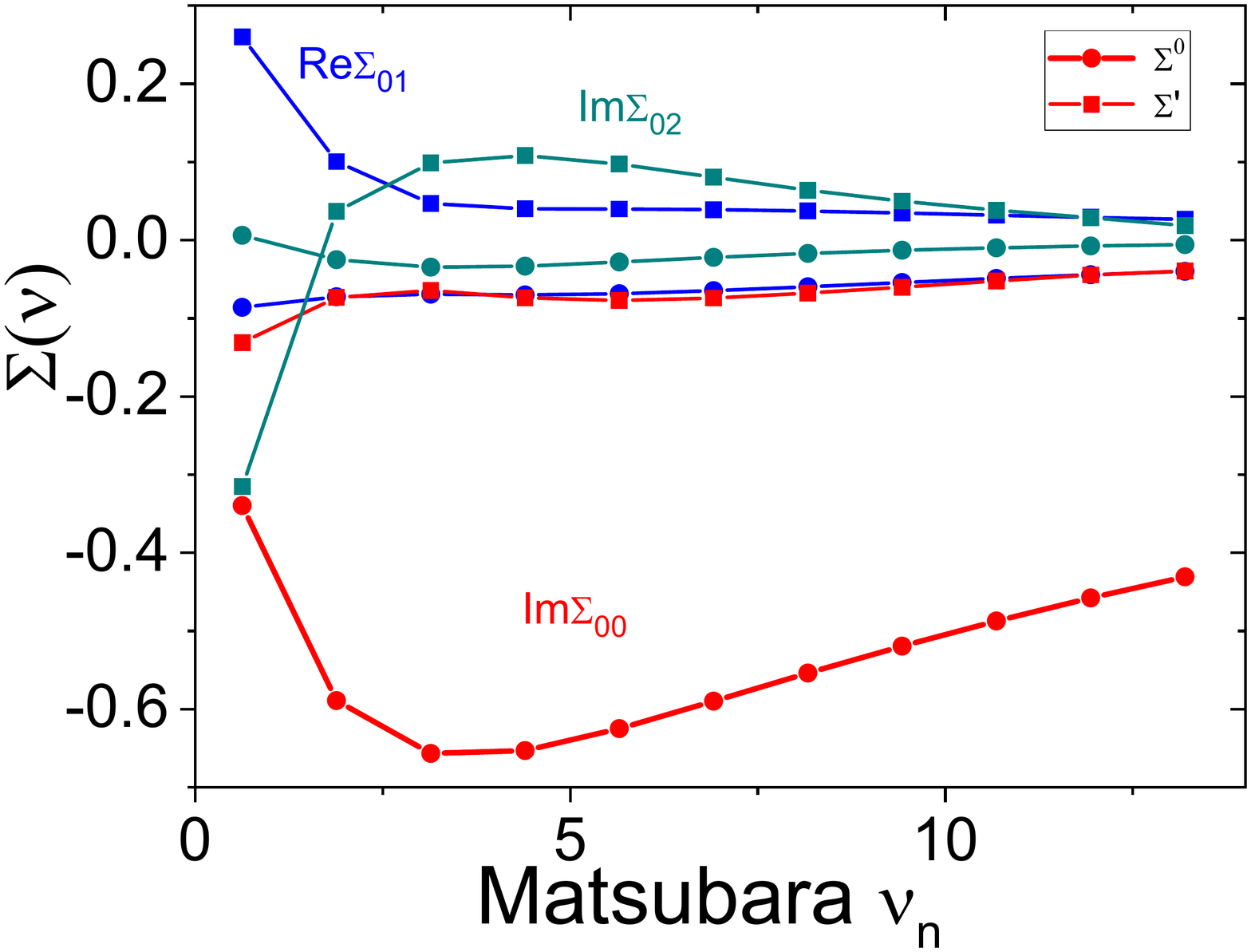}
\includegraphics[width=0.4\linewidth]{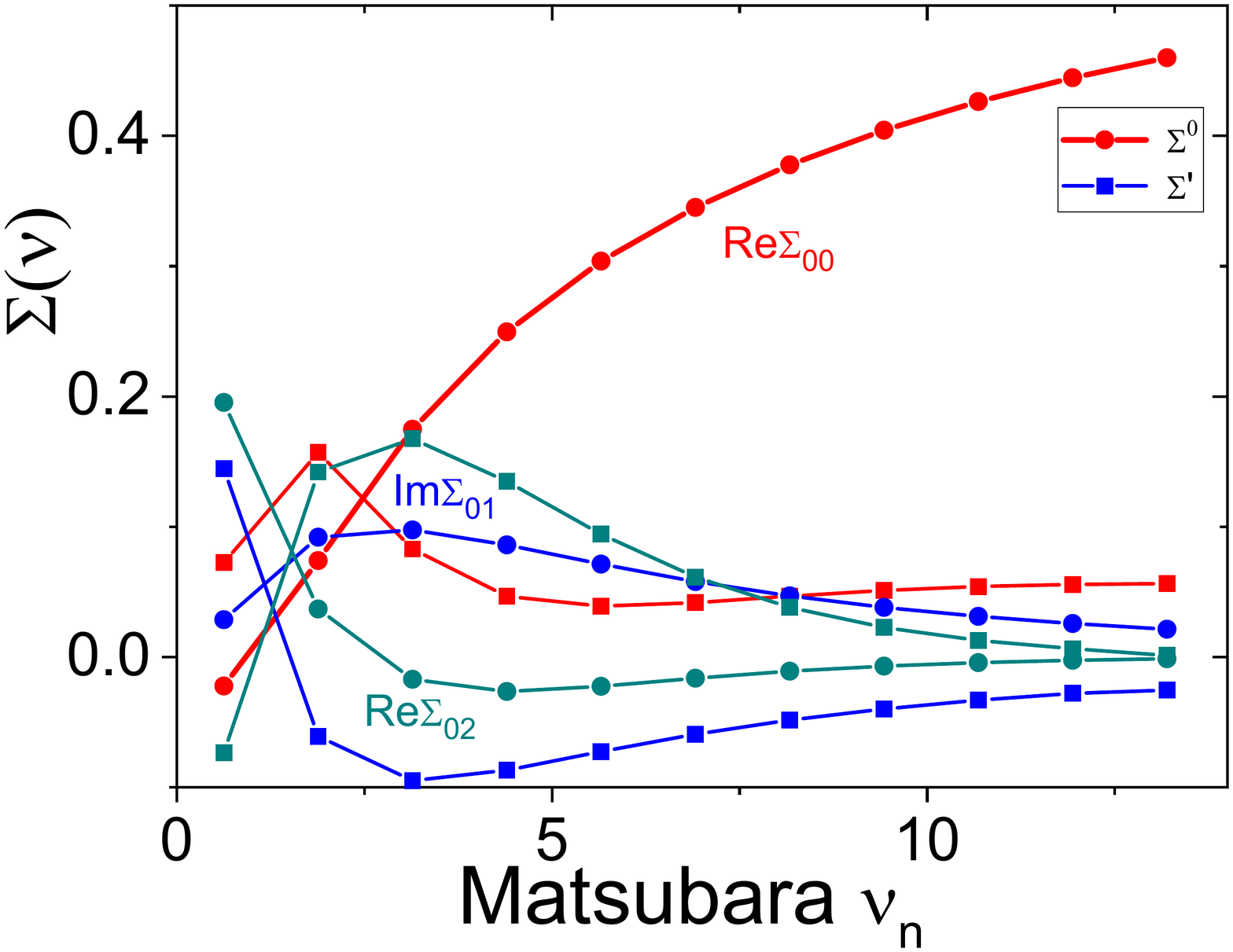}
\caption{Self-energy for the optimally doped case for the reference system ($\Sigma_0$, circle) and the non-local DF part ($\Sigma^\prime$, square) for $\beta=5$, $\mu=1.55$ from the plaquette DF-perturbation at the degenerate point.}
\label{fig:SigB5Mu155}
\end{figure}

We plot the frequency-dependent of $\Sigma'$ self-energy only within the plaquette extent ($\Sigma'$) and compare to the bare plaquette reference self-energy ($\Sigma_0$) [see Eq.~(\ref{DF_SE})] for two different temperatures in Fig.\ref{fig:SigB3Mu155} and Fig.\ref{fig:SigB5Mu155}.
Note the increase of ($\Sigma_{02}'$) (along $t'$) with decreasing temperature already for $\beta=5$. 


\section{ED for the 4 $\times $ 4 cluster}
\label{App_ED}
\begin{figure}[t!]
\includegraphics[width=0.6\linewidth]{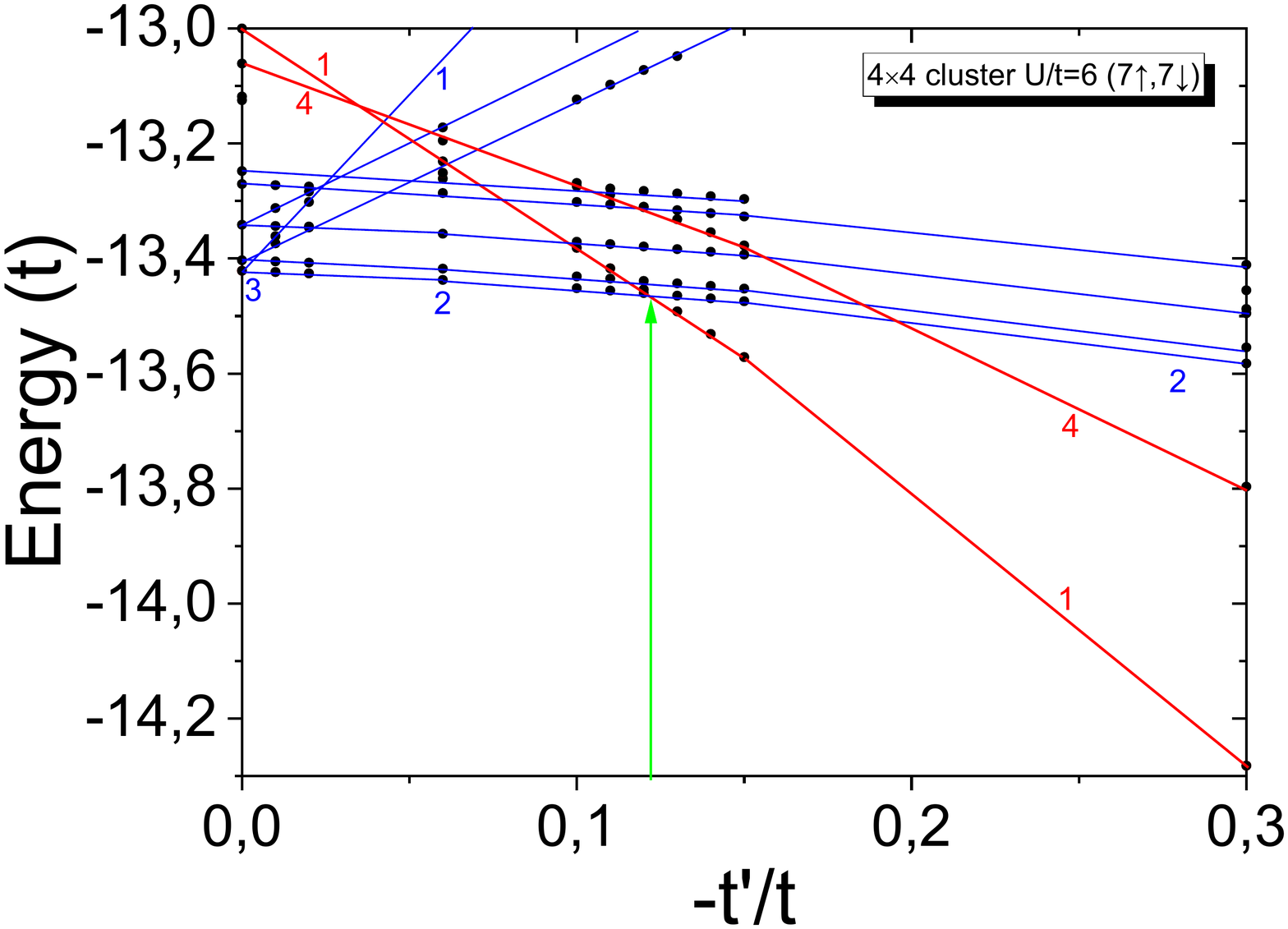}
\caption{Many body states of (4 $\times $ 4) periodic cluster for the sector (7$\uparrow$,7$\downarrow$) as a function of $t'$ for $U/t=6$. The degeneracy of few low-lying states are marked with the numbers. The green arrow indicate the critical $t'$ for ground state crossing.}
\label{fig:EgED7x7}
\end{figure}

\begin{figure}[t!]
\includegraphics[width=0.45\linewidth]{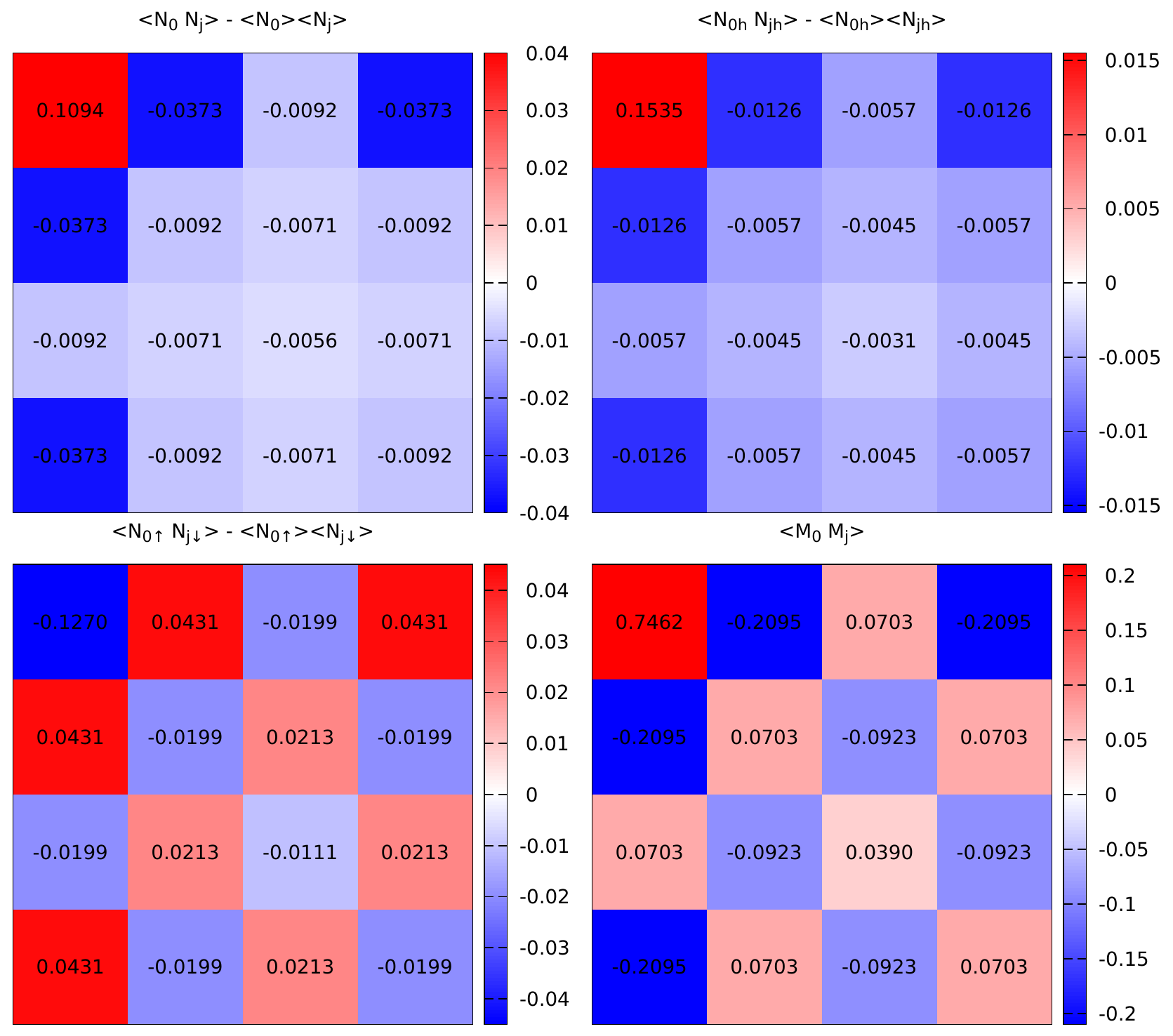}
\includegraphics[width=0.45\linewidth]{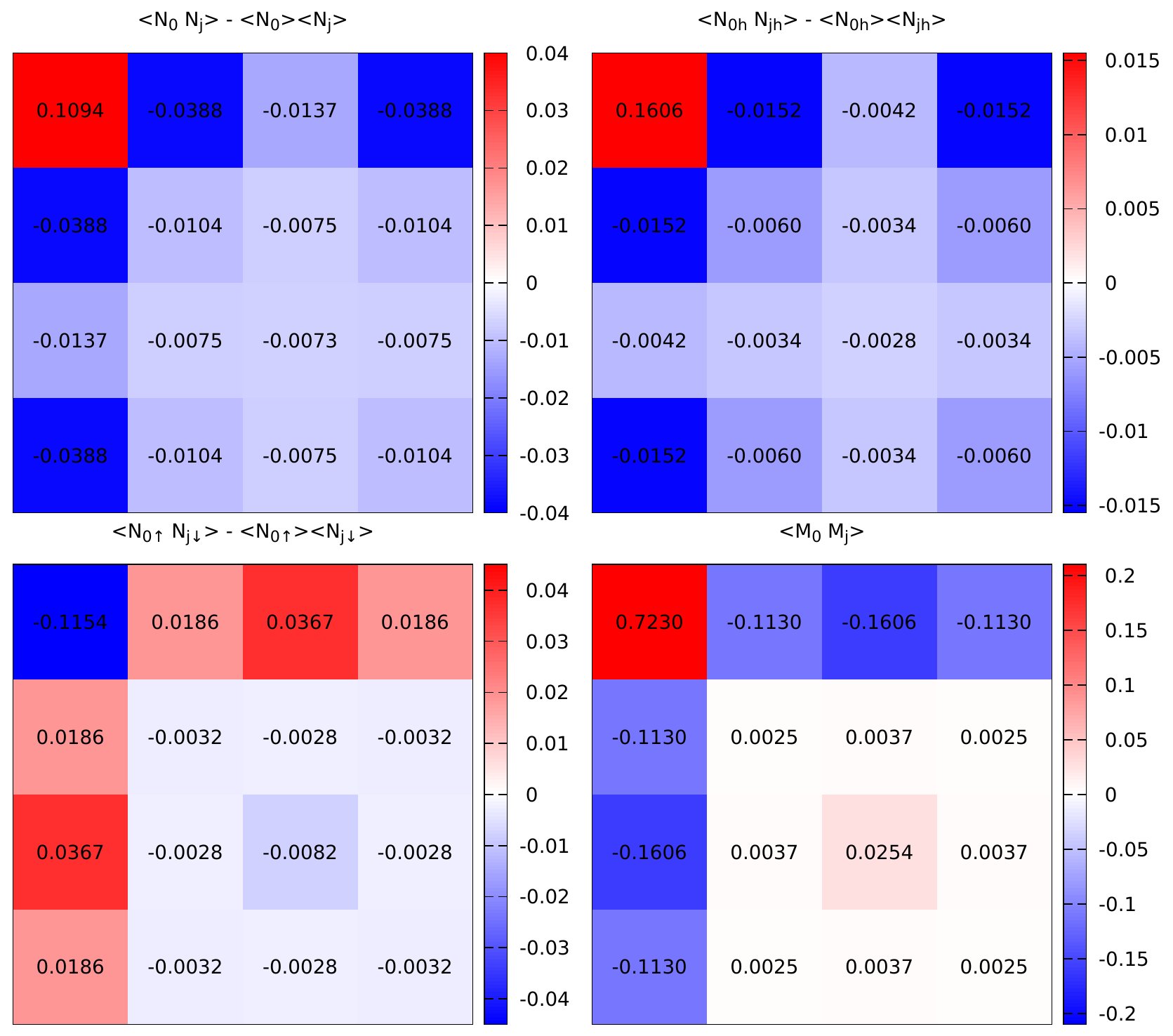}
\caption{Different Static correlators of the ground state of the (4 $\times $ 4) periodic cluster for the sector (7$\uparrow$,7$\downarrow$ )for $U/t=5.56$ and $t'/t=0$ (left) and $t'/t=-0.3$ (right).} 
\label{fig:MiMj0}
\end{figure}

We analyse low-lying many-body states of the (4 $\times $ 4) periodic cluster for the sector (7$\uparrow$,7$\downarrow$) in 
Fig.~\ref{fig:EgED7x7} and indicate the degeneracy of a few important states with numbers. The ground state for $t'=0$ is three-fold
degenerate due to $2^4$ super-cube symmetry\cite{Dagotto_2222}. As a function of $t'$ this state splits to a ground state doublet and a 
singlet which has higher energy. Around  $t'/t=-0.12$ the ground state of the sector (7$\uparrow$,7$\downarrow$) changes to a singlet 
(red line on the Fig.~\ref{fig:EgED7x7}) with a much lower energy and different symmetry related with a drastic change of the spin-spin correlations from antiferromagnetic-like for small $t'$ to basically nonmagnetic for larger $t'$ (Fig.~\ref{fig:MiMj0}). 

\begin{figure}[t!]
\includegraphics[width=0.4\linewidth]{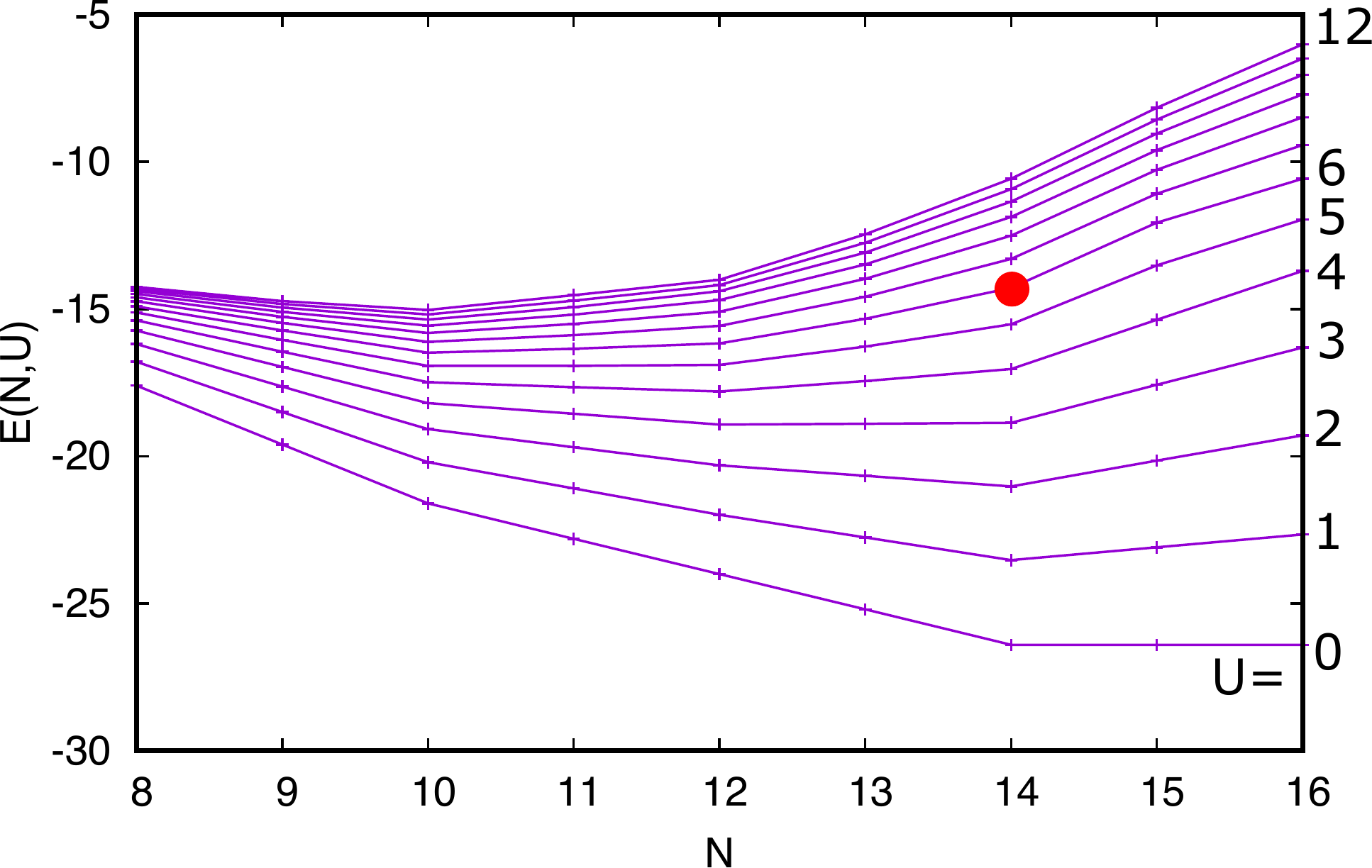}
\includegraphics[width=0.38\linewidth]{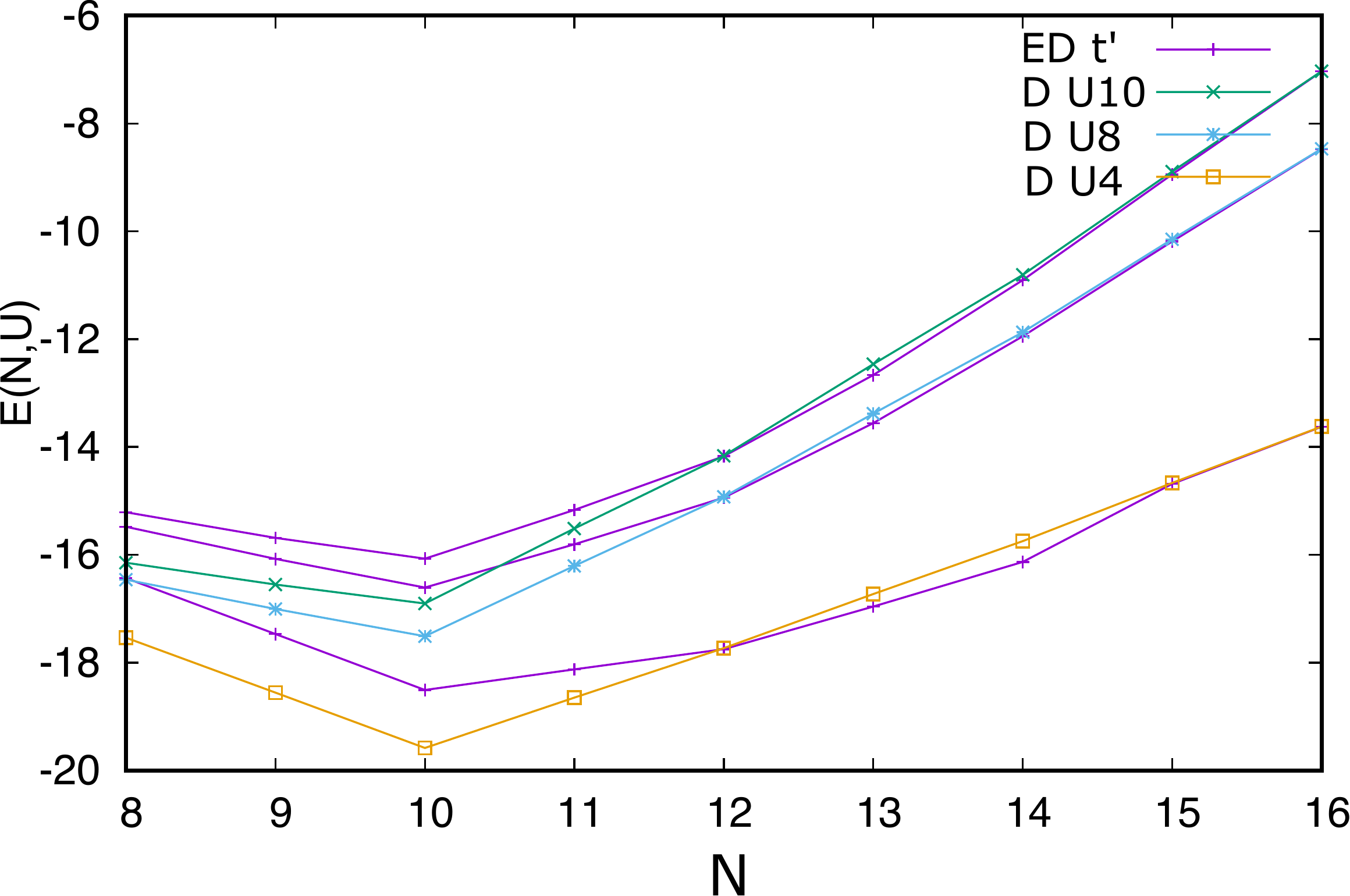}
\caption{Many body ground state energies of (4 $\times $ 4) periodic cluster for different sectors as function of $U$ for $t'/t=-0.3$ and $\mu=0$. The red points shows the largest effect of lowering the total energy for the sector $N=14$ ($7{\uparrow},7{\downarrow}$)
(left). In other words, $E(14,U)-E(16,U)$ is minimal for $U=6$. Comparison of the present ED calculations for $t'/t=-0.15$ with (ED $t'$) ED results of Dagotto et.al\cite{Dagotto_4x4} for U=4, 8, 10 from bottom to top (right).
}
\label{fig:EnED4x4}
\end{figure}

Fig.~\ref{fig:EnED4x4} (left panel) shows the many-body ground state energies for (4 $\times $ 4) periodic cluster for different sectors with $N$ electrons for different interaction strength $U=0\div 12$ (from bottom to top). One can clearly see that the largest
effect of lowering energy of $N=14$ sector compared to the half-field $N=16$ one (red dot) appears at $U=6$ where the pair-hole 
binding energy has the minimum for $t'/t=-0.3$ (Fig.~\ref{fig:Delta2h}). In the right panel of Fig.~\ref{fig:EnED4x4} we show a comparison
of our exact diagonalization of the 4 $\times $ 4 periodic cluster for $t'/t=-0.15$ with the ED results of Dagotto et.al\cite{Dagotto_4x4} for $t'=0$ and U=4, 8, 10. Note that our ED results for $t'=0$ exactly reproduce the ones of Dagotto et.al\cite{Dagotto_4x4}.
We note that the many-body energies for the half-field cluster $N=16$ is almost perfectly insensitive to small changes of $t'$ due to the
antifferomagnetic blocking of the $t'$ hoppings. The same effect is observed even for the one hole case ($N=15$), probably due to a large string-like $t-J$ blocking\cite{Dagotto_4x4}. However for the two-hole situation ($N=14$) there is clearly an appreciable lowering of ground state energy for
$U=4$ where the pair-hole binding energy has a minimum for $t'/t=-0.15$ (Fig.~\ref{fig:Delta2h}). 

In Fig.~\ref{fig:MiMj0} different static correlators are shown from ED calculations for the (4 $\times $ 4) periodic cluster
with $t'=0$ and $t'/t=-0.3$. The spin-spin correlators in the sector $N=14$ ($7{\uparrow},7{\downarrow}$) drastically change behaviour from almost antiferromagnetic at $t'=0$ to almost nonmagnetic structure with the weak negative correlations within the horizontal and vertical stripes. The charge-charge correlators seem to be robust against large changes of $t'$. The hole density correlators will be discussed below.

\begin{figure}[t!]
\includegraphics[width=0.6\linewidth]{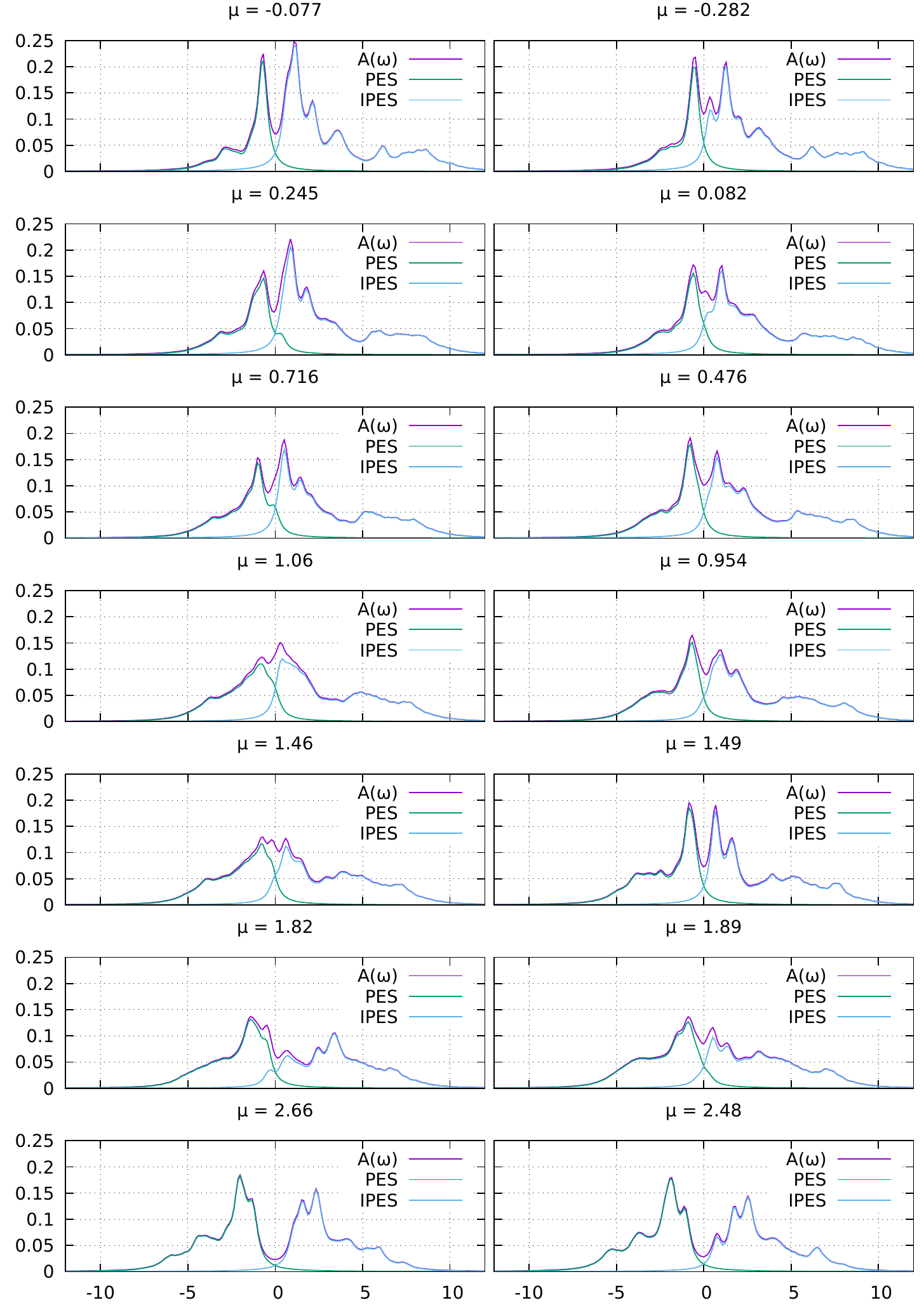}
\caption{DOS for (4 $\times $ 4) periodic cluster $\beta=10$
Spectral density A($\omega$) (as well as PES and IPES part) of periodic 4 $\times $ 4 cluster with U=5.56, $\beta=10$
and different NNN hopping t': 0.15 (left column) , 0.3 (right column).
Different  doping and corresponding sectors are shown from the top to bottom:  $\delta=$0.375 (5$\uparrow$,5$\downarrow$) ; $\delta=$0.3125 (6$\uparrow$,5$\downarrow$); $\delta=$0.25 (6$\uparrow$,6$\downarrow$); $\delta=$0.1875 (7$\uparrow$,6$\downarrow$); $\delta=$0.125 (7$\uparrow$,7$\downarrow$); $\delta=$0.0525 (8$\uparrow$,7$\downarrow$), $\delta=$0 (8$\uparrow$,8$\downarrow$).
}
\label{fig:DOS4x4}
\end{figure}

\begin{figure}[t!]
\includegraphics[width=0.3\linewidth]{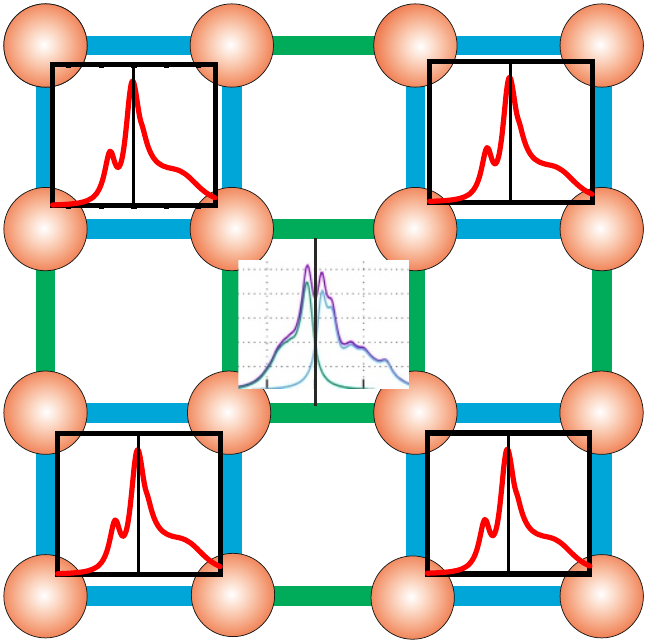}
\caption{Schematic view of pseudogap formation in  (4 $\times $ 4) periodic cluster from the peak DOS structure of individual 2 $\times $ 2 plaquettes.}
\label{fig:PGschematic}
\end{figure}

Fig.~\ref{fig:DOS4x4} shows the density of states for different sectors (hole concentrations) for ED calculations of (4 $\times $ 4) periodic cluster
with $t'/t=-0.15$ and $t'/t=-0.3$.
We can conclude that for $t'/t=-0.3$ and optimal $U=5.56$ all calculated sectors corresponding to doping $\delta=0.0525 \div 0.25$ have large
pseudogap DOS.
Simple pictorial view on such pseudogap formation presented in Fig.~\ref{fig:PGschematic} (left). If we consider (4 $\times $ 4) cluster buid from 
four interacting (2 $\times $ 2) plaquettes each of has sharp peak at Fermi level, then it is clear that through the resonant interactions 
the total DOS would have a pseudogap at $E_F$. This is similar to the Fano effect for Kondo-like impurity in the conducting bath.

\begin{figure}[t!]
\includegraphics[width=0.4\linewidth]{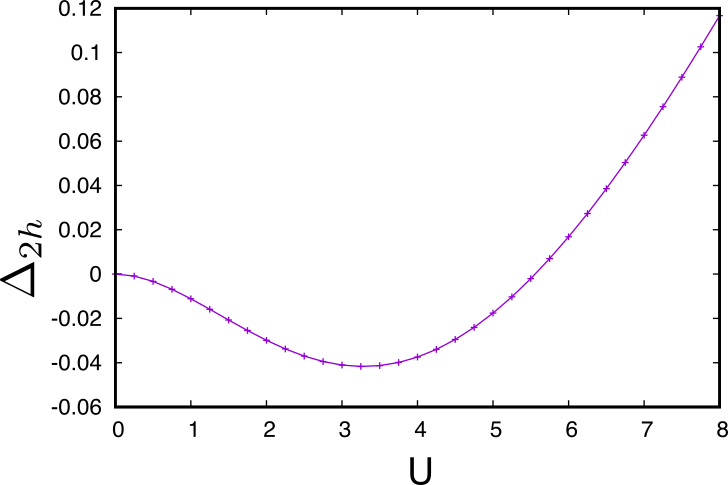}
\caption{Energy of two-hole binding for 2$\times$2 plaquette for $t'/t=-0.3$. Note that the energy scale is reduced by more than an order of magnitude compared to the $4\times 4$ plaquette, see Fig.~\ref{fig:Delta2h}. }
\label{fig:del2h2x2}
\end{figure}

We should point out that the optimal interaction $U/t\approx 6$ is smaller than the bandwidth $W/t=8$ and substantially below the strong coupling, effective $t-J$ model limit. Therefore, the huge hole-hole binding we found in the 4$\times$4 cluster at intermediate $U/t\approx 6$, with two holes located on different ``diagonal'' plaquettes, is very different from the so-called ``string-like'' effective hole-hole interactions in the $t-J$ model, where two holes are sitting with nearest-neighbor or next-nearest-neighbor distance~\cite{Dagotto_4x4}, i.e., in the same plaquette.
In the Fig.~\ref{fig:del2h2x2} we presented the pair-hole binding energy for the ED calculations of a  (2 $\times $ 2) periodic plaquette with $t'/t=-0.3$ as a function of $U$. The energy of the two-hole binding is much smaller than for (4 $\times $ 4) cluster with the same $t'$.
The  two-hole binding energy in a single 2$\times$2 plaquette is very similar to the results of Ref.~\cite{Auerbach} at $t'=0$. This indicates that it is not favourable to put two holes in a single plaquette. Thus, the pairing is a phenomenon that emerges in the lattice of plaquettes, as we could also see in the dual Bethe-Salpeter equation.

\begin{figure}[t!]
\includegraphics[width=0.4\linewidth]{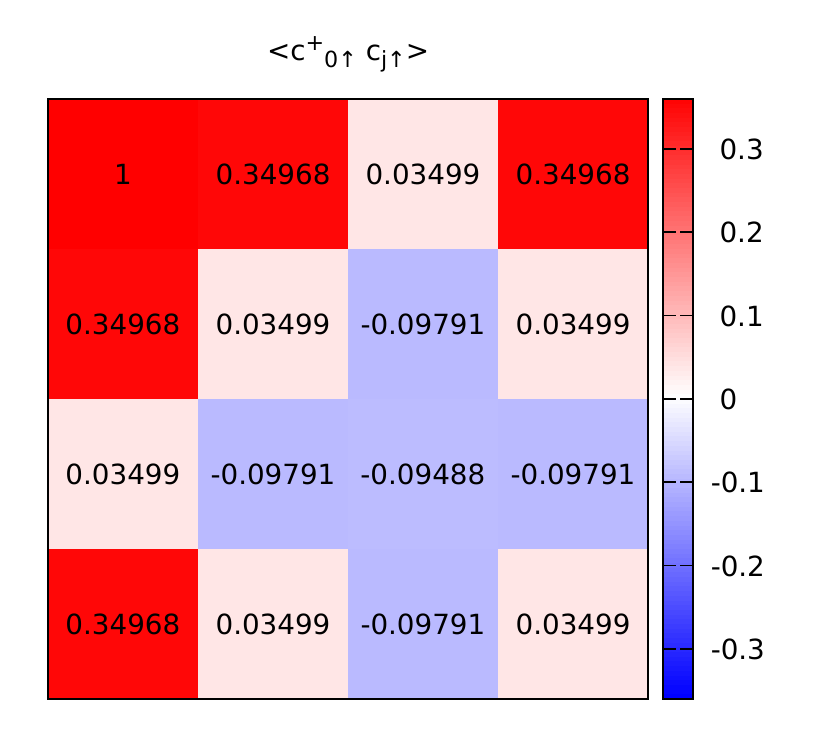}
\includegraphics[width=0.4\linewidth]{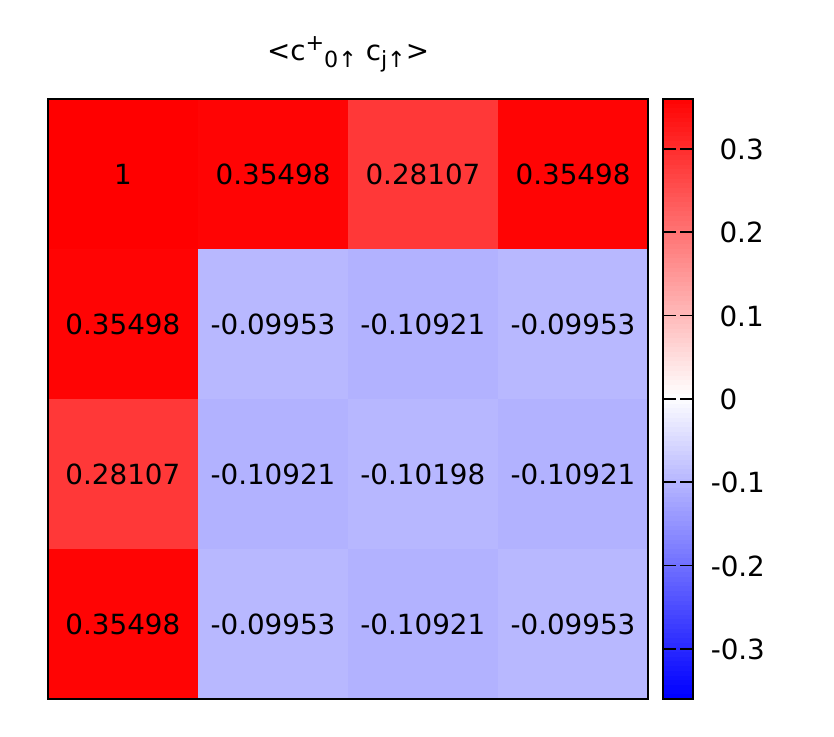}
\caption{Static correlators: $\langle 7\uparrow,7\downarrow|\hat{c}^\dagger_0 \hat{c}_j|7\uparrow,7\downarrow \rangle$ 
of (4 $\times $ 4) periodic cluster for the sector for $U/t=5.56$ and $t'/t=0$ (left) and $t'/t=-0.3$ (right).} 
\label{fig:cdicj}
\end{figure}

\begin{figure}[t!]
\includegraphics[width=0.4\linewidth]{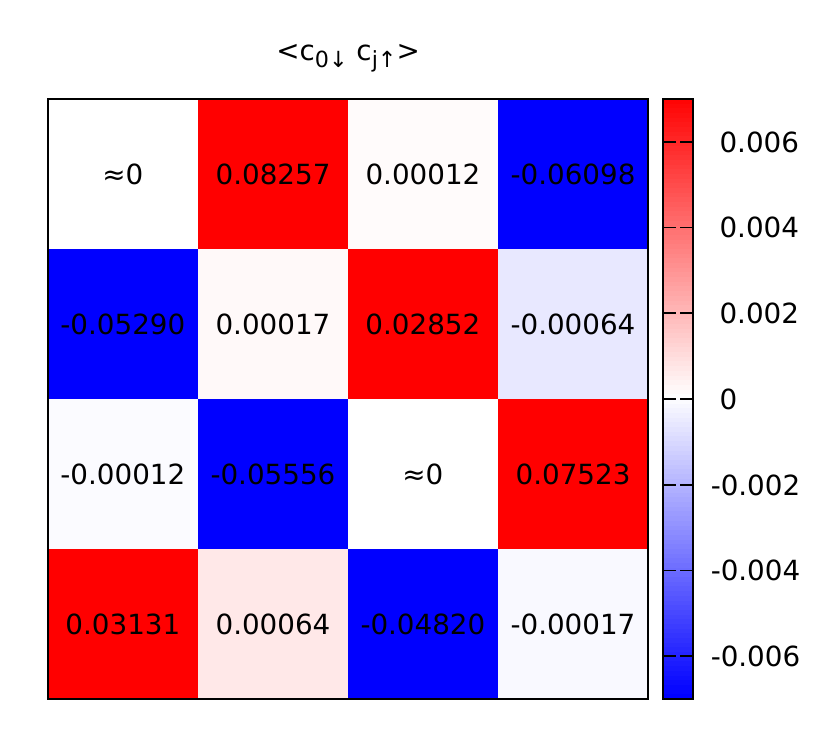}
\includegraphics[width=0.4\linewidth]{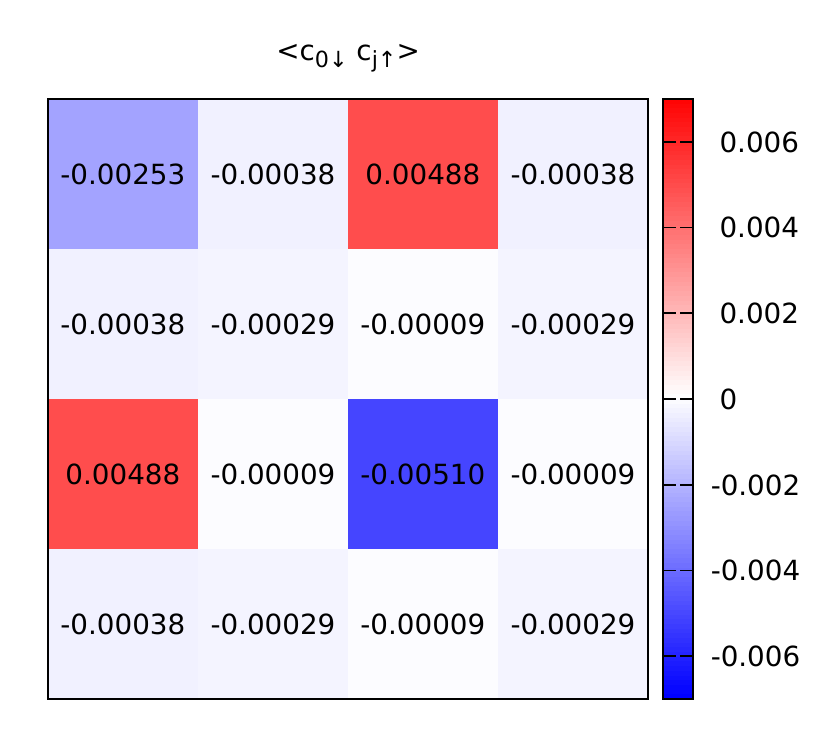}
\caption{Static correlators: 
and $\langle 7\uparrow,7\downarrow|\hat{c}_0 \hat{c}_j|8\uparrow,8\downarrow \rangle$ 
of (4 $\times $ 4) periodic cluster for the sector for $U/t=5.56$ and $t'/t=0$ (left) and $t'/t=-0.3$ (right).} 
\label{fig:cicj}
\end{figure}

It is also instructive to compare the changes in the static hopping correlator 
$\langle \hat{c}^\dagger_0 \hat{c}_j \rangle$ within the sector $(7\uparrow,7\downarrow)$ for different $t'$ (see Fig.~\ref{fig:cdicj}). 
While in the case of $t'=0$ all next-nearest hoppings  are very small,
including of $t'/t=-0.3$ produces "long-range" hopping correlators in all directions which
highlights the role of kinetic stabilization of the two-hole states. 

\begin{figure}[t!]
\includegraphics[width=0.8\linewidth]{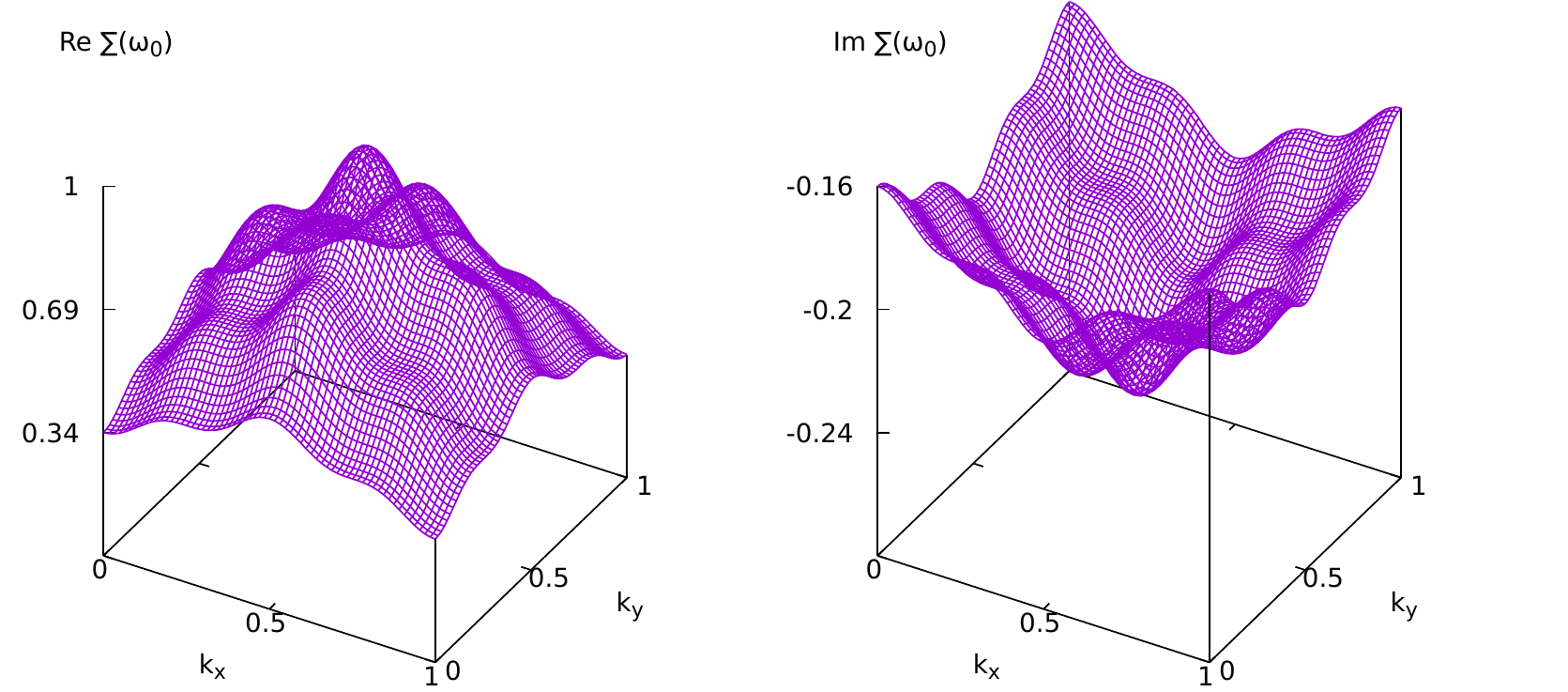}
\caption{Real (left) and imaginary (right) part of self-energy $\Sigma_{{\mathbf{k}},\nu=\pi T}$  in Brillouin zone from exact diagonalization of 4$\times$4 cluster for $U=5.56$, $t'/t=-0.3$, $\beta=10$ and $\delta=$0.125 in the (7$\uparrow$,7$\downarrow$) sector. }
\label{fig:SigReIm:ED}
\end{figure}

Fig.~\ref{fig:SigReIm:ED} shows the $\mathbf{k}$-dependent self-energy from exact diagonalization for of 4$\times$4 cluster for $\delta=$0.125 (7$\uparrow$,7$\downarrow$) with local and first three non-local elements of $\Sigma_{ij}(\nu=\pi T)$ in the real space. We cut the more long-range elements of  $\Sigma_{ij}(\nu)$ in the Fourier transform due to periodic boundary condition. The general shape of the self-energy
agree well with results of plaquette DF-perturbation (Fig.~\ref{fig:SigReIm}). 

\section{Hole-hole correlation in  $4\times4$ plaquette}

To investigate the important issue of the hole-hole interaction in the lattice, we use two approaches to the notion of a hole in the $4\times4$ plaquette. The first one is to consider a hole as an absence of electrons, {\it i.e.}, an empty site is viewed as a hole. The hole density operator at site $i$ is then given by
\begin{equation}
    n^h_{i}=(1-n_{i\uparrow})(1-n_{i\downarrow}).
\end{equation}
We then investigate the hole-hole correlation function $\langle n^{h}_{i}n^{h}_{j}\rangle-\langle n^h_i\rangle \langle n^h_j\rangle $ in the ground state of the $(7{\uparrow},7{\downarrow})$ sector as a function of the displacement $i-j$. The average hole density $\langle n^h_i\rangle$ is obviously given by $\langle n^h_i\rangle=1-\langle n\rangle+d$, where $\langle n\rangle$ is the average electron density, which in the given sector is just $7/8=0.875$, and $d=\langle n_{i\uparrow}n_{i\downarrow} \rangle$ is the double occupancy. The results for two different values of $t'$ are shown in Fig.~\ref{fig:MiMj0}.

The analysis of the hole-hole correlation function in this sector shows that the two holes occupy two different plaquettes in  4$\times$4 cluster with a slight tendency towards  ``diagonal'' plaquettes, or in other words the holes prefer to be as far from each other as the system permits.
Energetically, this configuration of the holes makes the $t'$ hopping along the diagonals very efficient (see also Fig.~\ref{fig:PGschematic} (right)). Thus, it is the kinetic energy associated with $t'$ which drives such a strong hole-condensation for this concentration $\delta=0.125$ (2 holes on 16 sites), which is not far from the optimal hole concentration for cuprates.

The second approach is in the spirit of Landau's Fermi liquid theory. The hole is then viewed as the result of an annihilation operator acting on the half-filled ($(8{\uparrow},8{\downarrow})$ sector) ground state $|\psi_{16;0}\rangle$. Correspondingly a state with two holes would be the result of two annihilation operators acting on that state: $|\psi_{14;ij}\rangle=A_{ij}c_{i\uparrow}c_{j\downarrow}|\psi_{16;0}\rangle$, with $A_{ij}$ being the normalization factor chosen in such a way that the norm of this state is unity. Then we calculate the overlap $C_{ij}$ between the ground state of the $(7_{\uparrow},7_{\downarrow})$ sector $|\psi_{14;0}$ and $|\psi_{14;ij}\rangle$ to see how well the two-hole state describes the true ground state. The results are shown in Fig.~\ref{fig:cicj}. Here we have to understand that for $t'=0$ the ground state of the $4\times4$ plaquette is 3 fold degenerate. This is an accidental degeneracy that occurs because the $4\times4$ periodic lattice without $t'$ is equivalent to a $2\times2\times2\times2$ hypercubic lattice\cite{Dagotto_2222}. This accidental degeneracy is unphysical in the sense that it is absent in larger two-dimensional clusters. Other effects of this property one can see on the left panel of fig.~\ref{fig:MiMj0} noting that observables when $i$ and $j$ are nearest neighbors along the diagonal are identical with those when $i$ and $j$ are next nearest neighbors along the horizontal or vertical direction. On a 4-dimensional hypercube those pairs of sites are equivalent.

Due to this degeneracy and to the fact that $C_{ij}$ is not an observable we have to take the results for $t'=0$ with a grain of salt. The results are obviously dependent on the linear combination of the three ground states we choose to calculate the overlap (fig.~\ref{fig:MiMj0} shows one such combination produced randomly by the ED solver). Still, we can see that $C_{ij}$ tends to be largest if $i$ and $j$ have different spins on the N\'eel state. This is a clear indication that antiferromagnetic fluctuations are well preserved in the $(7{\uparrow},7{\downarrow})$ sector with no NNN hopping. On the other hand for $t'=0.3$ the largest overlap is found for the pairs (1,3) and (1,11), sites being numbered from 1 to 16 from left to right and then from top to bottom, in agreement with the understanding that large NNN hopping completely destroys the antiferromagnetic order.

Another interesting observation arises when we calculate the sum $\sum_{ij}C_{ij}^2$ for different values of $t'$. This value shows how well the $|\psi_{14;0}\rangle$ state is described in terms of the two holes states $|\psi_{14;ij}\rangle$. It turns out that while for $t'=0$ this value is reasonably large ($1.25$, one should be surprised it is larger than one as the states $\Psi_{14;ij}$ are no orthogonal), for $t'=0.3$ it is very low ($0.0013$). This indicates that the second approach to the notion of hole, in terms of the Fermi liquid theory is hardly appropriate for large $t'$, in other words the holes in that regime are very incoherent.

\section{Lehmann representation for one-particle and  two-particle  Green's functions}

The one-particle Green's function for a finite  fermionic system with time-independent Hamiltonian and many body spectrum $\hat H |i \rangle = E_i |i\rangle $ has the following Lehmann representation in the Matsubara space:
\begin{eqnarray*}
    g^{\sigma}_{12}(\nu) = \frac{1}{Z}
        \sum_{ij}\frac{\langle i|\hat c_{1\sigma}|j\rangle\langle j|\hat c^+_{2\sigma}|i\rangle}
            {i\nu +E_i-E_j }(e^{-\beta E_i} + e^{-\beta E_j})
\end{eqnarray*}
where $Z = \sum_i e^{-\beta E_i}$.

For the  two-particle  Green's function (2PGF) we introduce first four "auxilary" fermionic frequencies ($\omega_1 \div \omega_4$ ) and define 2PGF in Matsubara space as following\cite{Superpert}:

\begin{eqnarray}
\kappa^{\sigma\sigma'}_{1234}({\omega_1 \omega_2 \omega_3})=\frac{1}{\beta^2}\int_0^\beta\! d\tau_1
\int_0^\beta\! d\tau_2 \int_0^\beta\! d\tau_3\, e^{i ( \omega_1\tau_1 +\omega\tau_2+\omega_3\tau_3)}
\langle T_\tau c_{1\sigma}(\tau_1)c_{2\sigma'}(\tau_2)c^\dagger_{4\sigma'}(\tau_3)c^\dagger_{3\sigma}(0)\rangle\, \ .
\label{eqn::chi_omega}
\end{eqnarray}

Here time translation invariance of the imaginary time 2PGF has been used. Note that here the frequencies in the exponential corresponding to annihilation and creation operators have the same sign in contrast to the usual definition for the Fourier transform. Correspondingly, energy conservation requires $\omega_1+\omega_2+\omega_3+\omega_4=0$. By restricting the range of integration such that time ordering is explicit, one obtains $3!$ different terms. These can be brought into the same form by permuting the operators \emph{and} corresponding frequencies. By the anticommutation relations, each term picks up the sign of the permutation. After introducing the sum over eigenstates, the 2PGF can be written as

\begin{eqnarray}
{\kappa^{\sigma\sigma'}_{1234}({\omega_1 \omega_2 \omega_3})=\frac{1}{Z}\sum_{ijkl}\sum_{\Pi}\! \phi (E_i,E_j,E_k,E_l,\omega_{\Pi_1},\omega_{\Pi_2},\omega_{\Pi_4})}
\sgn(\Pi) \langle i|\mathcal{O}_{\Pi_1}|j\rangle\,\langle j|\mathcal{O}_{\Pi_2}|k\rangle\,\langle k|\mathcal{O}_{\Pi_4}|l\rangle\,\langle l|c^\dagger_{3\sigma}|i\rangle\ 
\end{eqnarray}

where the first sum is over the eigenstates and the second over all permutations $\Pi$ of the indices $\{123\}$. We further have defined $\mathcal{O}_1=c_{1\sigma}$,  $\mathcal{O}_2=c_{2\sigma'}$   and $\mathcal{O}_4=c^\dagger_{4\sigma'}$  and e.g. $\Pi_1$ denotes the permutation of the first index. Here the different choice of convention for the Fourier transform simplifies the notation since otherwise the sign of the frequency associated with the creation operator would have to be permuted.
The function $\phi$ is given by the integral
\begin{eqnarray}
\phi(E_i,E_j,E_k,E_l,\omega_1,\omega_2,\omega_3) =\int_0^\beta\! d\tau_1  \int_0^{\tau_1}\! d\tau_2 \int_0^{\tau_2}\! d\tau_3\, e^{-\beta E_i+{(E_i-E_j)\tau_1} +(E_j-E_k)\tau_2 +{(E_k-E_l)\tau_3} +{i(\omega_1\tau_1+\omega_2\tau_2+\omega_3\tau_3)} }
\end{eqnarray}
The latter expression can be evaluated by taking care of the delta functions that arise from equal energies, with the final result\cite{Superpert}:

\begin{eqnarray}
&&\phi (E_i,E_j,E_k,E_l,\omega_1,\omega_2,\omega_3)=
\frac{1}{i\omega_3+E_k-E_l} \times\nonumber \\
&&\left[\frac{1-\delta_{\omega_2,-\omega_3}\delta_{E_j,E_l}}{i(\omega_2+\omega_3)+E_j-E_l}\left(
\frac{e^{-\beta E_i}+e^{-\beta E_j}}{i\omega_1+E_i-E_j}-
\frac{e^{-\beta E_i}+e^{-\beta E_l}}{i(\omega_1+\omega_2
+ \omega_3)+E_i-E_l}\right) \right.\nonumber \\
&&\left. +\delta_{\omega_2,-\omega_3}\delta_{E_j,E_l}
\left(\frac{e^{-\beta E_i}+e^{-\beta E_j}}{\left( i\omega_1+E_i-E_j\right)^2}-\beta\frac{e^{-\beta E_j}}{i\omega_1+E_i-E_j}\right) 
-\frac1{i\omega_2+E_j-E_k}  \times \right.\nonumber\\  
&&\left. \left( \frac{e^{-\beta E_i}+e^{-\beta E_j}}{i\omega_1+E_i-E_j}-\left( 1-\delta
_{\omega_1,-\omega_2}\delta_{E_i,E_k}\right)
\frac{e^{-\beta E_i}-e^{-\beta E_k}}{i(\omega_1+\omega
_2)+E_i-E_k}+\beta e^{-\beta E_i}\delta_{\omega
_1,-\omega_2}\delta_{E_i,E_k}\right)\right]\ .
\label{eqn::2piLehmann}
\end{eqnarray}
Finally connection to standard Matsubara frequencies in 2PGF used in this work is as follows:
\begin{equation}
\omega_1=\nu, \quad \omega_2=\omega-\nu, \quad \omega_3=\omega - \nu'
\end{equation}

There is an efficient open source implementation \texttt{pomerol}~\cite{pomerol} for extracting the two-particle Green's function in ED.

\begin{table}[h!]
\centering
\begin{tabular}{||c|| c| c| c| c| c||} 
 \hline
 \# States & 256 & 22 & 13 & 9 & 6 \\ [0.5ex] 
 \hline\hline
 $\kappa^P_{0110}(\nu _{0} , -\nu _{0},\omega=0) $ &10.75 & 10.51 & 9.62 &9.58 &8.02 \\ [1ex] 
 \hline
\end{tabular}
\caption{The main non-local contribution to two-particle plaquette Green function at the degenerate point for PP singlet channel as function of number of included states in the Lehmann representation for $\beta=10$. See Fig.~\ref{fig:DOS_CPT} for other parameters.}
\label{table}
\end{table}
Using the Lehmann representation Eq.~(\ref{eqn::chi_omega}) we can analyze effects of the 6-fold degenerate ground states for the "critical" point on the plaquette on the values of the two-particle Green function. In the 4-fold sum over all many body states $ijkl$
one can reduce only one summation using the low-temperature limit related with a cutoff due to the Boltzmann factors in Eq.~(\ref{eqn::2piLehmann}).
In order to understand effects of ground state degeneracy, we can formally reduce
sum over ${\it all}$ many body states over the lowest N-states.
In the Table ~\ref{table} we presented the main non-local contribution to two-particle plaquette Green function for PP singlet channel $\kappa^P_{1221}(\nu _{0} , -\nu _{0},\omega=0)$ as function of number of included states in the Lehmann representation. If we include all $256$ states for $\beta=10$ in the $2 \times 2$ plaquette then we get exactly the value of particle-particle singlet susceptibility ($10.75$) in the Fig. ~(\ref{fig:Xi1221}) for this inverse temperature. We note that there is no disconnected part and susceptibility is equal to $\kappa^P$ in this channel. It is interesting than using only the 6 degenerate ground states states gives about $75\%$ of the total value for $\kappa^P_{1221}(\nu _{0} , -\nu _{0},\omega=0)$ which is a "main contribution" for $d_{x^2-y^2}$ superconductivity.
Moreover taking into account only 3 additional states at the energy $E_n=0.42$ from $N=4$ triplet, the summation over 9 low-energy many-body states in plaquette gives
about $90\%$ of the total values. This test clearly shows importance of the ground states degeneracy in the special point of plaquette with $\delta=0.25$ for divergence of renormalized vertex for lower temperature in the dual-fermion perturbation theory.

\end{document}